\documentclass[twocolumn]{aastex6}
\usepackage[varg]{txfonts}
\usepackage{graphicx}

\newcommand\aastex{AAS\TeX}

\shorttitle{\aastex\ Quiescent and active phases in Be stars}
\shortauthors{Granada et al.}
\begin{document}
\title{Quiescent and active phases in Be stars: A WISE snapshot of young galactic open clusters}
\author{A. Granada\altaffilmark{1,2}, C. E. Jones\altaffilmark{1}, T. A. A. Sigut\altaffilmark{1}, T. Semaan\altaffilmark{3}, C. Georgy
\altaffilmark{3}, G. Meynet\altaffilmark{3} and S. Ekstr\"om\altaffilmark{3}}
\email{granada@fcaglp.unlp.edu.ar}
\altaffiltext{1}{Department of Physics and Astronomy, The University of Western Ontario, London, ON Canada N6A 3K7.}
\altaffiltext{2}{Now at Instituto de Astrof\'isica La Plata, CCT La Plata, CONICET-UNLP, Paseo del Bosque S/N, B1900FWA La Plata, Argentina.}
\altaffiltext{3}{Geneva Observatory, University of Geneva, Maillettes 51, CH-1290 Sauverny, Switzerland.}
\begin{abstract}
Through the modelling of near-infrared photometry of star plus disk systems with the codes {\sc bedisk}/{\sc beray}, we successfully describe the WISE photometric characteristics of Be stars in five young open clusters, NGC 663, NGC 869, NGC 884, NGC 3766 and NGC 4755, broadly studied in the literature.
WISE photometry allows previously known Be stars to be detected and to find new Be candidates which could be confirmed spectroscopically. 
The location of Be stars in the WISE colour-magnitude diagram, separates them in two groups; active (Be stars hosting a developed circumstellar disk) and quiescent objects (Be stars in a diskless phase), and this way, we can explore how often stars are observed in these different stages. The variability observed in most active variable Be stars is compatible with a disk dissipation phase. We find that 50 percent of Be stars in the studied open clusters are in an active phase. We can interpret this as Be stars having a developed circumstellar disk half of the time. 
The location of Be stars with a developed disk in the CMD require mass loss rates in agreement with values recently reported in the literature. For these objects, we expect to have a tight relation between the equivalent width of the H$\alpha$ line and the mass of the disk, if the inclination is known. Also, near-infrared photometry of Be stars in stellar clusters has the potential of being useful to test whether there is a preferential viewing angle.
\end{abstract}
\section{Introduction}\label{Intro}
Be stars as a group are among the most rapidly rotating stars, both in terms of their spin angular velocity rate ($\Omega_{\rm eq}/\Omega_{\rm crit}$) and their equatorial velocity ($V_{\rm eq}$). 
The defining observed characteristics of these main-sequence (MS), intermediate-mass stars include the presence of hydrogen and metallic lines in emission in their spectra, an infrared colour excess, as well as short and long-term photometric and spectroscopic variability. These observed characteristics, together with radio observations, polarimetric signatures and interferometric data, can be globally explained within the framework of the viscous decretion disk model \citep[for a recent review on this subject, we refer the reader to][]{Rivinius2013}. However, despite all the progress in understanding the characteristics of Be stars and their circumstellar disks, the underlying mechanism(s) triggering the formation of such a circumstellar envelope remains elusive.

 According to \citet{Sana2012},  over 70$\%$ of stars with masses larger than 8~M$_{\sun}$  exchange mass with a companion during some period of their evolution. Therefore a significant number of binaries is expected among B stars, in particular among earlier spectral types. For instance, the bimodal distribution of rotational velocities for the single early B-type stars found in the 30 Doradus region by \citet{Dufton2013}, could be partly due to evolutionary effects related to binarity. An incidence of 30$\%$ of binarity among Be stars was found by \citet{Oudmaijer2010}, consistent with the incidence among normal, MS B stars. According to these authors, binarity may not be a key aspect involved in the appearance of the Be phenomenon; however, when the companion is close enough to the Be star, it could affect the circumstellar disk, for instance,  by truncating it or triggering disk oscillations  \citep[e.g.][]{Okazaki2002,Oktariani2016}. In other binary systems, episodes of mass transfer, or even mergers, could lead to the formation of a rapidly rotating star that could potentially become a Be star.   

Understanding how the existence of Be stars depends on metallicity, spectral type, and evolutionary stage can help to understand the mechanism(s) involved in the appearance of the Be phenomenon. Open clusters constitute ideal laboratories to study the conditions in which Be stars form and evolve. We can assume that cluster stars come from the same primordial cloud and share a common spatial location, proper motions, initial chemical composition, and age. 

It has been long known that the detected fraction of Be stars with colour excess increases with wavelength \citep[e.g.][]{Dougherty1994}, 
as expected if the excess emission comes from free-free and bound-free processes occurring in their circumstellar disks. That is why 
the near-infrared (near-IR) spectral regions are particularly useful in detecting and confirming Be stars. IR surveys such as {\bf \it Spitzer} (the fourth and final of the NASA Great Observatories program, an infrared space telescope launched in 2003) and AKARI (an infrared astronomy satellite developed by Japan Aerospace Exploration Agency, in cooperation with institutes of Europe and Korea, launched in 2006)
have shown that the near-IR spectral region allows photometric detection and confirmation of Be stars \citep{Ita2010,Bonanos2010}.

The Wide-Field Infrared Survey Explorer \citep[WISE,][]{Wright2010}, which surveyed the sky in four bandpasses between 3.4\,$\mu$m and 22\,$\mu$m, provides a better understanding of the infrared sky. The AllWISE source catalogue \citep{Cutri2013} gives observations with good angular resolution which are suitable to study the Be stellar population in open clusters.

In the present article, we use the IR photometry provided by the AllWISE source catalogue \citep{Cutri2013} for a group of five open clusters with ages between 10 and 30 Myr, well known for hosting Be stars 
and which have been extensively studied in the literature.
We explore how these cluster Be stars are distributed in the WISE colour-magnitude diagram (CMD) 
and investigate whether the location of Be stars in these plots can provide global information on the characteristics of the circumstellar disks and activity cycles for these stars.
In order to interpret the observed near-IR characteristics of Be stars in open clusters, we generate a grid of synthetic WISE magnitudes and colours for star-plus-disk systems using the {\sc beray} code \citep{Sigut2011a}.

This paper is organized as follows: in Section~2 we describe the observations and selection method. Section~3 presents our results, Section~4 describes our disk model predictions, and conclusions are provided in Section~5.

\section{WISE photometry of young open clusters}

\subsection{WISE observations}

For the present work, we use data available at the AllWISE source catalogue \citep{Cutri2013}. This program extended the work of the successful WISE survey \citep{Wright2010}, by combining data from different phases of the mission.
AllWISE provides astrometry and photometry in four bandpasses in 3.4$\mu$m (W1), 4.6$\mu$m (W2), 12$\mu$m (W3) and 22$\mu$m (W4), for nearly 750 million objects, with a better sensitivity than the WISE All-Sky Release Catalogue. Faint source flux biases were corrected, and a more robust estimation of the background level was obtained. 

The angular resolution in  W1, W2, W3 and W4 bands are 6.1", 6.4", 6.5", and 12.0", respectively \citep{Wright2010}. As described in the AllWISE Data Processing documents, the AllWISE Source Catalog is intended to be a highly reliable and complete set of single, unique detections for compact objects on the sky, so targets in severely crowded regions are not considered in the AllWISE Source Catalogue. That is why in the present study, even though we might miss some targets, we are not affected by severely crowded regions.
 
For the W1 and W2 filters the saturation limits are W1$=$8 mag and W2$=$7 mag, and the limiting magnitudes, for which the background level becomes an issue in crowded cluster regions, occurs at W1$=$14 mag and W2$=$13.5 \citep{Cutri2013}.
Therefore, it is important for our work to select clusters
in which MS B stars have a brightness below the saturation limit and above the limiting magnitude. This way, we ensure that most of the cluster members have small errors in the W1 and W2 bands. It is worth stating that some targets located in rather crowded cluster regions, or with a nearby bright star, could suffer from a poor background determination, which may lead to an underestimation of the brightness and larger error bars. 

Most stars with brightness between the saturation limit and background level in the W1 and W2 bands, have a brightness that is fainter than the background sky level in the W4 band. That is why we focus only in the W1 and W2 data and, eventually, the W3 data for those objects with good quality observations.

\subsection{The selected sample of open clusters}

As we are interested in studying the general behaviour of Be stars, we selected five Galactic, young, open clusters broadly studied in the literature: NGC 663, NGC 3766, NGC 4755, and the double cluster NGC 869--NGC 884. Not only are they known for being particularly rich in Be stars, but also the B type stars within these clusters 
have a brightness below the saturation limit and above the limiting magnitude as discussed in the previous section. 

Even though various authors give different values for the cluster parameters (e.g. \citet{Phelps1994, Pigulski2001, Fabregat2005} for NGC 663, \citet{Slesnick2002, Keller2001, Marco2001, Maciejewski2007} for NGC 869 and NGC 884, \citet{Aidelman2012, Piatti1998, McSwain2005b} for NGC 3766 or \citet{Aidelman2012, Balona1994, Sanner2001} for NGC 4755), we chose to use those reported by \citet{Kharchenko2013}, which provides a homogeneous database. For all the clusters under study the parameters given by these authors agree with others in the literature. The names, ages, distance modulus, reddening, and radius assumed for the selected clusters are listed in Table~\ref{TablaCumulos}. 

The estimates of the errors given by \citet{Kharchenko2013} for E(B-V), age, distance and radius are 7\%, 39\%, 11\% and 25\%, respectively. The error estimate in distance of 11\% corresponds to an error of the absolute distance modulus of 0.275 mag.

To select the probable, early-type, cluster MS members, we used the cluster radius given by \citet{Kharchenko2013} and the 2MASS photometry provided together with the WISE photometry \citep{Cutri2013}. We converted the observed J magnitude and (J-H) colour to absolute values using the tabulated distances and E(B-V) excesses of Table \ref{TablaCumulos} and the empirical relations for the extinction A$_{\rm J}$ and excess E(J-H) given by \citet{Yuan2013}. The extinction coefficients given by these authors agree with the average values obtained by \citet{Davenport2014} as determined from 5$\times$10$^5$ stars.

\subsection{Synthetic Populations with SYCLIST}
Even though a detailed cluster parameter determination is beyond the scope of the present article, we generated synthetic stellar populations with SYCLIST, the  Geneva population synthesis code \citep{Georgy2014}, with the aim of determining the regions in the CMD where we expect to have {\bf MS} and red supergiant stars for different cluster ages. We seek to check whether the parameters chosen from the literature are adequate to describe the observations. We decided to build stellar populations (as described in the next paragraph), instead of using typical isochrones, in order to account for the effects of stellar rotation in the evolution, as well as some observational effects.

Synthetic populations of $50000$ stars with masses between 1.7\,M$_{\sun}$ and 15\,M$_{\sun}$ at the ZAMS were created following the typical Salpeter initial mass function (IMF) and the initial rotational velocity distribution given by \citet{Huang2010a}. The inclination angles (angle between the line of sight and the stellar rotation axis) are assumed to follow a random distribution. For these synthetic clusters, a fraction of $30\%$ was adopted for unresolved binaries, which produces broadening in the low mass range where we have mainly equal mass binaries. We considered this value following \citet{Oudmaijer2010}, who found that the incidence of binaries among Be stars to be $30\%$.  We used the colour-effective temperature calibration from \citet{Worthey2011}. Because we do not intend to do detailed model fitting to the observations, we have not introduced artificial errors in magnitude and colour to the stars in our synthetic clusters.

Figures~\ref{NGC663_2017}, \ref{NGC869_2017}, \ref{NGC884_2017}, \ref{NGC3766_2017} and \ref{NGC4755_2017} (left panels) show the synthetic populations (red dots) corresponding to the ages given by \citet{Kharchenko2013} together with the cluster observations, in the 2MASS M$_J$ versus M$_{\rm J}$-M$_{\rm H}$ CMD. All observed objects within the given cluster radius are indicated as gray crosses. 

For NGC~3766, blue circles correspond to a synthetic population with a younger age than the one provided by \citet{Kharchenko2013}, as has been proposed by different authors for this cluster  \citep{Moitinho1997,Tadross2001,Aidelman2012}, and also appears in the WEBDA\footnote{http://www.univie.ac.at/webda/webda.html}. The younger age is in better agreement with the existence of RSG stars and the large number of Be stars found in this cluster \citep{Granada2013a}. 

 The synthetic populations indicate that objects with J$\le$1.5 and intrinsic colour J-H$<$0.15, correspond to  MS stars of spectral type earlier than A0, and to blue supergiant (BSG) stars.
Then, from all the stars within the given cluster radius, we selected objects within these magnitude and colour limits. By doing this, we removed foreground and background non-cluster members, pre-MS stars \citep[see e.g.][]{Bonatto2006}, as well as red giants and red supergiant (RSG) stars. These objects are plotted as black points in Figures~\ref{NGC663_2017} to \ref{NGC4755_2017} (left panel).

For all the clusters, the region in the CMD occupied by MS stars, indicated with black points, is correctly traced by the models of ages given in Table \ref{TablaCumulos}, including the position of the cluster turn-off and the location of RSG stars.  

\subsection{The selection of WISE B-type stars}

\begin{figure*}
\figurenum{1}
\gridline{\fig{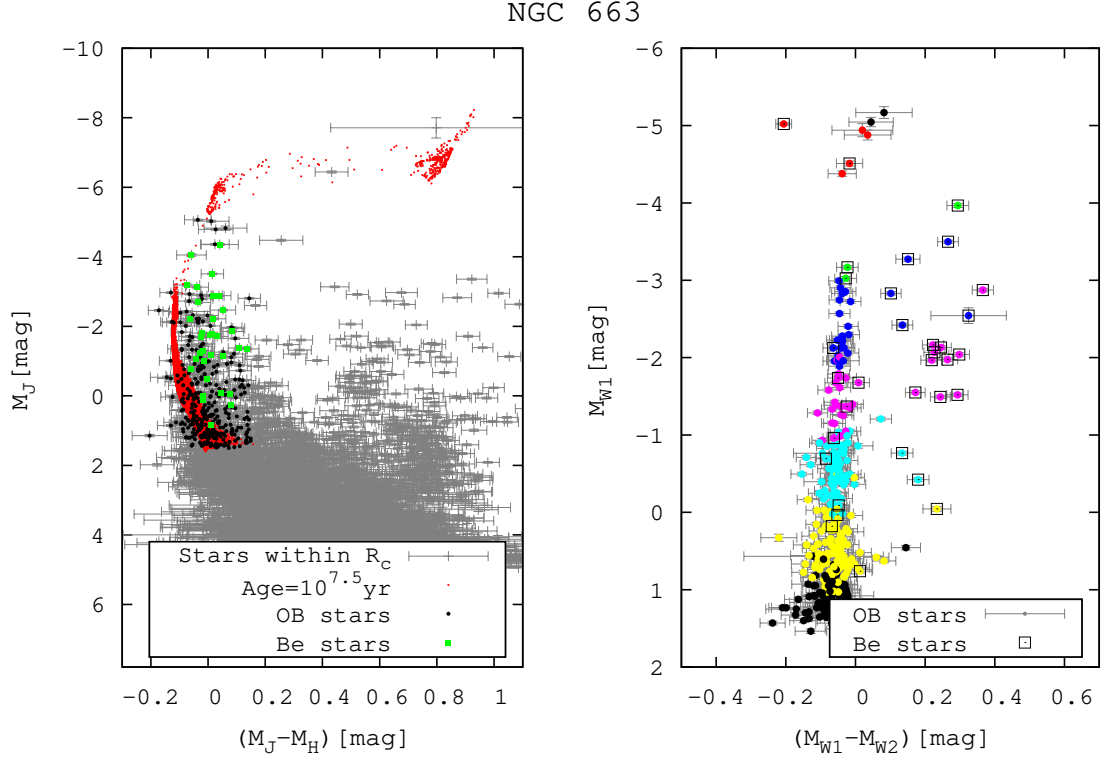}{0.85\textwidth}{}}
\caption{Colour magnitude diagrams of the cluster NGC 663. Left: M$_{\rm J}$ versus M$_{\rm J}$-M$_{\rm H}$ plot. The gray symbols indicate all the objects within the cluster field. Black symbols correspond to cluster early type MS stars and green small squares indicate known Be stars. Red points correspond to a synthetic stellar population computed with SYCLIST. Right: WISE  M$_{\rm W1}$ versus M$_{\rm W1}$-M$_{\rm W2}$ for cluster members. Different colours correspond to different M$_{\rm J}$ magnitudes, a proxy of spectral type. Black open squares indicate known Be stars. \label{NGC663_2017}}
\end{figure*}

\begin{figure*}
\figurenum{2}
\gridline{\fig{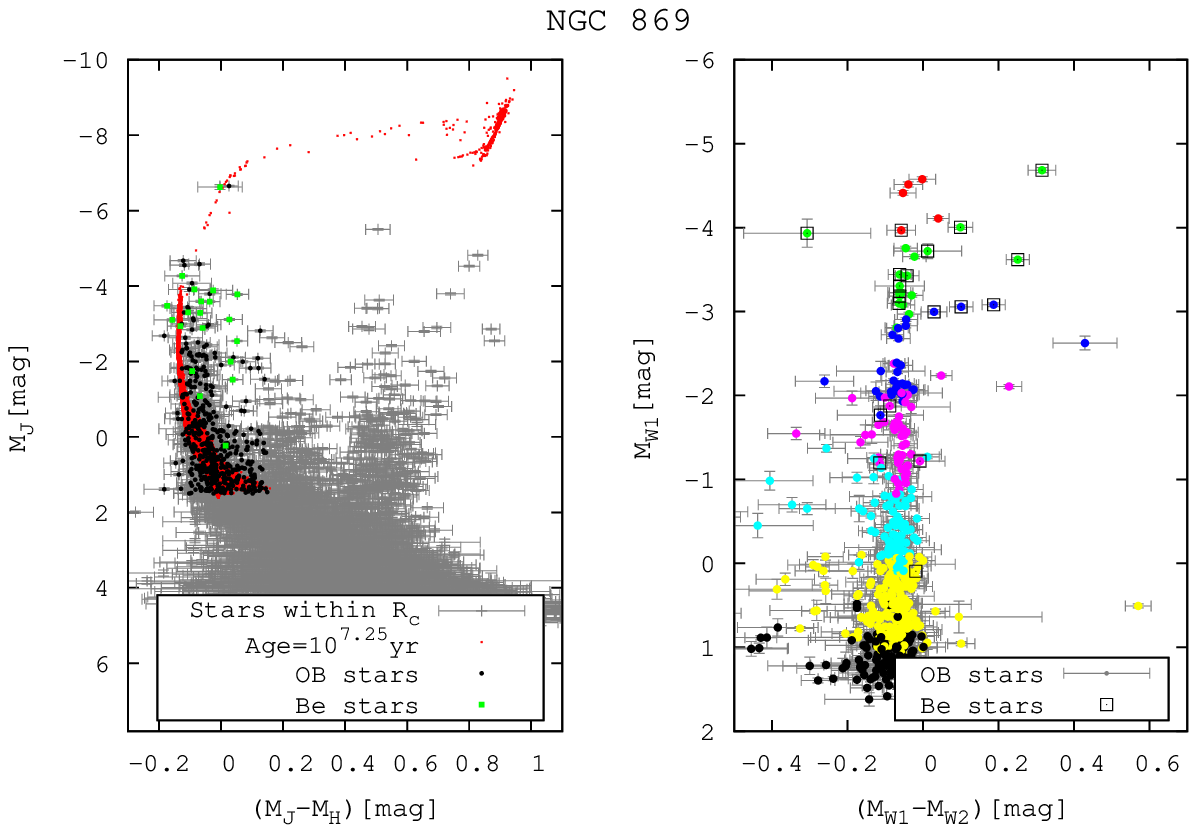}{0.85\textwidth}{}}
\caption{The same as Fig. \ref{NGC663_2017} for the cluster NGC 869. \label{NGC869_2017}}
\end{figure*}

\begin{figure*}
\figurenum{3}
\gridline{\fig{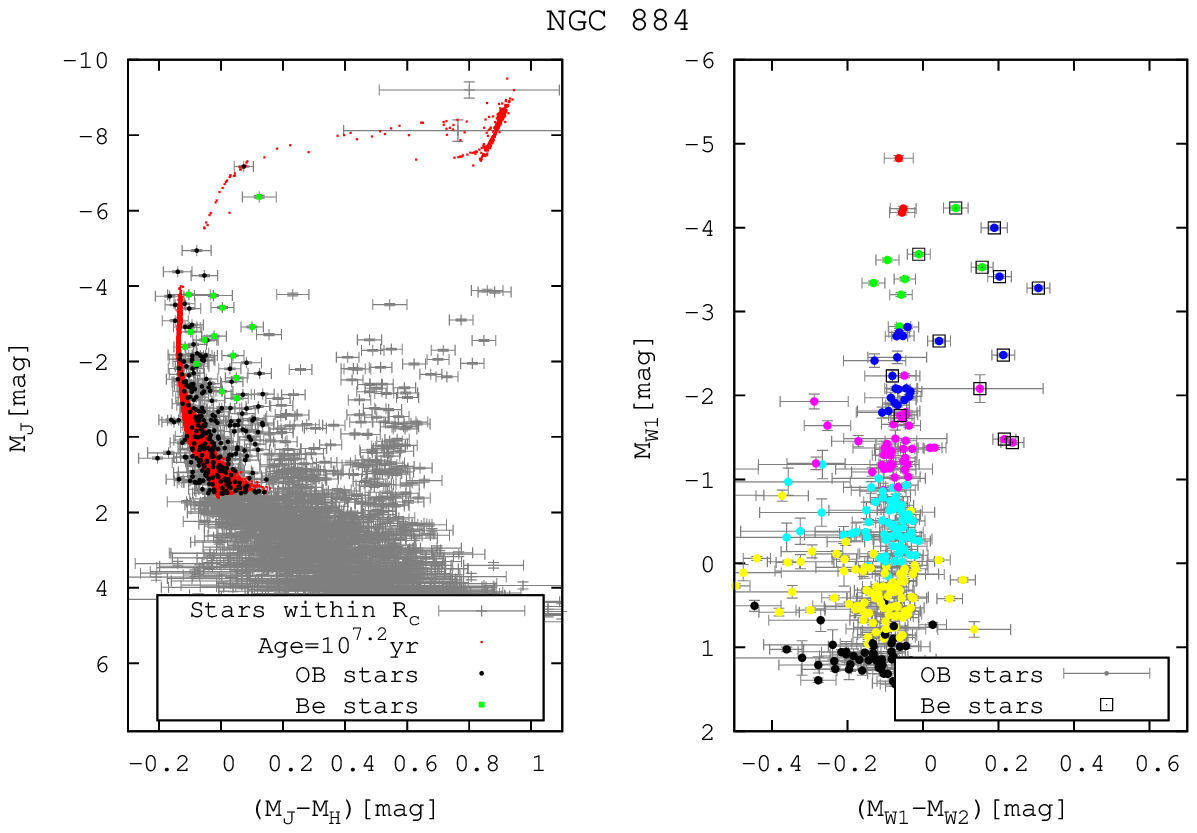}{0.85\textwidth}{}}
\caption{The same as Fig. \ref{NGC663_2017} for the cluster NGC 884 \label{NGC884_2017}}
\end{figure*}

\begin{figure*}
\figurenum{4}
\gridline{\fig{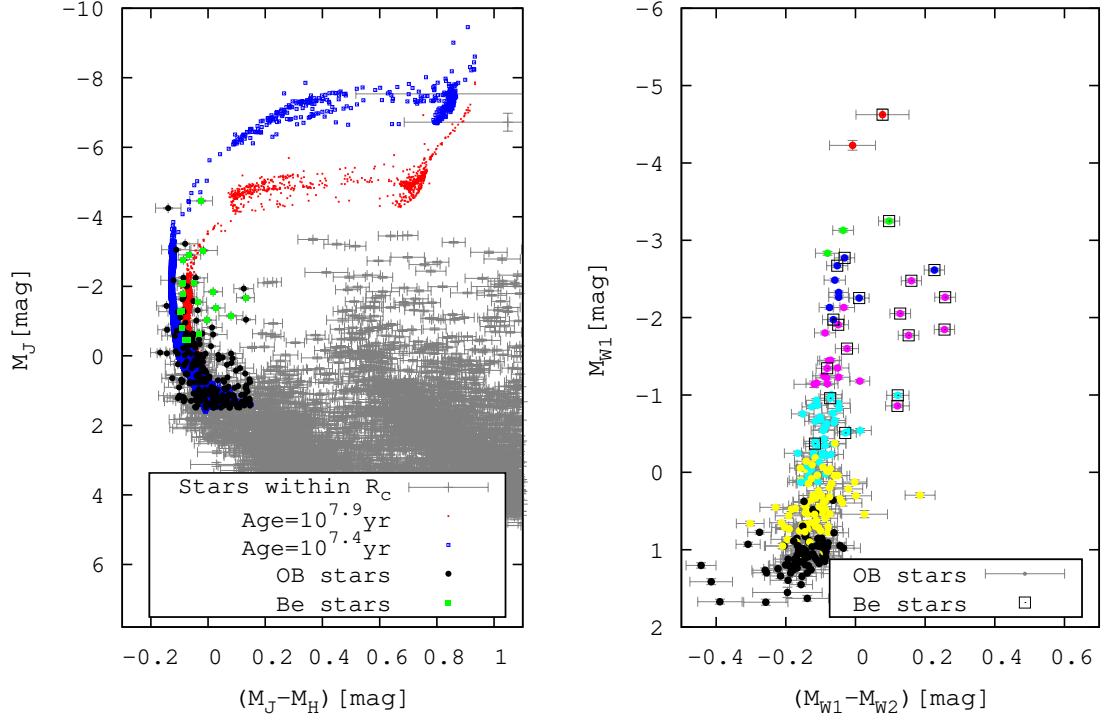}{0.85\textwidth}{}}
\caption{The same as Fig. \ref{NGC663_2017} for the cluster NGC 3766 \label{NGC3766_2017}}
\end{figure*}

\begin{figure*}
\figurenum{5}
\gridline{\fig{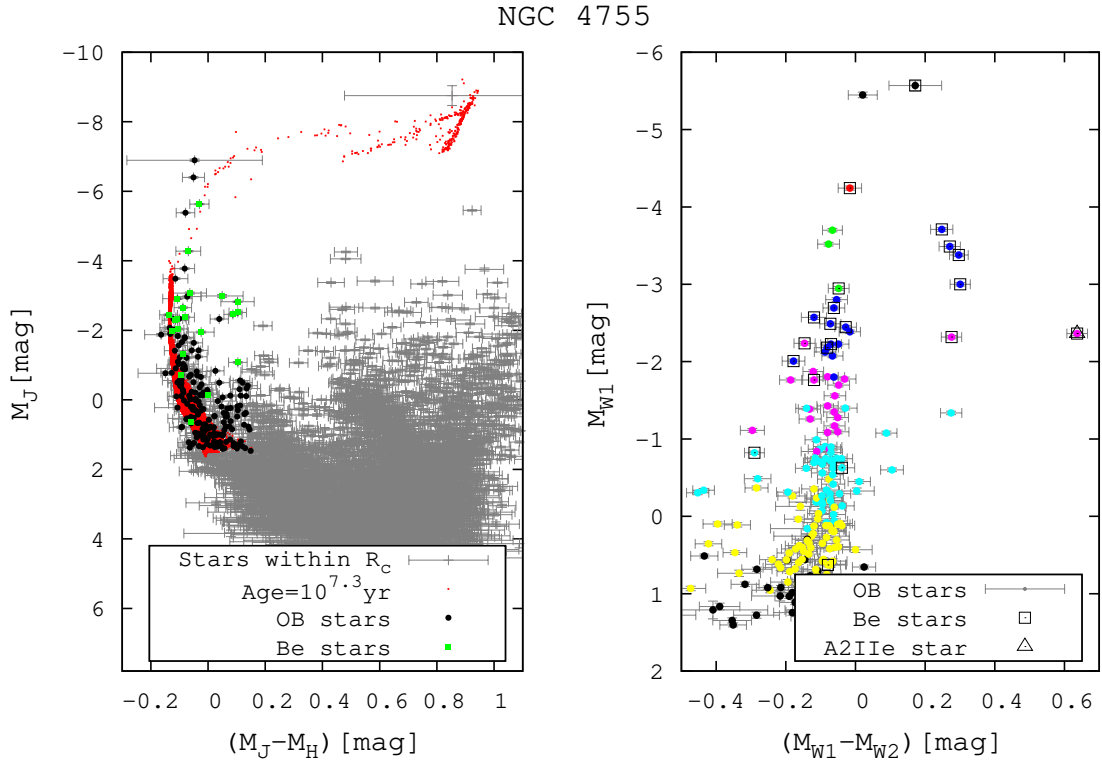}{0.85\textwidth}{}}
\caption{The same as Fig. \ref{NGC663_2017} for the cluster NGC 4755 \label{NGC4755_2017}}
\end{figure*}

As mentioned in the previous subsection, the left panels of Figures~\ref{NGC663_2017} to \ref{NGC4755_2017} show the 2MASS CMD of absolute magnitude M$_J$ versus the intrinsic colour J-H for synthetic clusters and observations. 

Our selection of targets, stars that are likely OB MS cluster members, determined using the procedure described above, are shown as black and green symbols.

For the stars in each cluster, we used the empirical relations given by \citet{Yuan2013} for the extinction $A(W1)=0.19\,E(B-V)$ and colour excess $E(W1-W2)=0.036\,E(B-V)$, as well as the colour excesses and  distance modulus ($\mu_0$) given by \citet{Kharchenko2013} to convert the observed W1 magnitude and W1-W2 colour to absolute magnitudes and intrinsic colours. 

An error of 0.1 mag in E(B-V), larger than the estimates given for the clusters under study, leads to a difference smaller than 0.004\,mag in the determination of W1-W2, while the typical observational error for this colour is larger than 0.02 mag. The effect of such error in colour excess in WISE magnitudes is also small, particularly in comparison to the error of 0.275\,mag introduced in the determination of cluster distances. However, these errors in the determination of $\mu_0$ neither produce significant changes in the  spectral type nor affect the main results of this work regarding the colour excesses of Be stars.

The WISE CMD for the clusters are presented in the right panels of Figures~\ref{NGC663_2017} to \ref{NGC4755_2017}. The colour of the points in these figures indicate different ranges of absolute J magnitude. From our models, we obtain that MS stars with effective temperatures between 10000\,K and 30000\,K, corresponding to the B-type range, have absolute J magnitudes between -4 and 1: we subdivide this range in J magnitude as -4$\leq$ J$<$-3 (green points), -3$\leq$ J$<$-2 (blue points), -2$\leq$ J$<$-1 (magenta points), -1$\leq$J$<$0 (cyan points), and 0$\leq$ J$<$1 (yellow points). Red and black points with J$<$-4 correspond to O type stars and supergiants. Objects with J$>$1 are indicated with black full circles. Open squares indicate the known Be stars. The Be nature of these objects has been previously determined either from spectroscopic observations \citep[e.g.][]{McSwain2005b,McSwain2009a,Marsh2012,Huang2010a,Mathew2011}, or H$\alpha$ narrow band photometry \citep[e.g.][]{Pigulski2001}.
 

\begin{table}[h]
\caption{Selected Open Clusters}             
\label{TablaCumulos}                
\begin{center}
\begin{tabular}{c l c c l }        
\hline\hline                 
Name  &  Age$^*$  &  $\mu_{0}$$^*$   &  E(B-V)$^*$ & R$_{cl}$$^*$ \\
      &[log(yr)]& [mag]      & [mag]   & [arcmin] \\
\hline                        
NGC 663 & 7.50 & 11.61 & 0.700 & 14.4 \\ 
NGC 869 & 7.28 & 11.81 & 0.520 & 14.4$^{\dagger\dagger}$ \\
NGC 884 & 7.20 & 11.85 & 0.560 & 10.5$^{\dagger\dagger}$ \\
NGC 3766& 7.40$^{\dagger}$ & 11.13 & 0.208 & 13.8 \\
NGC 4755& 7.30 & 11.47 & 0.396 & 13.5 \\
\hline\hline                            
\end{tabular}\\
\end{center}

{\tiny $*$ \citet{Kharchenko2013} estimate errors for E(B-V), age, distance and radius of 7\%, 39\%, 11\% and 25\%, respectively. The error estimate in distance of 11\% corresponds to an error of the absolute distance modulus of 0.275 mag.}

{\tiny $\dagger$ This cluster age, that adequately describes the cluster turnoff and presence of red supergiant stars using SYCLIST, was not taken from \citet{Kharchenko2013}.}

{\tiny $\dagger\dagger$ The cluster radii considered for these two clusters correspond to the angular radius of the central part (R1) by \citet{Kharchenko2013}, instead of the angular radius of the cluster (R2). This is to avoid cluster overlapping.}

\end{table} 

\begin{figure*}
\figurenum{6}
\gridline{\fig{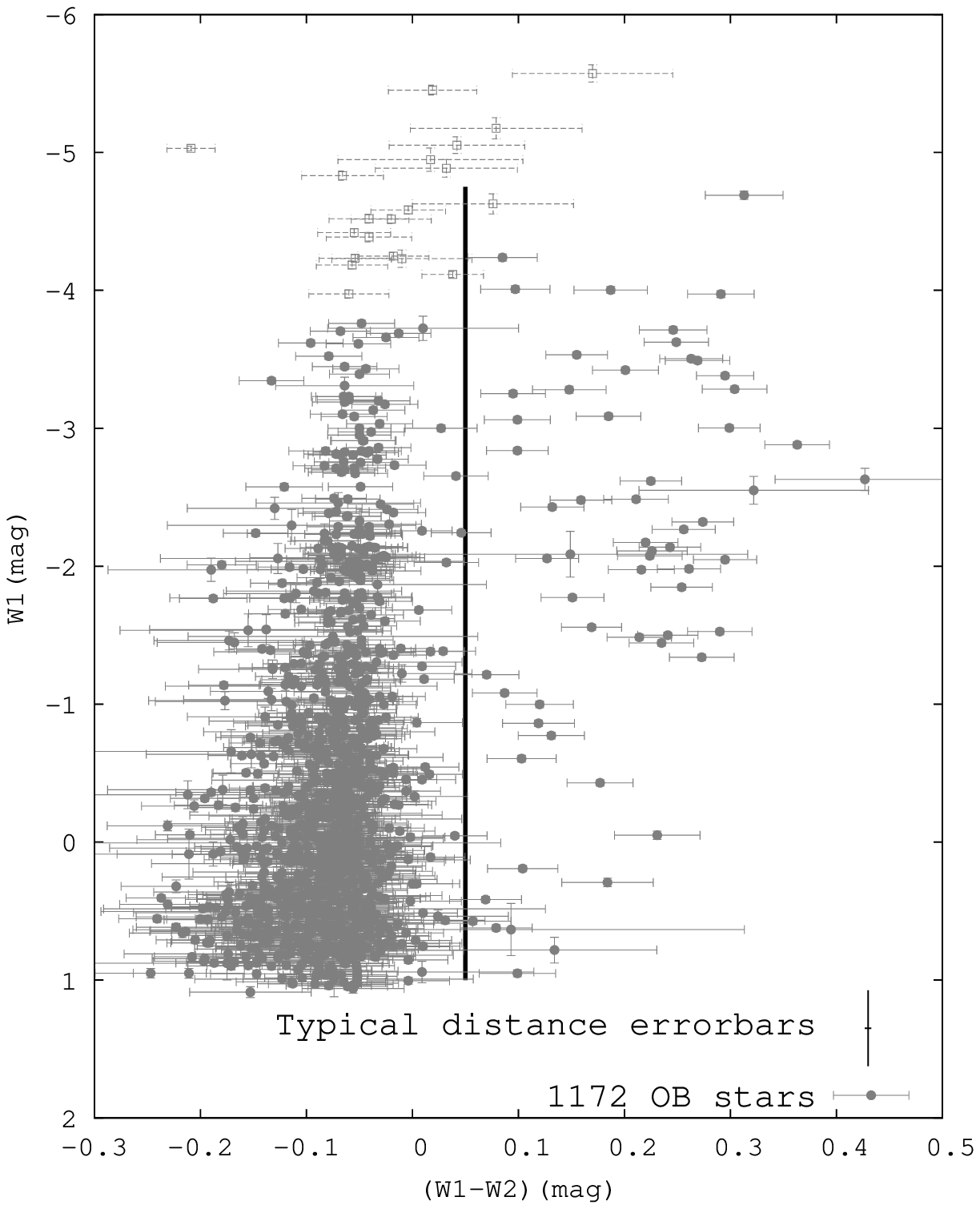}{0.465\textwidth}{(a)}
          \fig{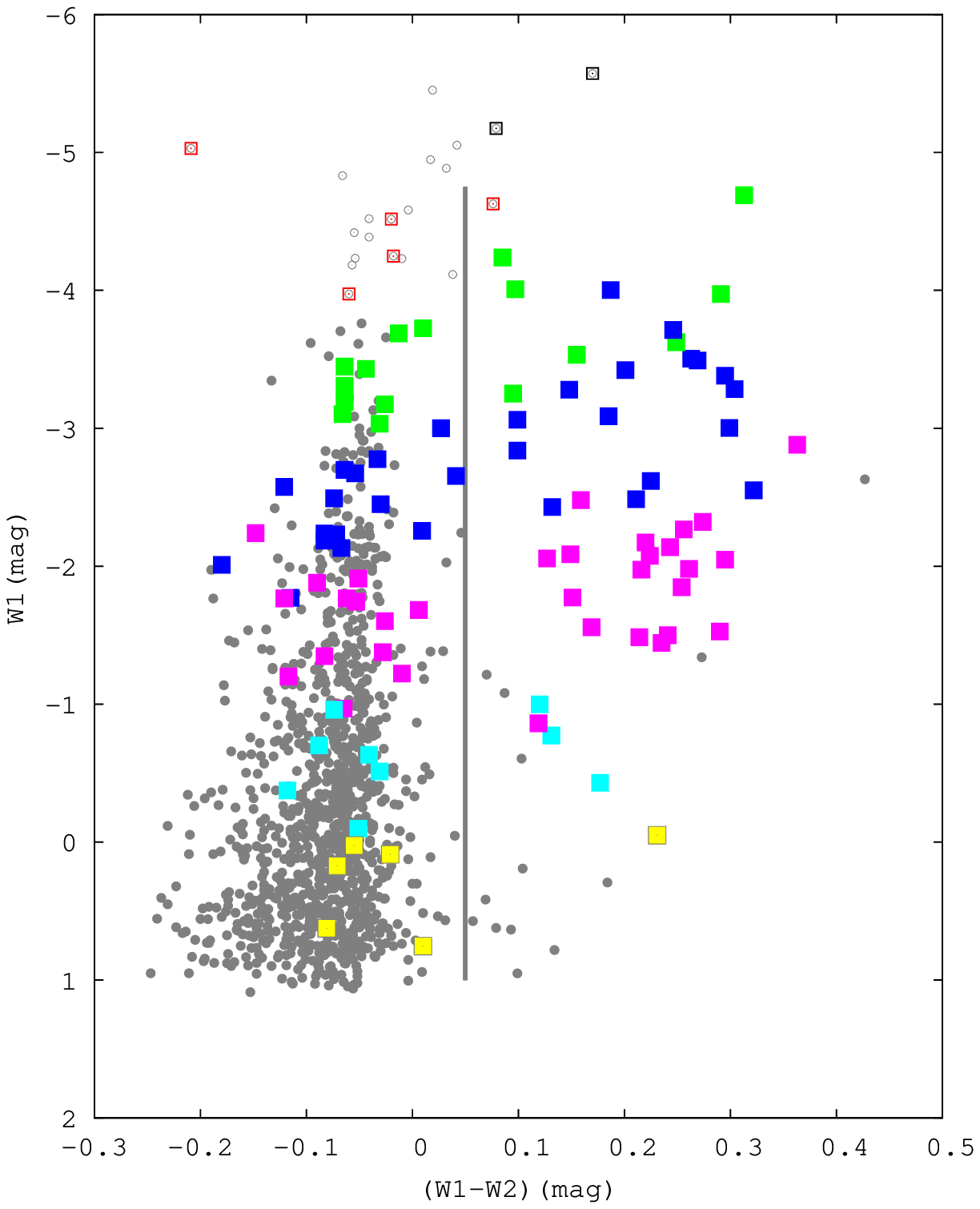}{0.465\textwidth}{(b)}
          }
\caption{Colour magnitude diagram of the 5 clusters. a) All datapoints are shown with error bars and the vertical line indicates the colour limit for {\it naked} stars ; b) All known Be stars are indicated with coloured squares with the different colours representing different absolute 
J magnitudes, an indication of spectral type.
\label{5Clusters_a}}
\end{figure*}

The WISE CMD using all OB stars from the five open clusters (with absolute magnitude J$<$1 and intrinsic colour J-H$<$0.15) is presented in Figure \ref{5Clusters_a}. Due to the low quality of the data of most objects with  W1-W2$<$-0.25, we removed stars with W1-W2 colour beyond this limit, and this is why the B star sample is incomplete, particularly towards later spectral types. Stars with W1-W2$\geq$0.5 are likely not MS B stars, but Class II young stellar objects \citep{Koenig2012}, so we removed them from the sample of B stars as well. Panel a) of Figure \ref{5Clusters_a} shows all the stars with gray points with errorbars, and panel b) highlights stars with a previous Be classification with coloured squares. Full squares indicate stars of B spectral type (yellow, cyan, magenta, blue and green points) and open squares correspond to earlier-type emission line stars (red and black). This sample of B and Be stars will be analyzed in the following section. 

Table \ref{BeStars} lists the 95 stars studied in this work with a previous Be classification. Their coordinates are tabulated in columns 1 and 2, their WISE and 2MASS intrinsic magnitudes, colours and errors are listed in columns 3 to 14, a number indicating the WISE variability of the star in column 15 (0 where no variability flag could be assigned, 1 for stars that are stable, 2 for those without a clear variability signature and 3 for stars that are variable). The spectral type (0 corresponds to B0, 1 to B1 and so on) and luminosity class are indicated in columns 16 and 17, respectively, column 18 gives the name of the object as available in SIMBAD together with references relevant to the Be star classification of the object. In the last column, we give the Be class assigned to the object in this work, either early, mid or late Be star.

\onecolumngrid
\newpage
\begin{deluxetable*}{ c  c  c c c c c c c c c c c c c c c p{3.5cm} c  } 
\tabletypesize{\tiny}
\tablewidth{0pt}
\tablecolumns{19}
\tablecaption{95 Be Stars analyzed in this work.\label{BeStars}}
\tablehead{
    \colhead{RA} & \colhead{declination} & \colhead{M$_{\rm W1}$} & \colhead{eM$_{\rm W1}$} & \colhead{M$_{\rm W12}$}&\colhead{eM$_{\rm W12}$}&\colhead{M$_{\rm W23}$} &\colhead{eM$_{\rm W23}$}&\colhead{M$_{\rm J}$}&\colhead{eM$_{\rm J}$} &  \colhead{M$_{\rm H}$}&\colhead{eM$_{\rm H}$} &  \colhead{M$_{\rm K}$}&\colhead{eM$_{\rm K}$} &  \colhead{Var} &\colhead{ST} &\colhead{LC} & \colhead{SIMBAD name $\&$ remarks} &\colhead{Be class} }
\startdata
($^o$)&($^o$)&(mag)&(mag)&(mag)&(mag)&(mag)&(mag)&(mag)&(mag)&(mag)&(mag)&(mag)&(mag)&&&&&\\\hline
26.525441	&	61.227595	&	-2.55	&	0.10	&	0.32	&	0.11	&	1.21	&	0.05	&	-2.47	&	0.03	&	-2.52	&	0.04	&	-2.78	&	0.02	&	0	&	0.5	&	5	&	EM* VES 616,  $^{(a)}$	&	early\\
26.558441	&	61.228838	&	-0.77	&	0.02	&	0.13	&	0.03	&	0.50	&	-	&	-0.48	&	0.02	&	-0.48	&	0.03	&	-0.66	&	0.02	&	2	&	5	&	5	&	V* V979 Cas,  $^{(a)}$	&	late\\
26.508573	&	61.250567	&	-0.05	&	0.03	&	0.23	&	0.04	&	1.52	&	0.10	&	0.28	&	0.04	&	0.19	&	0.03	&	0.05	&	0.03	&	0	&	5	&	5	&	NGC 663 84,  $^{(a)}$ *	&	late\\
26.584334	&	61.239349	&	-1.56	&	0.02	&	0.17	&	0.03	&	1.07	&	0.06	&	-1.71	&	0.02	&	-1.74	&	0.03	&	-1.90	&	0.02	&	0	&	2	&	5	&	EM* GGA 98,  $^{(a)}$	&	mid\\
26.497126	&	61.212676	&	-3.28	&	0.03	&	0.15	&	0.04	&	0.64	&	0.04	&	-2.69	&	0.03	&	-2.65	&	0.03	&	-2.66	&	0.02	&	0	&	0.5	&	5	&	BD+60 332A,  $^{(a)}$	&	early\\
26.483769	&	61.212565	&	-2.17	&	0.02	&	0.22	&	0.03	&	0.86	&	0.04	&	-1.85	&	0.03	&	-1.94	&	0.03	&	-2.15	&	0.02	&	3	&	2	&	5	&	EM* GGA 97,  $^{(a)}$	&	mid\\
26.612098	&	61.236493	&	-1.38	&	0.02	&	-0.03	&	0.03	&	0.48	&	0.07	&	-1.23	&	0.03	&	-1.21	&	0.04	&	-1.20	&	0.02	&	1	&	-	&	-	&	EM* GGA 99, $^{(b)}$ *	&	mid $\dagger$\\
26.619315	&	61.230679	&	-0.70	&	0.02	&	-0.09	&	0.03	&	0.25	&	0.15	&	-0.78	&	0.02	&	-0.72	&	0.03	&	-0.81	&	0.02	&	1	&	3	&	5	&	NGC 663 14,  $^{(a)}$	&	late\\
26.627616	&	61.241449	&	-2.08	&	0.02	&	0.22	&	0.03	&	0.78	&	0.04	&	-1.29	&	0.02	&	-1.27	&	0.03	&	-1.36	&	0.02	&	1	&	1	&	5	&	V* V984 Cas,  $^{(a)}$	&	mid\\
26.615266	&	61.207074	&	-2.84	&	0.02	&	0.10	&	0.03	&	0.75	&	0.04	&	-2.86	&	0.02	&	-2.88	&	0.03	&	-2.92	&	0.02	&	1	&	5	&	-	&	V* V983 Cas, $^{(c)}$ *	&	early\\
26.648359	&	61.227523	&	-1.98	&	0.02	&	0.26	&	0.03	&	1.01	&	0.04	&	-1.35	&	0.03	&	-1.49	&	0.03	&	-1.70	&	0.02	&	0	&	1	&	5	&	EM* GGA 101,  $^{(a)}$	&	mid\\
26.449515	&	61.273316	&	-0.10	&	0.02	&	-0.05	&	0.03	&	0.60	&	0.14	&	-0.05	&	0.02	&	-0.13	&	0.03	&	-0.08	&	0.02	&	1	&	-	&	-	&	2MASSJ01454789+6116239, $^{(b)}$	&	late $\dagger$\\
26.415063	&	61.216452	&	-1.50	&	0.02	&	0.24	&	0.03	&	1.09	&	0.04	&	-1.19	&	0.02	&	-1.20	&	0.03	&	-1.33	&	0.02	&	0	&	2	&	5	&	EM* GGA 94,  $^{(a)}$	&	mid\\
26.601678	&	61.177033	&	0.17	&	0.03	&	-0.07	&	0.03	&	1.42	&	0.14	&	0.11	&	0.02	&	0.13	&	0.04	&	0.10	&	0.02	&	1	&	6	&	5	&	NGC 663 61,  $^{(a)}$	&	late\\
26.672702	&	61.220864	&	-2.13	&	0.02	&	-0.07	&	0.03	&	0.10	&	0.07	&	-2.22	&	0.03	&	-2.15	&	0.03	&	-2.12	&	0.02	&	0	&	4	&	-	&	EM* GGA 622, $^{(b)}$	&	early\\
26.612723	&	61.167195	&	-3.03	&	0.02	&	-0.03	&	0.03	&	-0.04	&	0.07	&	-3.13	&	0.03	&	-3.09	&	0.03	&	-3.00	&	0.02	&	1	&	5	&	3	&	BD+60 340, $^{(c)}$	&	early\\
26.689256	&	61.199832	&	-0.97	&	0.02	&	-0.07	&	0.03	&	-1.15	&	-	&	-1.05	&	0.03	&	-1.01	&	0.03	&	-0.93	&	0.02	&	1	&	-	&	-	&	EM* VES 623, $^{(d)}$	&	mid $\dagger$\\
26.443298	&	61.155813	&	-2.14	&	0.02	&	0.24	&	0.03	&	0.87	&	0.03	&	-1.79	&	0.02	&	-1.81	&	0.03	&	-1.93	&	0.02	&	0	&	1	&	5	&	EM* GGA 95,  $^{(a)}$	&	mid\\
26.641521	&	61.150654	&	-3.17	&	0.02	&	-0.03	&	0.03	&	0.32	&	0.05	&	-3.19	&	0.04	&	-3.11	&	0.03	&	-3.09	&	0.02	&	1	&	2	&	5	&	BD+60 334, $^{(b)}$	&	early\\
26.748328	&	61.208206	&	-0.43	&	0.02	&	0.18	&	0.03	&	1.41	&	0.09	&	-0.08	&	0.02	&	-0.13	&	0.03	&	-0.23	&	0.02	&	1	&	5	&	5	&	EM* VES 624,  $^{(a)}$	&	late\\
26.611840	&	61.128259	&	-3.51	&	0.02	&	0.26	&	0.03	&	0.85	&	0.03	&	-2.89	&	0.02	&	-2.92	&	0.04	&	-3.15	&	0.02	&	1	&	0.5	&	5	&	BD+60 341,  $^{(a)}$	&	early\\
26.407547	&	61.133112	&	-1.98	&	0.02	&	0.22	&	0.03	&	0.75	&	0.05	&	-1.01	&	0.03	&	-1.00	&	0.04	&	-1.18	&	0.02	&	1	&	2	&	5	&	EM* GGA 93,  $^{(a)}$ *	&	mid\\
26.765592	&	61.292237	&	-1.53	&	0.02	&	0.29	&	0.03	&	1.13	&	0.04	&	-1.14	&	0.03	&	-1.19	&	0.03	&	-1.29	&	0.02	&	3	&	2	&	5	&	V* V986 Cas,  $^{(a)}$	&	mid\\
26.634543	&	61.119229	&	-1.74	&	0.02	&	-0.05	&	0.03	&	0.29	&	0.07	&	-1.81	&	0.03	&	-1.78	&	0.03	&	-1.73	&	0.02	&	1	&	2	&	5	&	Cl* NGC 663 L604$^{(b)}$	&	mid\\
26.645246	&	61.107708	&	0.03	&	0.02	&	-0.06	&	0.03	&	1.99	&	0.10	&	0.01	&	0.03	&	0.03	&	0.04	&	0.07	&	0.02	&	1	&	5	&	5	&	NGC 663 180,  $^{(a)}$	&	late\\
26.822992	&	61.221549	&	0.75	&	0.03	&	0.01	&	0.04	&	0.27	&	-	&	0.85	&	0.03	&	0.84	&	0.03	&	0.80	&	0.03	&	1	&	6	&	5	&	NGC 663 151,  $^{(a)}$	&	late\\
26.325104	&	61.115690	&	-3.97	&	0.03	&	0.29	&	0.03	&	0.79	&	0.02	&	-3.51	&	0.02	&	-3.53	&	0.03	&	-3.63	&	0.02	&	0	&	1	&	5	&	BD+60 325,  $^{(a)}$	&	early\\
26.750870	&	61.356541	&	-2.43	&	0.02	&	0.13	&	0.03	&	0.63	&	0.03	&	-2.22	&	0.03	&	-2.23	&	0.04	&	-2.34	&	0.02	&	1	&	1	&	3	&	EM* GGA 104, $^{(b)}$	&	early\\
26.861504	&	61.145601	&	-2.88	&	0.02	&	0.36	&	0.03	&	0.99	&	0.04	&	-1.69	&	0.03	&	-1.66	&	0.03	&	-1.59	&	0.02	&	2	&	0.5	&	5	&	EM* GGA 108,  $^{(a)}$	&	mid\\
26.914048	&	61.305712	&	-2.05	&	0.02	&	0.30	&	0.03	&	0.96	&	0.04	&	-1.38	&	0.03	&	-1.49	&	0.03	&	-1.68	&	0.02	&	1	&	1	&	5	&	EM* GGA 109,  $^{(a)}$	&	mid\\
26.739843	&	61.027863	&	-1.68	&	0.02	&	0.01	&	0.03	&	0.13	&	0.06	&	-1.76	&	0.02	&	-1.76	&	0.03	&	-1.89	&	0.02	&	3	&	3	&	-	&	EM* GGA 103 $^{(b)}$	&	mid\\
34.737496	&	57.128716	&	-1.20	&	0.02	&	-0.12	&	0.03	&	-0.20	&	0.14	&	-1.08	&	0.05	&	-1.02	&	0.05	&	-1.06	&	0.04	&	1	&	3	&	-	&	[KPK99]J021856.82+570742.5, $^{(e)}$ *	&	mid\\
34.740807	&	57.138137	&	-1.88	&	0.02	&	-0.09	&	0.03	&	-0.18	&	0.09	&	-1.75	&	0.03	&	-1.65	&	0.03	&	-1.60	&	0.02	&	1	&	2	&	4	&	NSV 776	$^{(f)}$&	mid\\
34.750714	&	57.145720	&	-3.00	&	0.03	&	0.03	&	0.03	&	0.25	&	0.04	&	-2.94	&	0.02	&	-2.81	&	0.04	&	-2.79	&	0.02	&	1	&	1	&	5	&	V* V614 Per, $^{(e)}$	&	early\\
34.724407	&	57.139512	&	-3.19	&	0.02	&	-0.06	&	0.03	&	0.01	&	0.04	&	-3.30	&	0.03	&	-3.20	&	0.03	&	-3.16	&	0.02	&	1	&	0.5	&	5	&	BD+56 515, $^{(e)}$	&	early\\
34.806949	&	57.128771	&	-3.63	&	0.02	&	0.25	&	0.03	&	0.71	&	0.03	&	-3.12	&	0.03	&	-3.14	&	0.03	&	-3.29	&	0.02	&	1	&	1	&	5	&	BD+56 529,  $^{(a)}$	&	early\\
34.864416	&	57.138239	&	-3.45	&	0.02	&	-0.06	&	0.03	&	0.06	&	0.03	&	-3.60	&	0.03	&	-3.53	&	0.03	&	-3.43	&	0.02	&	1	&	0.5	&	5	&	HD 14162, $^{(e)}$	&	early\\
34.870057	&	57.117908	&	-3.09	&	0.02	&	0.19	&	0.03	&	0.44	&	0.03	&	-2.54	&	0.02	&	-2.60	&	0.03	&	-2.76	&	0.02	&	2	&	0.5	&	5	&	EM* GGA 156,  $^{(a)}$	&	early\\
34.699721	&	57.067276	&	-4.01	&	0.03	&	0.10	&	0.03	&	0.57	&	0.03	&	-3.89	&	0.04	&	-3.86	&	0.05	&	-3.87	&	0.02	&	1	&	0.5	&	5	&	BD+56 511,  $^{(a)}$	&	early\\
34.815054	&	57.188465	&	0.09	&	0.02	&	-0.02	&	0.03	&	0.01	&	-	&	0.23	&	0.03	&	0.21	&	0.03	&	0.19	&	0.02	&	1	&	10	&	-	&	2MASSJ02191561+5711185, $^{(e)}$	&	late\\
34.624291	&	57.150883	&	-3.10	&	0.02	&	-0.07	&	0.03	&	0.05	&	0.04	&	-3.29	&	0.02	&	-3.22	&	0.03	&	-3.15	&	0.02	&	1	&	0.5	&	1	&	V* V611 Per, Teff=25400K, logg=3.38, $^{(g)}$	&	early\\
34.860474	&	57.078364	&	-4.69	&	0.03	&	0.31	&	0.04	&	0.99	&	0.03	&	-3.78	&	0.03	&	-3.83	&	0.02	&	-4.10	&	0.02	&	1	&	3	&	5	&	 BD+56 534, Teff=26400K, logg= 3.55,  $^{(a)}$  	&	early\\
34.870366	&	57.190083	&	-1.22	&	0.06	&	-0.01	&	0.07	&	0.33	&	0.08	&	-1.52	&	0.02	&	-1.56	&	0.03	&	-1.68	&	0.02	&	3	&	0.5	&	5	&	NGC 869 1278, Teff=25100K, logg=4.24,  $^{(a)}$	&	mid\\
34.636556	&	57.211031	&	-3.31	&	0.06	&	-0.06	&	0.07	&	0.04	&	0.03	&	-3.48	&	0.03	&	-3.30	&	0.03	&	-3.32	&	0.02	&	1	&	1	&	5	&	BD+56 502, Teff=26700K, logg=3.91, $^{(g)}$	&	early\\
34.949225	&	57.111018	&	-1.77	&	0.11	&	-0.12	&	0.11	&	0.65	&	0.05	&	-2.00	&	0.02	&	-2.03	&	0.03	&	-2.10	&	0.02	&	2	&	2	&	-	&	2MASSJ02194783+5706395, Teff=20709K, logg=3.94, $^{(f)}$	&	early\\
34.562878	&	57.171130	&	-3.06	&	0.02	&	0.10	&	0.03	&	0.60	&	0.03	&	-2.89	&	0.03	&	-2.83	&	0.03	&	-2.89	&	0.02	&	1	&	3	&	-	&	BD+56 489, $^{(e)}$	&	early\\
34.700842	&	57.240094	&	-3.73	&	0.09	&	0.01	&	0.09	&	0.30	&	0.03	&	-3.10	&	0.03	&	-2.95	&	0.05	&	-2.91	&	0.04	&	3	&	1	&	-	&	NAME BD+56 509AB, $^{(e)}$,  *	&	early\\
34.700054	&	57.285527	&	-3.43	&	0.02	&	-0.04	&	0.03	&	0.00	&	0.03	&	-3.59	&	0.02	&	-3.55	&	0.03	&	-3.43	&	0.02	&	1	&	2	&	5	&	V* V665 Per, $^{(e)}$	&	early\\
35.573678	&	57.123500	&	-3.69	&	0.02	&	-0.01	&	0.03	&	0.55	&	0.03	&	-3.77	&	0.02	&	-3.67	&	0.07	&	-3.69	&	0.02	&	1	&	2	&	3	&	V* V622 Per, $^{(h)}$	&	early\\
35.510306	&	57.155664	&	-2.49	&	0.02	&	0.21	&	0.03	&	0.67	&	0.04	&	-2.16	&	0.02	&	-2.20	&	0.03	&	-2.35	&	0.02	&	3	&	1	&	5	&	NGC 869 2242,  $^{(a)}$	&	early\\
35.709491	&	57.147412	&	-1.49	&	0.02	&	0.21	&	0.03	&	0.92	&	0.04	&	-1.21	&	0.02	&	-1.21	&	0.03	&	-1.26	&	0.02	&	3	&	3	&	-	&	2MASSJ02225028+5708506, $^{(e)}$	&	mid\\
35.519004	&	57.177461	&	-2.24	&	0.07	&	-0.08	&	0.07	&	0.05	&	0.06	&	-2.39	&	0.02	&	-2.27	&	0.03	&	-2.25	&	0.02	&	1	&	2	&	5	&	EM* MWC 39, $^{(e)}$	&	early\\
35.481492	&	57.099632	&	-1.77	&	0.02	&	-0.06	&	0.03	&	-0.17	&	0.09	&	-1.94	&	0.02	&	-1.86	&	0.03	&	-1.77	&	0.02	&	1	&	3	&	5	&	[CHI2010] h Per M2623, $^{(e)}$	&	mid\\
35.470567	&	57.166364	&	-4.00	&	0.03	&	0.19	&	0.04	&	0.51	&	0.03	&	-2.79	&	0.02	&	-2.69	&	0.03	&	-2.61	&	0.02	&	1	&	1	&	5	&	BD+56 566, Teff=24800K, logg=3.79,  $^{(a)}$	&	early\\
35.430794	&	57.125830	&	-3.53	&	0.02	&	0.16	&	0.03	&	1.18	&	0.03	&	-3.43	&	0.02	&	-3.44	&	0.03	&	-3.44	&	0.02	&	3	&	1	&	3	&	BD+56 563, Teff=21900, logg=3.72, $^{(g)}$	&	early\\
35.767418	&	57.127469	&	-2.65	&	0.02	&	0.04	&	0.03	&	0.66	&	0.03	&	-2.57	&	0.02	&	-2.52	&	0.03	&	-2.53	&	0.02	&	3	&	2	&	-	&	EM* GGA 163, $^{(e)}$	&	early\\
35.700408	&	57.200277	&	-3.28	&	0.02	&	0.30	&	0.03	&	0.93	&	0.03	&	-2.67	&	0.02	&	-2.65	&	0.03	&	-2.72	&	0.03	&	2	&	0.5	&	5	&	EM* MWC 711,  $^{(a)}$, *	&	early\\
35.435269	&	57.181195	&	-2.09	&	0.17	&	0.15	&	0.17	&	0.89	&	0.04	&	-1.57	&	0.02	&	-1.62	&	0.03	&	-1.72	&	0.02	&	1	&	3	&	5	&	EM* GGA 162,  $^{(a)}$	&	mid\\
35.353822	&	57.197898	&	-1.44	&	0.02	&	0.24	&	0.03	&	0.94	&	0.04	&	-1.03	&	0.02	&	-1.08	&	0.03	&	-1.21	&	0.02	&	1	&	2	&	-	&	EM* GGA 161, $^{(i)}$	&	mid\\
35.594727	&	57.284735	&	-4.24	&	0.03	&	0.09	&	0.03	&	0.62	&	0.03	&	-3.75	&	0.03	&	-3.73	&	0.06	&	-3.69	&	0.02	&	1	&	1	&	3	&	BD+56 582, $^{(i)}$	&	early\\
35.887723	&	57.075727	&	-3.42	&	0.02	&	0.20	&	0.03	&	0.56	&	0.03	&	-2.92	&	0.02	&	-3.02	&	0.03	&	-3.21	&	0.02	&	3	&	-	&	-	&	TYC 3694-1331-1, Teff=22535K, logg=3.94, $^{(j)}$	&	early\\
174.090981	&	-61.608303	&	-0.96	&	0.02	&	-0.07	&	0.03	&	-0.72	&	0.44	&	-0.81	&	0.03	&	-0.72	&	0.03	&	-0.70	&	0.03	&	1	&	-	&	-	&	CPD-60 3149, Teff=16890K, logg=3.84, $^{(i)}$	&	late\\
174.091240	&	-61.624731	&	-1.00	&	0.02	&	0.12	&	0.03	&	0.60	&	0.07	&	-0.64	&	0.03	&	-0.61	&	0.03	&	-0.70	&	0.02	&	1	&	-	&	-	&	CPD-60 3144, Teff=17687K, logg=3.61, $^{(i)}$     	&	late\\
174.058513	&	-61.626579	&	-1.35	&	0.02	&	-0.08	&	0.03	&	-0.02	&	0.08	&	-1.55	&	0.02	&	-1.51	&	0.02	&	-1.52	&	0.02	&	1	&	2	&	5	&	CPD-60 3133, Teff=17519K, logg=3.69, $^{(i)}$	&	mid\\
174.042312	&	-61.627764	&	-2.62	&	0.02	&	0.23	&	0.03	&	0.81	&	0.03	&	-2.11	&	0.02	&	-2.06	&	0.02	&	-2.05	&	0.02	&	3	&	1.5	&	5	&	CPD-60 3126, Teff=18564K, logg=3.53  , $^{(i)}$	&	early\\
174.049414	&	-61.597301	&	-2.78	&	0.02	&	-0.03	&	0.03	&	0.19	&	0.05	&	-2.91	&	0.02	&	-2.84	&	0.03	&	-2.77	&	0.02	&	1	&	2	&	5	&	CPD-60 3128, Teff=18817K, logg=3.31, $^{(i)}$  	&	mid\\
174.039818	&	-61.593906	&	-2.26	&	0.02	&	0.01	&	0.03	&	0.34	&	0.04	&	-2.07	&	0.02	&	-1.98	&	0.02	&	-1.96	&	0.02	&	3	&	-	&	-	&	CPD-60 3125, Teff=18274K, logg=3.49, $^{(i)}$	&	early\\
174.088804	&	-61.587768	&	-1.91	&	0.02	&	-0.05	&	0.03	&	-0.31	&	0.12	&	-1.79	&	0.03	&	-1.70	&	0.03	&	-1.75	&	0.02	&	1	&	2	&	5	&	CPD-60 3147, Teff=18883K, logg=3.23, $^{(i)}$	&	mid\\
174.004467	&	-61.621609	&	-1.77	&	0.02	&	0.15	&	0.03	&	1.07	&	0.03	&	-1.28	&	0.02	&	-1.19	&	0.03	&	-1.15	&	0.03	&	3	&	-	&	-	&	CPD-60 3108, Teff=17792K, logg=3.75, $^{(i)}$	&	mid\\
174.033412	&	-61.643960	&	-2.27	&	0.02	&	0.26	&	0.03	&	0.80	&	0.03	&	-1.67	&	0.02	&	-1.80	&	0.02	&	-1.99	&	0.02	&	1	&	-	&	-	&	CPD-60 3122, Teff=13254K, logg=3.29, $^{(i)}$	&	mid\\
173.981095	&	-61.603857	&	-2.67	&	0.02	&	-0.05	&	0.03	&	0.08	&	0.06	&	-2.76	&	0.02	&	-2.68	&	0.03	&	-2.67	&	0.02	&	1	&	2	&	4	&	HD 100856, Teff=18725, logg=3.34, $^{(i)}$	&	early\\
174.131523	&	-61.573839	&	-3.25	&	0.02	&	0.10	&	0.03	&	0.42	&	0.03	&	-3.03	&	0.02	&	-3.02	&	0.04	&	-3.11	&	0.02	&	3	&	-	&	-	&	HD 306791,Teff=18399K, logg=3.30, $^{(i)}$	&	early\\
174.174609	&	-61.631627	&	-0.37	&	0.02	&	-0.12	&	0.03	&	-0.33	&	0.29	&	-0.46	&	0.03	&	-0.38	&	0.03	&	-0.38	&	0.03	&	1	&	-	&	-	&	CPD-60 3165, Teff=15945K, logg=3.95, $^{(i)}$	&	late\\
174.205566	&	-61.605880	&	-0.51	&	0.02	&	-0.03	&	0.03	&	0.09	&	0.08	&	-0.45	&	0.03	&	-0.38	&	0.02	&	-0.35	&	0.02	&	3	&	-	&	-	&	CPD-60 3174, Teff=15650, logg=3.78, $^{(i)}$	&	late\\
174.051578	&	-61.545558	&	-2.48	&	0.02	&	0.16	&	0.03	&	0.54	&	0.03	&	-1.27	&	0.03	&	-1.17	&	0.07	&	-1.15	&	0.05	&	1	&	-	&	-	&	CD-60 3626, Teff=17834K, logg=3.82, $^{(i)}$	&	mid\\
174.022890	&	-61.701685	&	-2.06	&	0.02	&	0.13	&	0.03	&	0.53	&	0.03	&	-1.85	&	0.02	&	-1.86	&	0.02	&	-2.00	&	0.02	&	1	&	-	&	-	&	HD 306797, Teff=16301K, logg=3.51, $^{(i)}$	&	mid\\
173.842030	&	-61.536173	&	-0.86	&	0.02	&	0.12	&	0.03	&	0.90	&	0.23	&	-1.03	&	0.02	&	-1.03	&	0.02	&	-1.08	&	0.02	&	3	&	-	&	-	&	HD 306793, Teff=18995K, logg=4.02, $^{(i)}$	&	mid\\
173.813207	&	-61.699876	&	-1.85	&	0.02	&	0.25	&	0.03	&	0.94	&	0.04	&	-1.38	&	0.02	&	-1.41	&	0.03	&	-1.57	&	0.02	&	3	&	-	&	-	&	HD 306657, Teff=19580K, logg=4.00, $^{(i)}$	&	mid\\
174.453112	&	-61.751439	&	-1.60	&	0.03	&	-0.03	&	0.03	&	-0.50	&	0.11	&	-1.15	&	0.03	&	-1.23	&	0.03	&	-1.39	&	0.02	&	1	&	-	&	-	&	CPD-60 3087, $^{(k)}$ 	&	mid\\
193.432798	&	-60.374656	&	-2.45	&	0.02	&	-0.03	&	0.03	&	1.77	&	0.03	&	-2.29	&	0.03	&	-2.18	&	0.04	&	-2.18	&	0.02	&	2	&	1.5	&	5	&	V* CT Cru, $^{(i)}$	&	early\\
193.446492	&	-60.372198	&	-3.71	&	0.02	&	0.25	&	0.03	&	0.97	&	0.03	&	-2.99	&	0.04	&	-3.04	&	0.06	&	-3.22	&	0.03	&	1	&	1.5	&	5	&	2MASS J12534725-6022200, Teff=25040K, logg=3.91, $^{(i)}$	&	early\\
193.465538	&	-60.366237	&	-2.01	&	0.03	&	-0.18	&	0.03	&	2.34	&	0.04	&	-2.04	&	0.02	&	-1.94	&	0.03	&	-1.83	&	0.02	&	2	&	1	&	5	&	V* CX Cru, $^{(l)}$	&	early\\
193.356909	&	-60.366710	&	-1.77	&	0.02	&	-0.12	&	0.03	&	-0.45	&	0.08	&	-1.97	&	0.02	&	-1.84	&	0.02	&	-1.82	&	0.02	&	1	&	1	&	5	&	CPD-59 4532, $^{(l)}$	&	mid\\
193.466655	&	-60.371088	&	-2.58	&	0.03	&	-0.12	&	0.04	&	0.40	&	0.05	&	-2.66	&	0.02	&	-2.57	&	0.02	&	-2.53	&	0.02	&	1	&	2	&	-	&	V* EI Cru, $^{(l)}$	&	early\\
193.470570	&	-60.358505	&	-2.24	&	0.02	&	-0.15	&	0.03	&	0.60	&	0.04	&	-1.95	&	0.03	&	-1.93	&	0.04	&	-1.94	&	0.02	&	1	&	2	&	5	&	V* CZ Cru, $^{(l)}$	&	mid\\
193.467558	&	-60.374383	&	-2.49	&	0.02	&	-0.07	&	0.03	&	0.47	&	0.03	&	-2.45	&	0.02	&	-2.31	&	0.02	&	-2.30	&	0.02	&	2	&	1.5	&	5	&	V* CY Cru, $^{(l)}$	&	early\\
193.412417	&	-60.395461	&	-3.49	&	0.02	&	0.27	&	0.03	&	0.82	&	0.03	&	-2.82	&	0.02	&	-2.93	&	0.06	&	-3.16	&	0.02	&	1	&	2	&	4	&	CPD-59 4546, $^{(m)}$	&	early\\
193.397882	&	-60.396305	&	-2.19	&	0.02	&	-0.08	&	0.03	&	-0.01	&	0.04	&	-2.38	&	0.02	&	-2.30	&	0.02	&	-2.20	&	0.02	&	1	&	1.5	&	5	&	V* BT Cru, $^{(l)}$	&	early\\
193.464869	&	-60.388024	&	-3.38	&	0.02	&	0.30	&	0.03	&	0.73	&	0.03	&	-2.51	&	0.02	&	-2.62	&	0.02	&	-2.83	&	0.02	&	1	&	2	&	4	&	CPD-59 4559, $^{(l)}$	&	early\\
193.469057	&	-60.399304	&	-0.63	&	0.03	&	-0.04	&	0.04	&	-0.59	&	0.25	&	-0.15	&	0.02	&	-0.15	&	0.03	&	-0.18	&	0.02	&	1	&	-	&	-	&	CPD-59 4561, $^{(i)}$	&	late\\
193.445743	&	-60.309989	&	-2.23	&	0.02	&	-0.07	&	0.03	&	0.15	&	0.04	&	-2.31	&	0.02	&	-2.21	&	0.02	&	-2.18	&	0.02	&	1	&	1.5	&	5	&	V* CV Cru, $^{(l)}$	&	early\\
193.489734	&	-60.416141	&	-2.70	&	0.02	&	-0.06	&	0.03	&	-0.01	&	0.05	&	-2.91	&	0.02	&	-2.81	&	0.05	&	-2.73	&	0.02	&	1	&	1	&	5	&	V* BW Cru, $^{(l)}$	&	early\\
193.220361	&	-60.296769	&	0.63	&	0.03	&	-0.08	&	0.04	&	-0.18	&	-	&	0.65	&	0.02	&	0.71	&	0.03	&	0.71	&	0.03	&	1	&	-	&	-	&	Cl* NGC 4755 ESL 101, $^{(n)}$	&	late\\
193.159809	&	-60.338676	&	-2.32	&	0.02	&	0.27	&	0.03	&	0.87	&	0.03	&	-1.33	&	0.02	&	-1.24	&	0.02	&	-1.20	&	0.02	&	1	&	-	&	-	&	HD 312076, $^{(o)}$	&	mid $\dagger$\\
193.150957	&	-60.307060	&	-3.00	&	0.02	&	0.30	&	0.03	&	0.90	&	0.03	&	-2.46	&	0.02	&	-2.55	&	0.02	&	-2.75	&	0.02	&	1	&	0	&	-	&	HD 312075, $^{(o)}$	&	early\\
\enddata
\tablecomments{a. \citet{Mathew2011}, b.\citet{Pigulski2001}, c. \citet{Kohoutek1997}, d. \citet{Coyne1978}, e. \citet{Abad1995}, f. \citet{Slesnick2002}, g. \citet{Marsh2012}, h. \citet{Hog2000}, i. \citet{McSwain2009a}, j. \citet{Huang2010a}, k. \citet{Slettebak1985}, l. \citet{Sanner2001}, m. \citet{Jaschek1982}, n. \citet{McSwain2005b}, o. \citet{Wray1966}. The symbol * in column 18 indicates that there is a source within 8asec of the target, for most of them significantly weaker. The symbol $\dagger$ in column 19 indicates that this is the first time an estimate of the spectral type is given for the star.}
\end{deluxetable*}

\section{Results}

\subsection{WISE colour-magnitude diagram}

Similar to \citet{Bonanos2010}, we define {\it photometric Be stars} as B type stars having colour excesses within a certain range.
In the case of WISE colours, \citet{Nikutta2014} defined {\it naked} stars as stellar objects without a protostellar dusty disk with W1-W2$<$0.8. This limit prevents contamination from faint IR sources. For the whole group of {\it naked} stars not severely affected by extinction and dominated by MS or {\it normal} stars, objects with no evidence of a circumstellar disk, these authors find a mean value $\mu_{W_{12}}$=-0.04 for the colour W$_{12}$=W1-W2, with a standard deviation $\sigma_{W_{12}}$=0.03. 

For our sample of B stars with absolute J magnitude values between -4 and 1, 
the mean colour is $\mu_{W_{12}}$=-0.066 with a standard deviation of $\sigma_{W_{12}}$=0.002, whereas for the subgroup of early and mid B stars (-4$<$J$<$-1), the values are $\mu_{W_{12}}$=-0.023 and $\sigma_{W_{12}}$=0.006.
We chose to use the conservative criteria given by \citet{Nikutta2014} according to which most {\it naked} or {\it normal} stars have W1-W2$<$0.05. This limit is represented with a vertical black line in Figure \ref{5Clusters_a} (a) and represents the limit between stars behaving as the majority (or normal) and those B-type stars having an infrared excess.  
 
Figure \ref{5Clusters_a} (b) shows  all the stars in our sample of five clusters as gray circles and known Be stars as coloured squares. As described before, different colours correspond to different J magnitude bins.  
We can see that Be stars cluster mainly in two regions of this plot, a large group is found together with {\it naked} or {\it normal} B stars in the region with W1-W2 between -0.25 and 0.05, and another group in the region with W1-W2 between 0.1 and 0.3. 

\begin{figure}
\figurenum{7}
\gridline{\fig{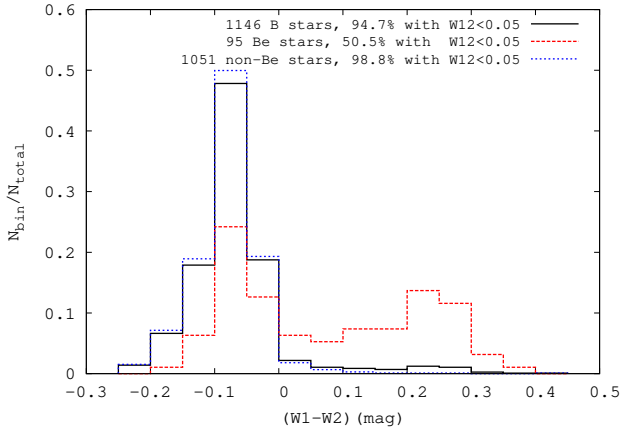}{0.465\textwidth}{}}
\caption{Fraction of stars in each W1-W2 colour bin. The black continuous line shows the distribution of all B stars in our sample, the red line shows the distribution of Be stars and the blue line shows the distribution of objects that are not known to be Be stars.\label{Histo_Be}}
\end{figure}

Indeed, as can be seen in Figure~\ref{Histo_Be}, we find that $98.8\%$ of the non-Be objects or {\it normal stars} have W1-W2$<$0.05. We define Be star candidates as objects that have no Be star classification but have W1-W2$\geq$0.05.

\begin{figure}
\figurenum{8}
\gridline{\fig{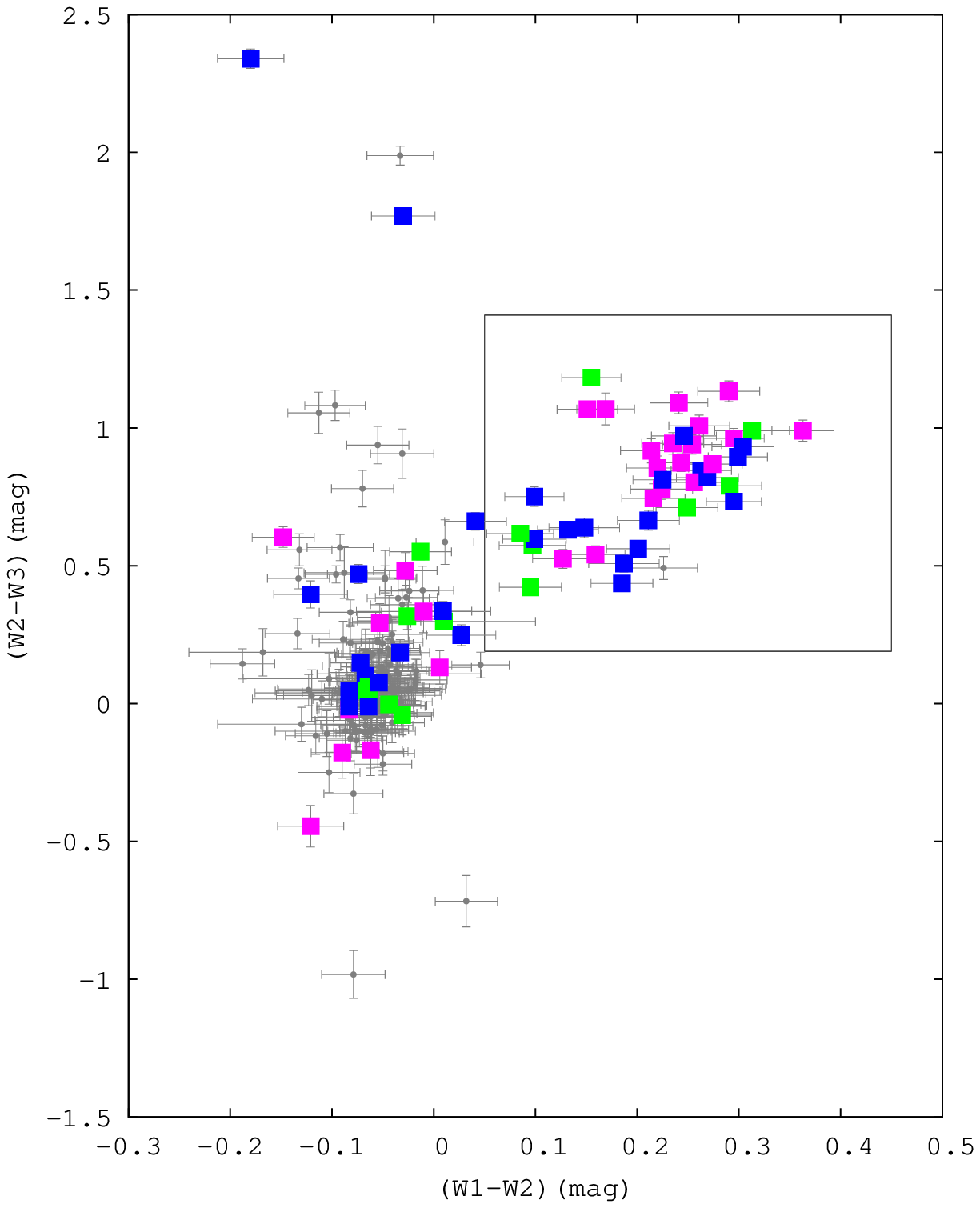}{0.465\textwidth}{}}
\caption{Colour-colour diagram W2-W3 versus W1-W2 for mid and early B stars. The quality of W3 band of late B type stars is usually poor so we did not consider these objects. Gray points indicate all early and mid B-type stars. Coloured symbols indicate Be stars. The region occupied by Be stars in an active phase,  W1-W2$\geq$0.05, is indicated with a black rectangle. \label{CC_all}}
\end{figure}

We split our sample of B stars in absolute magnitude bins corresponding to early (-4$\leq$J$<$-2, green and blue points in Figure \ref{5Clusters_a} (b)), mid (-2$\leq$J$<$-1, magenta points) and late B stars (-1$\leq$J$<$1 cyan and yellow points). Red  and black points in Figure \ref{5Clusters_a} (b) correspond to O type stars and supergiants, with J$<$-4. The presence of B stars of luminosity class III could eventually pollute the different bins. However, as we do not expect to have a significant number of giants given the ages of the clusters under study, we do not take this possibility into account in the present article. This may be of interest when older clusters are studied.

Then, we have 144 early B stars of which 47 are known Be stars (32.6$\%$). 
Among the 166 mid B stars, 33 are known Be stars (19.9$\%$) 
and within the 836 late B stars, 15 (1.8$\%$) 
are known Be stars. If we look at the early and mid B stars together, 25.8$\%$ are Be stars. The fraction of Be stars decreases significantly when the sample contains later type stars. 

This dependency of the fraction of Be stars on the spectral type range under consideration has been extensively reported in the literature, and is the reason why different authors find large Be fractions when observing the upper MS of open clusters \citep[e.g.][]{Grebel1992, Grebel1997, Maeder1999a, Keller2001, McSwain2005b, Martayan2010a, Iqbal2013}. These works do not properly account for the late B and Be-type stars.

When looking at our overall sample ($\rm -4<J<1$), the fraction of Be stars is 8.5$\%$.
Because of the lack of completeness, in particular of the late B-type sample, the proportion of Be stars might certainly be lower for the whole B sample.

Of all early Be stars, 49$\%$ have $\rm W1-W2\geq0.05$; among mid Be stars the percentage increases to 60.6$\%$ and for late Be stars decreases to $26.7\%$. The number of Be candidates 
in our sample is 1 in the early Be group, 1 in the mid Be group, and 12 in the late Be group. 

In the mid and early Be star groups, 43 out of the 45 stars with an excess $\rm W1-W2\geq0.05$ are known Be stars (95.6$\%$).  
We assume that we know all the Be stars in these spectral ranges, for the open clusters studied in this article. This indicates that there may remain only a few objects to be discovered as mid and early Be stars within these clusters.

On the contrary, in the late Be group, the large proportion of new candidates indicates that for late B stars the classification as Be stars can be more difficult, and a significant number of these objects may have eluded a previous detection. In Table \ref{Interesting_objects} we list the coordinates, absolute W1 magnitude, W1-W2 (indicated as W12), W2-W3 (indicated as W23), as well as other relevant data for all the Be candidates and other interesting objects as well.

\begin{table*}[h]
\caption{Be candidates and other interesting objects that deserve further study}             
\label{Interesting_objects}   
\centering                          
\begin{tabular}{c c c c c c p{4cm}}        
\hline\hline
Cluster&     R.A.    &  $\delta$  &  M$_{\rm W1}$  &W12&W23&Simbad Name/Relevant Data\\ \hline
  \multicolumn{7}{|c|}{Candidate Be} \\
  \hline
       &01 45 55.876&+61 12 33.69& 0.573&0.057&-&MV 34016$^{(1)}$\\
NGC 663&01 45 45.559&+61 10 55.47&-1.214&0.070&-&G 111$^{(2)}$\\
       &01 47 28.119&+61 22 45.95& 0.622&0.079&-&-\\\hline
       &02 18 34.121&+57 13 50.24&-2.630&0.427&-&HG 731$^{(3)}$, V$_{\rm proj}$=165kms$^{-1\,(4)}$, T$_{\rm eff}$=20561K$^{(4)}$,  logg=3.944$^{(4)}$, SB1$^{(3)}$.\\
&02 19 03.292&+57 08 20.34&-2.111&0.226&0.492&W2$^{(4)}$, B2V, T$_{ eff}$=20566K$^{(5)}$, logg=3.96$^{(5)}$,V$_{\rm proj}$=171kms$^{-1 (5)}$.\\
       &02 19 51.091&+57 17 34.14& 0.634&0.093&-&NGC 869 1455\\
NGC 869/884&02 20 24.749&+57 06 53.07&0.952&0.099&-&LAV 1432$^{(6)}$\\
       &02 21 33.416&+57 12 01.62&0.783&0.134&-&LAV 1757$^{(6)}$\\
       &02 21 44.694&+57 04 09.23&0.192&0.104&-&LAV 1825$^{(6)}$\\
       &02 23 17.891&+57 14 23.71&0.417&0.069&-&LAV 2353$^{(6)}$\\\hline
NGC 3766&11 36 30.153&-61 39 47.62&0.293&0.184&1.130&MG\,173$^{(7)}$, V$_{\rm proj}$=171kms$^{-1\,(8)}$, T$_{\rm eff}$=14210K$^{(8)}$,  logg=4.21$^{(8)}$, He weak.\\\hline
        &12 53 37.775&-60 17 45.24&-1.340&0.273&2.331&SB 133$^{(10)}$\\
NGC 4755&12 53 29.469&-60 21 16.56&-0.605&0.103&1.562&ESL 56$^{(11)}$,  B3Vn.\\
        &12 53 52.896&-60 23 06.78&-1.081&0.087&0.451&ESL 51$^{(11)}$, B3Vn.\\
  \hline
  \multicolumn{7}{|c|}{Interesting objects with W12$<$0.05 but large W23} \\
  \hline
    &12 53 43.872&-60 22 28.76&-2.449&-0.030&1.769&CT Cru, Variable Star of $\beta$ Cep type, V$_{\rm proj}$=195km$s^{-1\,(12)}$, B1.5V\\
NGC 4755&12 53 48.177&-60 21 54.32&-1.779&-0.033&1.988&CPD-59 4556. Eclipsing binary ($\beta$ Lyr type), B1V.\\
    &12 53 51.729&-60 21 58.45&-2.010&-0.180&2.340&CX Cru, Variable Star of $\beta$ Cep type, V$_{\rm proj}$=278km$s^{-1\,(12)}$, B1V.\\
\hline\hline                       
\end{tabular}\\
References: (1)\citet{Moffat1974}, 
(2)\citet{Gushee1919},
(3)\citet{Huang2006a}, 
(4)\citet{Marsh2012},\\
(5)\citet{Huang2010a},
(6)\citet{Lavdovsky1961},
(7)\citet{McSwain2005b},
(8)\citet{McSwain2009a},\\
(9)\citet{Wildey1964},
(10)\citet{Sanner2001},
(11)\citet{Evans2005},
(12)\citet{Hunter2009}.\\
All data available in SIMBAD astronomical database \citep{Wenger2000}.
\end{table*} 

The presence of a circumstellar disk may not always be evident 
if the disk is not dense or large enough and/or the spectroscopic observations are not done with sufficient resolution. In this sense, the WISE photometric classification can identify new targets to look for the Be phenomenon, which could be studied later spectroscopically.

Figure \ref{CC_all} represents the WISE colour-colour diagram W2-W3 versus W1-W2. We included only early and mid B type stars because of the low quality of  W3 for most of late B-type stars. All Be stars with W1-W2$>$0.05 reside within a well determined region, indicated as a black rectangle in the plot. Most Be stars with W1-W2$<$0.05 behave as normal B stars, most of which have W2-W3$<$0.6, with the remarkable exception of two Be stars with W2-W3$>$1.5, and a third non Be star. These very red colours are definitely not characteristic of Be stars, and interestingly both stars are not only Be stars but also  have a $\beta$ Cephei classification. 
The non-Be star that behaves similar to these two objects is an eclipsing binary.
The colours of these objects agree  with those presented by \citet{Nikutta2014} for a central star with a temperature of 10\,000K surrounded by an optically thin dusty shell with a rather flat density distribution. This kind of density profile places more dust at large radial distances, where the temperature is lower. In this case, a small increase in optical depth can significantly enhance long wavelength emission and therefore increasing W2-W3 while having a value of close to 0 for the warmer colour W1-W2. This observable signature could be evidence of the $\beta$ Cephei behaviour, as pulsations could enhance the shifting of dust at large radial distances. All three objects deserve further study. The coordinates and other  characteristics of both Be stars and the third interesting star are listed in Table~\ref{Interesting_objects}. 

\begin{figure}
\figurenum{9}
\gridline{\fig{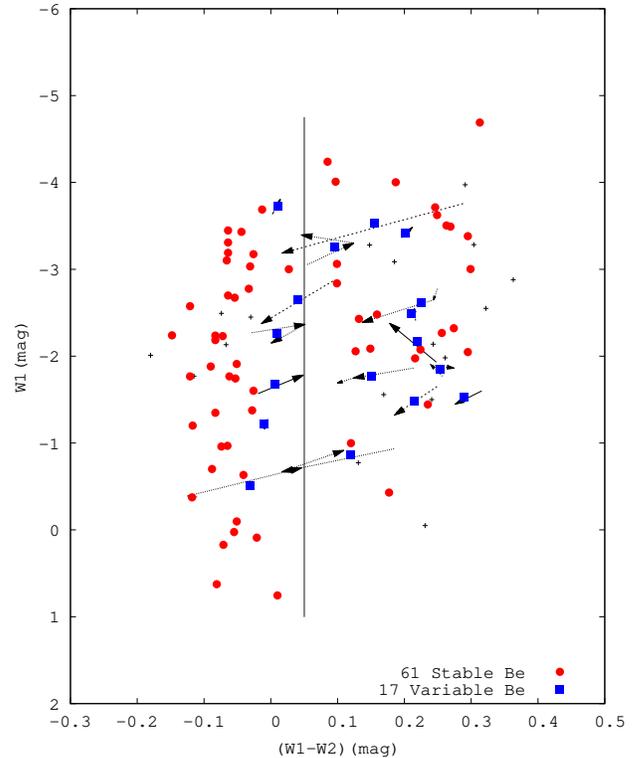}{0.465\textwidth}{}}
\caption{CMD of Be stars. The different colours indicate different WISE variability flags: Red circles indicate stable Be stars while blue squares correspond to objects that are variable objects. The remaining stars do not have a clear variability signature or have low quality data so a variability flag could not be identified. The arrows indicate the extent of variability between different observing epochs of the blue squares.\label{Variability_Be}}
\end{figure}
\subsection{WISE variability of Be stars}
In order to probe the variability of our sample of Be stars, we investigate the variability flags given in the ALLWISE catalogue \citep{Cutri2013}. Figure \ref{Variability_Be} shows the WISE CMD for all Be stars in our sample. The different symbols and colours represent Be stars that have different variability flags in their WISE magnitudes \citep{Hoffman2012}. 

As explained in the AllWISE data release\footnote{http://wise2.ipac.caltech.edu/docs/release/allwise/}, the AllWISE source catalog contains one set of calibrated magnitudes per object. With this aim, during the AllWISE science data processing, single-exposure images obtained in different epochs were coadded, allowing the detection of sources and the measurement of the position, apparent motion and photometry for each of them. Profile-fit photometry was performed simultaneously in the four WISE bands, and also for each filter and single-exposure image, so not only one set of calibrated magnitudes per object was obtained, but also multiepoch photometry is available for each object and in each band. During this data processing, flux variability in a band was evaluated by analyzing the distribution of flux measurements of a source on the individual single-exposure frames, and hence a variability flag was assigned to each band.

Because the clusters under study are near the Galactic plane, the single-exposure images from which the observations of the Be stars presented in our article originate were performed in either two or three different epochs, separated by around 180 days. In each epoch, ten to sixteen individual observations were obtained within two to five days, all of them  available in the AllWISE Multiepoch Photometry Database\footnote{http://wise2.ipac.caltech.edu/docs/release/allwise/expsup/sec3$\_$1.html}.

In Figure \ref{Variability_Be}, the red circles indicate Be stars with variability flags compatible with no variation (flags 0-5) while blue squares correspond to objects that are very likely variable objects (flags 8-9). For the remaining objects, identified with black crosses, either the quality of the data did not allow any variability flag to be provided in W1 or W2 bands, or the variability flags did not provide a clear variability signature (flag 6 or 7).

We see that red and blue symbols occupy different regions of the CMD: while the 61 non-variable Be stars tend to reside either with the {\it normal} B stars, or with stars with a significant excess, the 17 variable Be stars tend to occupy a transition region between {\it normal} B stars and stars having a significant excess. 

\begin{figure}
\figurenum{10}
\gridline{\fig{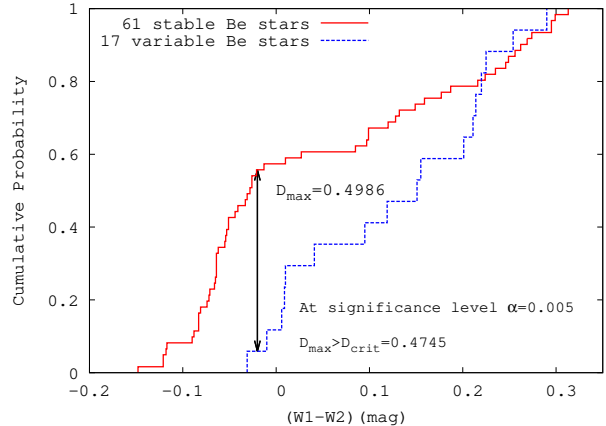}{0.465\textwidth}{}}
\caption{Cumulative distribution of stable and variable Be stars. The two samples are unlikely to come from the same distribution , because the the Kolmogorov–Smirnov statistic (D$_{\rm max}$) is larger than the corresponding critical value for the two-sample Kolmogorov-Smirnov test (D$_{\rm crit}$) .\label{Cumulative_var}}
\end{figure}
By doing a two-sample Kolmogorov-Smirnov test of these variable and non-variable
Be stars, we find that it is unlikely that these two samples come from the same distribution, as shown in Figure \ref{Cumulative_var}. The null hypothesis that both samples come from a population with the same distribution can be rejected at a significance level of $\alpha=0.005$.

We interpret the differences between the two samples in the WISE CMD in terms of quiescent ({\it normal} B) and active phases, in which the star hosts a circumstellar disk. A lack of variability is found either for Be stars in a quiescent (diskless phase), or for stars that have a developed disk that is not changing significantly. On the other hand, variable stars are Be
stars undergoing disk changes. Among the 95 Be stars of our sample, 17 are definitely variable stars, with light and colour changes observed during the 18 months of WISE observations. This constitutes 18\% of the sample.

To show the variability of the Be stars flagged as variables, we calculated the mean W1 and W2 magnitudes for each epoch and plotted the corresponding data in Figure \ref{Variability_Be}. The variability between consecutive epochs is indicated with an arrow pointing in the direction of the time evolution. The six variable stars that are in the quiescent state, undergo small color and magnitude changes, indicating that some minor mass loss episodes could occur in these objects. There is a significant change in colour and magnitude in most of the variable active stars.

As described in the following section, the variability observed in the most active variable Be stars is compatible with a disk dissipation phase. In order to interpret the nature of the variability behavior of Be stars, we have generated synthetic WISE colours using {\sc beray} circumstellar disk code \citep{Sigut2011a}.

\section{Disk Model Predictions}

In order to obtain the IR continuum flux of B1, B3 and B7 spectral type stars\footnote{The stellar parameters for these spectral types are taken from \citet{Cox2000}, consistent with previous work of the authors.} hosting circumstellar disks,
we used the codes {\sc bedisk} \citep{Sigut2007} and {\sc beray} \citep{Sigut2011a}. The former computes the temperature structure for a given disk density structure; the statistical equilibrium equations are solved to obtain the atomic level populations used for computation of the heating and cooling rates, which are balanced to fix the temperature.
The latter solves the radiative transfer equation along 10$^5$ rays through the star-plus-disk system for a given inclination angle. Rays that terminate on the stellar surface use an appropriately limb-darkened intensity for the boundary condition on the transfer equation. In this way, different observables can be computed, such as
line profiles, spectral energy distributions (SEDs), or monochromatic images projected on the sky. These codes have been broadly used for the interpretation of H$\alpha$ lines \citep{Silaj2010a,Ahmed2012,Sigut2013,Silaj2014}, near-IR spectroscopy \citep{Jones2009}, and interferometric observations \citep{Jones2008,Tycner2008,Sigut2015} of Be star disks. In the present work, SEDs are the focus of the modelling as these can be used to compute the required near-IR magnitudes and colours.

For the present work, we use an axisymmetric disk density distribution in which the radial density depends on two parameters: $\rho_{0}$, the density at the base of the disk, and $n$, 
the power-law exponent that determines how the density decreases with distance in the equatorial plane from the star. The disk density distribution in the cyclindrical co-ordinates $(R,Z)$ is
\begin{equation}
 \rho(R,Z)=\rho_0\left(\frac{r_0}{R}\right)^n e^{-\left(\frac{Z}{H}\right)^2}.
\end{equation}
Here, $r_0$ is the stellar radius and $H$ is the disk scale height, computed assuming the disk is in hydrostatic equilibrium in the vertical (i.e.\ $Z$) direction with an isothermal temperature of $0.6\,T_{\rm eff}$  \citep[see][]{Sigut2009}. This isothermal temperature is used only to fix the disk scale height, and this, coupled with the assumption of vertical hydrostatic equilibrium, produces a flaring disk with $H\propto R^{3/2}$. 

We computed spectral energy distributions assuming different values of $\rho_{0}$, between $10^{-12}\,{\rm g}\,{\rm cm^{-3}}$ and $10^{-10}\,{\rm g}\,{\rm cm^{-3}}$, and for different values of $n$, ranging between 2 and 4. These are typical values often considered for Be stars in the literature \citep{Rivinius2013}.
For the present calculations, we considered a disk size of 50 stellar radii, and a variety of inclination angles. 

For each model, we computed WISE magnitudes by convolving our synthetic energy distributions with each WISE filter, as described by \citet{Jarrett2011}. 
We also obtained 2MASS J magnitudes by using 2MASS filter definitions and the fluxes for zero magnitude from \citet{Cohen2003}.  
We present the computed magnitudes (W1, J, H, K) and WISE colours (W1-W2 indicated as W12 and W2-W3 indicated as W23) for three different spectral types for five different values of $n$, eight values of $\rho_0$, and five different inclination angles in Tables~\ref{Tab_n20}, \ref{Tab_n25}, \ref{Tab_n30}, \ref{Tab_n35} and \ref{Tab_n40}. In addition, we also provide for each model the predicted equivalent width of the H$\alpha$ line (EW$_{H\alpha}$), a quantity of special interest for Be stars as emission in the Balmer series (i.e.\ a negative value of EW$_{H\alpha}$) is the defining characteristic of the Be stars and a signature of disk emission.

\floattable
\begin{deluxetable}{ c  c | c c c c c c c| c c c c c c c | c c c c c c c} 
\tabletypesize{\tiny}
\rotate
\tablewidth{0pt}
\tablecolumns{23}
\tablecaption{Synthetic infrared magnitudes and colours computed for n=2.\label{Tab_n20}}
\tablehead{
    \colhead{$\rho_{0}/10^{-12}$} & \colhead{i} & \colhead{W1} &  \colhead{W12} &  \colhead{W23} &  \colhead{J} &  \colhead{H} &  \colhead{K} &  \colhead{EW$_{H\alpha}$} & \colhead{W1} &  \colhead{W12} &  \colhead{W23} &  \colhead{J} &  \colhead{H} &  \colhead{K} &  \colhead{EW$_{H\alpha}$} & \colhead{W1} &  \colhead{W12} &  \colhead{W23} &  \colhead{J} &  \colhead{H} &  \colhead{K} &  \colhead{EW$_{H\alpha}$} }
\startdata 
	(gcm$^{-3}$)&($^o$)&(mag)&(mag)&(mag)&(mag)&(mag)&(mag)&($\AA$)&(mag)&(mag)&(mag)&(mag)&(mag)&(mag)&($\AA$)&(mag)&(mag)&(mag)&(mag)&(mag)&(mag)&($\AA$)\\
		 	&	 	&	 &	&	&B1	&	 	&	 	&	 	&	 	&		&	 	&	B3		 	&	 	&	 	&	 	&		&	 	& &	B7	&	 	&	 	&		\\ \hline
&	18	&	-6.557	&	0.416	&	1.010	&	-5.519	&	-5.577	&	-5.996	&	-6.86	&	-4.924	&	0.332	&	0.834	&	-3.990	&	-4.114	&	-4.457	&	-10.47	&	-2.794	&	0.357	&	0.937	&	-1.853	&	-1.951	&	-2.343	&	-13.89	\\
			&	45	&	-6.380	&	0.408	&	0.984	&	-5.366	&	-5.415	&	-5.825	&	-9.95	&	-4.721	&	0.326	&	0.831	&	-3.814	&	-3.930	&	-4.264	&	-18.2	&	-2.655	&	0.351	&	0.910	&	-1.722	&	-1.814	&	-2.209	&	-19.62	\\
		100	&	60	&	-6.143	&	0.400	&	0.965	&	-5.163	&	-5.201	&	-5.596	&	-11.53	&	-4.474	&	0.323	&	0.836	&	-3.596	&	-3.703	&	-4.024	&	-23.92	&	-2.493	&	0.343	&	0.878	&	-1.577	&	-1.660	&	-2.055	&	-26.98	\\
			&	72	&	-5.748	&	0.403	&	0.993	&	-4.806	&	-4.826	&	-5.198	&	-11.16	&	-4.103	&	0.333	&	0.882	&	-3.254	&	-3.344	&	-3.653	&	-23.13	&	-2.259	&	0.330	&	0.842	&	-1.381	&	-1.449	&	-1.834	&	-25.19	\\
			&	84	&	-4.992	&	0.466	&	1.027	&	-3.950	&	-3.944	&	-4.348	&	-8.18	&	-3.514	&	0.400	&	0.986	&	-2.554	&	-2.622	&	-2.983	&	-20.41	&	-1.790	&	0.314	&	0.843	&	-0.989	&	-1.029	&	-1.383	&	-22.62	\\\hline
			&	18	&	-6.206	&	0.437	&	1.104	&	-5.137	&	-5.207	&	-5.633	&	-9.61	&	-4.691	&	0.362	&	0.878	&	-3.692	&	-3.825	&	-4.189	&	-12.87	&	-2.526	&	0.361	&	0.973	&	-1.606	&	-1.705	&	-2.075	&	-14.7	\\
			&	45	&	-6.046	&	0.430	&	1.074	&	-4.996	&	-5.058	&	-5.477	&	-14.21	&	-4.503	&	0.352	&	0.865	&	-3.529	&	-3.657	&	-4.011	&	-21.95	&	-2.383	&	0.357	&	0.950	&	-1.472	&	-1.565	&	-1.937	&	-20.75	\\
		75	&	60	&	-5.835	&	0.423	&	1.045	&	-4.812	&	-4.865	&	-5.272	&	-17.01	&	-4.268	&	0.342	&	0.861	&	-3.330	&	-3.449	&	-3.790	&	-29.31	&	-2.218	&	0.350	&	0.921	&	-1.328	&	-1.411	&	-1.779	&	-28.86	\\
			&	72	&	-5.478	&	0.420	&	1.042	&	-4.495	&	-4.533	&	-4.918	&	-16	&	-3.908	&	0.342	&	0.895	&	-3.019	&	-3.122	&	-3.441	&	-28.24	&	-1.981	&	0.338	&	0.886	&	-1.142	&	-1.209	&	-1.560	&	-27.31	\\
			&	84	&	-4.715	&	0.479	&	1.104	&	-3.659	&	-3.673	&	-4.070	&	-13.76	&	-3.272	&	0.407	&	1.032	&	-2.323	&	-2.400	&	-2.739	&	-26.79	&	-1.518	&	0.319	&	0.879	&	-0.788	&	-0.824	&	-1.125	&	-23.77	\\\hline
			&	18	&	-5.683	&	0.453	&	1.219	&	-4.605	&	-4.685	&	-5.108	&	-15.32	&	-4.311	&	0.406	&	0.999	&	-3.257	&	-3.391	&	-3.774	&	-16.63	&	-2.174	&	0.366	&	1.022	&	-1.292	&	-1.387	&	-1.726	&	-15.36	\\
			&	45	&	-5.541	&	0.450	&	1.196	&	-4.477	&	-4.549	&	-4.968	&	-22.66	&	-4.147	&	0.397	&	0.968	&	-3.113	&	-3.243	&	-3.617	&	-27.99	&	-2.026	&	0.363	&	1.004	&	-1.160	&	-1.248	&	-1.582	&	-21.77	\\
		50	&	60	&	-5.359	&	0.445	&	1.168	&	-4.315	&	-4.379	&	-4.789	&	-28.09	&	-3.942	&	0.385	&	0.939	&	-2.940	&	-3.063	&	-3.424	&	-37.79	&	-1.857	&	0.357	&	0.980	&	-1.025	&	-1.101	&	-1.423	&	-30.47	\\
			&	72	&	-5.055	&	0.441	&	1.144	&	-4.048	&	-4.099	&	-4.490	&	-26.79	&	-3.620	&	0.371	&	0.934	&	-2.677	&	-2.786	&	-3.121	&	-36.74	&	-1.621	&	0.344	&	0.948	&	-0.861	&	-0.921	&	-1.213	&	-29.42	\\
			&	84	&	-4.306	&	0.490	&	1.212	&	-3.272	&	-3.301	&	-3.676	&	-26.64	&	-2.951	&	0.416	&	1.079	&	-2.038	&	-2.117	&	-2.421	&	-36.43	&	-1.173	&	0.317	&	0.924	&	-0.570	&	-0.596	&	-0.818	&	-24.14	\\\hline
			&	18	&	-4.758	&	0.463	&	1.334	&	-3.731	&	-3.799	&	-4.194	&	-30.51	&	-3.540	&	0.450	&	1.249	&	-2.511	&	-2.613	&	-2.985	&	-23.48	&	-1.614	&	0.378	&	1.100	&	-0.834	&	-0.903	&	-1.172	&	-15.45	\\
			&	45	&	-4.641	&	0.461	&	1.325	&	-3.640	&	-3.698	&	-4.082	&	-46.06	&	-3.408	&	0.445	&	1.228	&	-2.413	&	-2.506	&	-2.862	&	-38.06	&	-1.466	&	0.373	&	1.090	&	-0.737	&	-0.794	&	-1.039	&	-22.61	\\
		25	&	60	&	-4.496	&	0.458	&	1.312	&	-3.528	&	-3.574	&	-3.945	&	-57.27	&	-3.249	&	0.439	&	1.199	&	-2.296	&	-2.379	&	-2.715	&	-50.81	&	-1.300	&	0.363	&	1.074	&	-0.639	&	-0.684	&	-0.899	&	-31.29	\\
			&	72	&	-4.268	&	0.453	&	1.295	&	-3.352	&	-3.385	&	-3.731	&	-60.1	&	-3.010	&	0.427	&	1.155	&	-2.127	&	-2.198	&	-2.498	&	-50.01	&	-1.085	&	0.338	&	1.042	&	-0.526	&	-0.557	&	-0.728	&	-31.7	\\
			&	84	&	-3.644	&	0.468	&	1.318	&	-2.822	&	-2.829	&	-3.109	&	-54.52	&	-2.429	&	0.420	&	1.186	&	-1.702	&	-1.742	&	-1.955	&	-46.17	&	-0.703	&	0.279	&	0.968	&	-0.341	&	-0.344	&	-0.438	&	-23.22	\\\hline
			&	18	&	-3.501	&	0.419	&	1.375	&	-2.972	&	-2.939	&	-3.105	&	-39.25	&	-2.349	&	0.425	&	1.393	&	-1.840	&	-1.822	&	-1.959	&	-27.6	&	-0.832	&	0.353	&	1.229	&	-0.464	&	-0.451	&	-0.535	&	-14.25	\\
			&	45	&	-3.425	&	0.408	&	1.361	&	-2.934	&	-2.893	&	-3.047	&	-49.49	&	-2.264	&	0.411	&	1.378	&	-1.802	&	-1.776	&	-1.896	&	-44.24	&	-0.741	&	0.331	&	1.199	&	-0.430	&	-0.410	&	-0.473	&	-21.62	\\
		10	&	60	&	-3.334	&	0.393	&	1.342	&	-2.887	&	-2.839	&	-2.977	&	-58.9	&	-2.164	&	0.393	&	1.360	&	-1.756	&	-1.724	&	-1.822	&	-57.53	&	-0.641	&	0.305	&	1.160	&	-0.393	&	-0.367	&	-0.406	&	-29.01	\\
			&	72	&	-3.205	&	0.370	&	1.311	&	-2.814	&	-2.759	&	-2.877	&	-64.74	&	-2.026	&	0.367	&	1.330	&	-1.688	&	-1.650	&	-1.719	&	-59.93	&	-0.519	&	0.265	&	1.092	&	-0.345	&	-0.314	&	-0.326	&	-31.41	\\
			&	84	&	-2.870	&	0.325	&	1.245	&	-2.592	&	-2.526	&	-2.599	&	-46.9	&	-1.723	&	0.311	&	1.235	&	-1.512	&	-1.466	&	-1.483	&	-45.67	&	-0.319	&	0.188	&	0.898	&	-0.252	&	-0.216	&	-0.193	&	-19.92	\\\hline
			&	18	&	-2.794	&	0.236	&	1.202	&	-2.759	&	-2.662	&	-2.659	&	-23.17	&	-1.802	&	0.237	&	1.199	&	-1.716	&	-1.642	&	-1.656	&	-18.95	&	-0.459	&	0.196	&	1.095	&	-0.395	&	-0.342	&	-0.344	&	-10.29	\\
			&	45	&	-2.754	&	0.222	&	1.170	&	-2.743	&	-2.642	&	-2.633	&	-21.65	&	-1.764	&	0.222	&	1.162	&	-1.702	&	-1.625	&	-1.631	&	-22.92	&	-0.421	&	0.179	&	1.046	&	-0.383	&	-0.328	&	-0.322	&	-13.37	\\
		5	&	60	&	-2.711	&	0.206	&	1.129	&	-2.724	&	-2.620	&	-2.604	&	-22.2	&	-1.721	&	0.206	&	1.117	&	-1.685	&	-1.605	&	-1.603	&	-28.06	&	-0.379	&	0.160	&	0.989	&	-0.369	&	-0.312	&	-0.297	&	-17.61	\\
			&	72	&	-2.653	&	0.188	&	1.065	&	-2.693	&	-2.586	&	-2.562	&	-22.65	&	-1.665	&	0.185	&	1.050	&	-1.658	&	-1.577	&	-1.565	&	-30.02	&	-0.328	&	0.136	&	0.904	&	-0.349	&	-0.291	&	-0.264	&	-19.61	\\
			&	84	&	-2.509	&	0.153	&	0.923	&	-2.589	&	-2.480	&	-2.442	&	-13.34	&	-1.537	&	0.147	&	0.900	&	-1.576	&	-1.493	&	-1.465	&	-18.88	&	-0.237	&	0.100	&	0.714	&	-0.295	&	-0.237	&	-0.197	&	-10.33	\\\hline
			&	18	&	-2.480	&	0.061	&	0.713	&	-2.687	&	-2.563	&	-2.494	&	-8.05	&	-1.525	&	0.071	&	0.701	&	-1.658	&	-1.562	&	-1.516	&	-8.31	&	-0.287	&	0.064	&	0.630	&	-0.370	&	-0.302	&	-0.271	&	-4.19	\\
			&	45	&	-2.466	&	0.054	&	0.677	&	-2.681	&	-2.556	&	-2.485	&	-5.83	&	-1.512	&	0.064	&	0.666	&	-1.652	&	-1.556	&	-1.508	&	-7.07	&	-0.275	&	0.058	&	0.592	&	-0.365	&	-0.297	&	-0.264	&	-3.97	\\
		2.5	&	60	&	-2.450	&	0.047	&	0.638	&	-2.673	&	-2.547	&	-2.474	&	-4.56	&	-1.497	&	0.057	&	0.626	&	-1.646	&	-1.548	&	-1.498	&	-7.11	&	-0.262	&	0.050	&	0.549	&	-0.360	&	-0.292	&	-0.256	&	-5.05	\\
			&	72	&	-2.429	&	0.040	&	0.591	&	-2.659	&	-2.533	&	-2.458	&	-3.61	&	-1.479	&	0.050	&	0.578	&	-1.636	&	-1.537	&	-1.485	&	-6.97	&	-0.246	&	0.042	&	0.496	&	-0.353	&	-0.284	&	-0.245	&	-5.67	\\
			&	84	&	-2.370	&	0.030	&	0.499	&	-2.609	&	-2.481	&	-2.404	&	-0.63	&	-1.433	&	0.041	&	0.488	&	-1.599	&	-1.500	&	-1.445	&	-2.15	&	-0.213	&	0.033	&	0.406	&	-0.329	&	-0.260	&	-0.217	&	-0.75	\\\hline
			&	18	&	-2.366	&	-0.023	&	0.169	&	-2.658	&	-2.525	&	-2.436	&	0.08	&	-1.415	&	-0.009	&	0.177	&	-1.631	&	-1.527	&	-1.460	&	0.64	&	-0.209	&	0.000	&	0.163	&	-0.354	&	-0.281	&	-0.234	&	2.23	\\
			&	45	&	-2.363	&	-0.024	&	0.156	&	-2.656	&	-2.523	&	-2.433	&	0.74	&	-1.412	&	-0.010	&	0.164	&	-1.630	&	-1.526	&	-1.458	&	1.49	&	-0.207	&	-0.001	&	0.152	&	-0.352	&	-0.280	&	-0.233	&	2.97	\\
		1	&	60	&	-2.359	&	-0.025	&	0.144	&	-2.653	&	-2.520	&	-2.430	&	1.25	&	-1.409	&	-0.011	&	0.153	&	-1.627	&	-1.523	&	-1.456	&	1.99	&	-0.204	&	-0.002	&	0.142	&	-0.351	&	-0.278	&	-0.231	&	3.29	\\
			&	72	&	-2.353	&	-0.027	&	0.132	&	-2.648	&	-2.515	&	-2.424	&	1.75	&	-1.404	&	-0.012	&	0.142	&	-1.624	&	-1.520	&	-1.452	&	2.43	&	-0.201	&	-0.003	&	0.132	&	-0.349	&	-0.276	&	-0.228	&	3.56	\\
			&	84	&	-2.331	&	-0.028	&	0.114	&	-2.628	&	-2.494	&	-2.404	&	2.69	&	-1.389	&	-0.014	&	0.127	&	-1.610	&	-1.506	&	-1.437	&	4.07	&	-0.192	&	-0.004	&	0.120	&	-0.340	&	-0.267	&	-0.220	&	5.16\\
\enddata
\end{deluxetable}

\floattable
\begin{deluxetable}{ c  c | c c c c c c c| c c c c c c c | c c c c c c c} 
\tabletypesize{\tiny}
\rotate
\tablewidth{0pt}
\tablecolumns{23}
\tablecaption{Synthetic infrared magnitudes and colours computed for n=2.5\label{Tab_n25}}
\tablehead{
    \colhead{$\rho_{0}/10^{-12}$} & \colhead{i} & \colhead{W1} &  \colhead{W12} &  \colhead{W23} &  \colhead{J} &  \colhead{H} &  \colhead{K} &  \colhead{EW$_{H\alpha}$} & \colhead{W1} &  \colhead{W12} &  \colhead{W23} &  \colhead{J} &  \colhead{H} &  \colhead{K} &  \colhead{EW$_{H\alpha}$} & \colhead{W1} &  \colhead{W12} &  \colhead{W23} &  \colhead{J} &  \colhead{H} &  \colhead{K} &  \colhead{EW$_{H\alpha}$} }
\startdata 
	(gcm$^{-3}$)&($^o$)&(mag)&(mag)&(mag)&(mag)&(mag)&(mag)&($\AA$)&(mag)&(mag)&(mag)&(mag)&(mag)&(mag)&($\AA$)&(mag)&(mag)&(mag)&(mag)&(mag)&(mag)&($\AA$)\\
		 	&	 	&	 &	&	&B1	&	 	&	 	&	 	&	 	&		&	 	&	B3		 	&	 	&	 	&	 	&		&	 	& &	B7	&	 	&	 	&		\\ \hline
			&	18	&	-5.163	&	0.359	&	1.030	&	-4.335	&	-4.413	&	-4.716	&	-15.09	&	-4.035	&	0.359	&	0.968	&	-3.171	&	-3.292	&	-3.578	&	-10.81	&	-2.110	&	0.323	&	0.913	&	-1.338	&	-1.436	&	-1.713	&	-8.67	\\
			&	45	&	-4.981	&	0.360	&	1.033	&	-4.153	&	-4.229	&	-4.531	&	-22.28	&	-3.845	&	0.359	&	0.953	&	-2.980	&	-3.100	&	-3.384	&	-17.35	&	-1.931	&	0.321	&	0.903	&	-1.167	&	-1.260	&	-1.536	&	-12.13	\\
		100	&	60	&	-4.763	&	0.362	&	1.038	&	-3.941	&	-4.012	&	-4.310	&	-30.51	&	-3.619	&	0.357	&	0.934	&	-2.761	&	-2.878	&	-3.158	&	-25.74	&	-1.726	&	0.318	&	0.888	&	-0.986	&	-1.070	&	-1.337	&	-18.07	\\
			&	72	&	-4.438	&	0.367	&	1.053	&	-3.635	&	-3.695	&	-3.983	&	-36.92	&	-3.290	&	0.353	&	0.920	&	-2.460	&	-2.566	&	-2.833	&	-29.14	&	-1.445	&	0.309	&	0.869	&	-0.778	&	-0.843	&	-1.078	&	-20.28	\\
			&	84	&	-3.686	&	0.419	&	1.157	&	-2.893	&	-2.925	&	-3.183	&	-38.32	&	-2.585	&	0.385	&	1.033	&	-1.815	&	-1.887	&	-2.111	&	-31.34	&	-0.952	&	0.282	&	0.852	&	-0.465	&	-0.490	&	-0.645	&	-14.59	\\\hline
			&	18	&	-4.867	&	0.353	&	1.038	&	-4.076	&	-4.147	&	-4.436	&	-17.29	&	-3.765	&	0.362	&	1.029	&	-2.927	&	-3.038	&	-3.317	&	-12.32	&	-1.915	&	0.323	&	0.936	&	-1.172	&	-1.264	&	-1.524	&	-9.01	\\
			&	45	&	-4.688	&	0.355	&	1.044	&	-3.897	&	-3.965	&	-4.253	&	-24.77	&	-3.580	&	0.364	&	1.023	&	-2.741	&	-2.851	&	-3.128	&	-19.01	&	-1.734	&	0.322	&	0.930	&	-1.003	&	-1.089	&	-1.344	&	-12.71	\\
		75	&	60	&	-4.474	&	0.356	&	1.052	&	-3.692	&	-3.755	&	-4.038	&	-33.78	&	-3.364	&	0.365	&	1.012	&	-2.532	&	-2.638	&	-2.909	&	-27.93	&	-1.528	&	0.319	&	0.919	&	-0.832	&	-0.907	&	-1.147	&	-18.73	\\
			&	72	&	-4.163	&	0.360	&	1.072	&	-3.406	&	-3.458	&	-3.727	&	-41.87	&	-3.055	&	0.364	&	0.998	&	-2.256	&	-2.349	&	-2.604	&	-32.26	&	-1.252	&	0.307	&	0.902	&	-0.647	&	-0.703	&	-0.901	&	-21.43	\\
			&	84	&	-3.450	&	0.398	&	1.169	&	-2.762	&	-2.779	&	-2.996	&	-39.82	&	-2.384	&	0.379	&	1.070	&	-1.703	&	-1.756	&	-1.943	&	-31.32	&	-0.786	&	0.267	&	0.866	&	-0.386	&	-0.402	&	-0.513	&	-14.18	\\\hline
			&	18	&	-4.464	&	0.342	&	1.033	&	-3.726	&	-3.787	&	-4.057	&	-19.92	&	-3.376	&	0.353	&	1.064	&	-2.597	&	-2.688	&	-2.953	&	-14.25	&	-1.652	&	0.322	&	0.965	&	-0.957	&	-1.037	&	-1.272	&	-9.1	\\
			&	45	&	-4.289	&	0.344	&	1.041	&	-3.560	&	-3.614	&	-3.881	&	-26.36	&	-3.193	&	0.356	&	1.070	&	-2.424	&	-2.510	&	-2.767	&	-21.05	&	-1.469	&	0.322	&	0.963	&	-0.800	&	-0.870	&	-1.091	&	-12.91	\\
		50	&	60	&	-4.084	&	0.345	&	1.051	&	-3.373	&	-3.418	&	-3.676	&	-34.79	&	-2.983	&	0.359	&	1.074	&	-2.234	&	-2.312	&	-2.557	&	-30.2	&	-1.264	&	0.318	&	0.956	&	-0.652	&	-0.709	&	-0.902	&	-18.71	\\
			&	72	&	-3.792	&	0.347	&	1.071	&	-3.131	&	-3.159	&	-3.393	&	-42.79	&	-2.693	&	0.360	&	1.075	&	-2.001	&	-2.063	&	-2.276	&	-35.43	&	-1.002	&	0.298	&	0.940	&	-0.502	&	-0.540	&	-0.683	&	-21.85	\\
			&	84	&	-3.148	&	0.353	&	1.146	&	-2.640	&	-2.628	&	-2.781	&	-34.84	&	-2.100	&	0.351	&	1.099	&	-1.591	&	-1.609	&	-1.733	&	-29.54	&	-0.587	&	0.237	&	0.867	&	-0.307	&	-0.308	&	-0.365	&	-13.13	\\\hline
			&	18	&	-3.771	&	0.320	&	1.003	&	-3.170	&	-3.186	&	-3.403	&	-19.73	&	-2.666	&	0.331	&	1.045	&	-2.046	&	-2.083	&	-2.285	&	-13.86	&	-1.193	&	0.319	&	0.996	&	-0.633	&	-0.671	&	-0.840	&	-7.61	\\
			&	45	&	-3.609	&	0.318	&	1.008	&	-3.059	&	-3.058	&	-3.254	&	-20.96	&	-2.487	&	0.329	&	1.060	&	-1.931	&	-1.952	&	-2.124	&	-18.44	&	-1.016	&	0.313	&	0.999	&	-0.530	&	-0.554	&	-0.687	&	-10.55	\\
		25	&	60	&	-3.425	&	0.313	&	1.013	&	-2.941	&	-2.925	&	-3.092	&	-24.35	&	-2.287	&	0.324	&	1.076	&	-1.811	&	-1.817	&	-1.951	&	-24.96	&	-0.830	&	0.297	&	0.995	&	-0.439	&	-0.449	&	-0.541	&	-15.05	\\
			&	72	&	-3.184	&	0.296	&	1.015	&	-2.801	&	-2.768	&	-2.893	&	-27.58	&	-2.038	&	0.307	&	1.086	&	-1.679	&	-1.668	&	-1.748	&	-29.34	&	-0.620	&	0.259	&	0.962	&	-0.351	&	-0.348	&	-0.390	&	-17.95	\\
			&	84	&	-2.736	&	0.243	&	0.981	&	-2.553	&	-2.489	&	-2.533	&	-17.43	&	-1.632	&	0.248	&	1.031	&	-1.477	&	-1.438	&	-1.437	&	-19.04	&	-0.333	&	0.177	&	0.806	&	-0.240	&	-0.216	&	-0.200	&	-9.24	\\\hline
			&	18	&	-2.910	&	0.248	&	0.950	&	-2.788	&	-2.705	&	-2.731	&	-8.79	&	-1.905	&	0.250	&	0.955	&	-1.740	&	-1.679	&	-1.718	&	-7.14	&	-0.579	&	0.236	&	0.962	&	-0.417	&	-0.379	&	-0.407	&	-3.28	\\
			&	45	&	-2.811	&	0.222	&	0.925	&	-2.749	&	-2.657	&	-2.664	&	-7.36	&	-1.806	&	0.222	&	0.923	&	-1.704	&	-1.635	&	-1.653	&	-6.98	&	-0.484	&	0.206	&	0.925	&	-0.386	&	-0.341	&	-0.348	&	-4.08	\\
		10	&	60	&	-2.713	&	0.190	&	0.882	&	-2.710	&	-2.611	&	-2.600	&	-7.03	&	-1.707	&	0.189	&	0.875	&	-1.667	&	-1.593	&	-1.590	&	-8.13	&	-0.394	&	0.171	&	0.870	&	-0.357	&	-0.307	&	-0.293	&	-6.03	\\
			&	72	&	-2.610	&	0.152	&	0.799	&	-2.665	&	-2.560	&	-2.532	&	-6.75	&	-1.603	&	0.147	&	0.793	&	-1.626	&	-1.548	&	-1.523	&	-8.95	&	-0.305	&	0.130	&	0.772	&	-0.325	&	-0.273	&	-0.239	&	-7.31	\\
			&	84	&	-2.448	&	0.096	&	0.624	&	-2.579	&	-2.468	&	-2.418	&	-3.27	&	-1.459	&	0.090	&	0.618	&	-1.559	&	-1.475	&	-1.427	&	-3.52	&	-0.204	&	0.079	&	0.567	&	-0.280	&	-0.224	&	-0.175	&	-1.84	\\\hline
			&	18	&	-2.535	&	0.095	&	0.735	&	-2.698	&	-2.579	&	-2.521	&	-3.2	&	-1.577	&	0.102	&	0.727	&	-1.668	&	-1.576	&	-1.543	&	-2.2	&	-0.326	&	0.096	&	0.698	&	-0.374	&	-0.311	&	-0.287	&	0.18	\\
			&	45	&	-2.493	&	0.074	&	0.667	&	-2.683	&	-2.561	&	-2.496	&	-1.95	&	-1.538	&	0.082	&	0.656	&	-1.655	&	-1.561	&	-1.519	&	-1.35	&	-0.290	&	0.076	&	0.622	&	-0.363	&	-0.298	&	-0.266	&	0.43	\\
		5	&	60	&	-2.456	&	0.055	&	0.588	&	-2.668	&	-2.543	&	-2.473	&	-1.24	&	-1.503	&	0.063	&	0.575	&	-1.642	&	-1.546	&	-1.498	&	-1.22	&	-0.257	&	0.057	&	0.539	&	-0.353	&	-0.286	&	-0.247	&	-0.03	\\
			&	72	&	-2.421	&	0.039	&	0.493	&	-2.649	&	-2.523	&	-2.448	&	-0.59	&	-1.470	&	0.047	&	0.478	&	-1.626	&	-1.529	&	-1.476	&	-1.04	&	-0.227	&	0.039	&	0.440	&	-0.341	&	-0.273	&	-0.229	&	-0.28	\\
			&	84	&	-2.362	&	0.018	&	0.352	&	-2.608	&	-2.481	&	-2.402	&	0.9	&	-1.422	&	0.027	&	0.341	&	-1.597	&	-1.499	&	-1.441	&	1.42	&	-0.192	&	0.023	&	0.310	&	-0.321	&	-0.253	&	-0.204	&	2.45	\\\hline
						&	18	&	-2.397	&	0.000	&	0.343	&	-2.666	&	-2.536	&	-2.452	&	-0.09	&	-1.444	&	0.013	&	0.342	&	-1.638	&	-1.537	&	-1.475	&	0.86	&	-0.233	&	0.020	&	0.325	&	-0.359	&	-0.288	&	-0.246	&	2.63	\\
			&	45	&	-2.384	&	-0.007	&	0.291	&	-2.660	&	-2.529	&	-2.443	&	0.68	&	-1.432	&	0.006	&	0.290	&	-1.634	&	-1.531	&	-1.468	&	1.58	&	-0.223	&	0.013	&	0.275	&	-0.355	&	-0.283	&	-0.240	&	3.14	\\
		2.5	&	60	&	-2.372	&	-0.013	&	0.245	&	-2.654	&	-2.522	&	-2.435	&	1.21	&	-1.421	&	0.001	&	0.245	&	-1.629	&	-1.526	&	-1.461	&	1.98	&	-0.213	&	0.008	&	0.230	&	-0.351	&	-0.279	&	-0.234	&	3.29	\\
			&	72	&	-2.359	&	-0.017	&	0.203	&	-2.645	&	-2.513	&	-2.425	&	1.73	&	-1.411	&	-0.003	&	0.205	&	-1.622	&	-1.519	&	-1.453	&	2.37	&	-0.204	&	0.004	&	0.191	&	-0.347	&	-0.275	&	-0.228	&	3.49	\\
			&	84	&	-2.336	&	-0.021	&	0.147	&	-2.626	&	-2.493	&	-2.404	&	2.56	&	-1.394	&	-0.007	&	0.154	&	-1.609	&	-1.505	&	-1.438	&	3.7	&	-0.193	&	0.001	&	0.146	&	-0.338	&	-0.266	&	-0.218	&	4.86	\\\hline
			&	18	&	-2.349	&	-0.035	&	0.051	&	-2.652	&	-2.518	&	-2.426	&	2.06	&	-1.400	&	-0.020	&	0.069	&	-1.626	&	-1.521	&	-1.452	&	2.91	&	-0.197	&	-0.009	&	0.075	&	-0.350	&	-0.277	&	-0.228	&	4.47	\\
			&	45	&	-2.346	&	-0.036	&	0.038	&	-2.650	&	-2.516	&	-2.423	&	2.39	&	-1.397	&	-0.021	&	0.057	&	-1.625	&	-1.520	&	-1.450	&	3.3	&	-0.195	&	-0.010	&	0.065	&	-0.349	&	-0.276	&	-0.226	&	4.83	\\
1	&	60	&	-2.342	&	-0.037	&	0.029	&	-2.648	&	-2.513	&	-2.421	&	2.67	&	-1.394	&	-0.022	&	0.048	&	-1.623	&	-1.518	&	-1.448	&	3.59	&	-0.193	&	-0.010	&	0.057	&	-0.348	&	-0.274	&	-0.225	&	5.07	\\
&	72	&	-2.339	&	-0.037	&	0.021	&	-2.644	&	-2.510	&	-2.417	&	2.97	&	-1.392	&	-0.022	&	0.042	&	-1.621	&	-1.516	&	-1.445	&	3.91	&	-0.191	&	-0.011	&	0.052	&	-0.347	&	-0.273	&	-0.224	&	5.33	\\
&	84	&	-2.330	&	-0.038	&	0.014	&	-2.636	&	-2.502	&	-2.409	&	3.35	&	-1.386	&	-0.023	&	0.036	&	-1.615	&	-1.510	&	-1.440	&	4.47	&	-0.188	&	-0.011	&	0.047	&	-0.343	&	-0.270	&	-0.220	&	6.06	\\
\enddata 
\end{deluxetable}

\floattable
\begin{deluxetable}{ c  c | c c c c c c c| c c c c c c c | c c c c c c c} 
\tabletypesize{\tiny}
\rotate
\tablewidth{0pt}
\tablecolumns{23}
\tablecaption{Synthetic infrared magnitudes and colours computed for n=3.\label{Tab_n30}}
\tablehead{
    \colhead{$\rho_{0}/10^{-12}$} & \colhead{i} & \colhead{W1} &  \colhead{W12} &  \colhead{W23} &  \colhead{J} &  \colhead{H} &  \colhead{K} &  \colhead{EW$_{H\alpha}$} & \colhead{W1} &  \colhead{W12} &  \colhead{W23} &  \colhead{J} &  \colhead{H} &  \colhead{K} &  \colhead{EW$_{H\alpha}$} & \colhead{W1} &  \colhead{W12} &  \colhead{W23} &  \colhead{J} &  \colhead{H} &  \colhead{K} &  \colhead{EW$_{H\alpha}$} }
\startdata 
	(gcm$^{-3}$)&($^o$)&(mag)&(mag)&(mag)&(mag)&(mag)&(mag)&($\AA$)&(mag)&(mag)&(mag)&(mag)&(mag)&(mag)&($\AA$)&(mag)&(mag)&(mag)&(mag)&(mag)&(mag)&($\AA$)\\
		 	&	 	&	 &	&	&B1	&	 	&	 	&	 	&	 	&		&	 	&	B3		 	&	 	&	 	&	 	&		&	 	& &	B7	&	 	&	 	&		\\ \hline				
			&	18	&	-4.343	&	0.264	&	0.806	&	-3.786	&	-3.829	&	-4.027	&	-9.74	&	-3.285	&	0.281	&	0.871	&	-2.683	&	-2.761	&	-2.948	&	-6.74	&	-1.700	&	0.282	&	0.851	&	-1.079	&	-1.160	&	-1.362	&	-4.43	\\
			&	45	&	-4.132	&	0.266	&	0.815	&	-3.579	&	-3.618	&	-3.814	&	-11.52	&	-3.067	&	0.285	&	0.883	&	-2.467	&	-2.543	&	-2.726	&	-8.8	&	-1.495	&	0.282	&	0.853	&	-0.887	&	-0.963	&	-1.159	&	-6.34	\\
		100	&	60	&	-3.886	&	0.267	&	0.828	&	-3.347	&	-3.380	&	-3.570	&	-15.03	&	-2.818	&	0.288	&	0.897	&	-2.229	&	-2.300	&	-2.475	&	-13.13	&	-1.264	&	0.280	&	0.851	&	-0.697	&	-0.761	&	-0.937	&	-9.89	\\
			&	72	&	-3.546	&	0.268	&	0.853	&	-3.052	&	-3.070	&	-3.238	&	-19.19	&	-2.483	&	0.291	&	0.914	&	-1.942	&	-1.996	&	-2.147	&	-17.45	&	-0.966	&	0.265	&	0.844	&	-0.509	&	-0.551	&	-0.677	&	-12.7	\\
			&	84	&	-2.882	&	0.258	&	0.917	&	-2.579	&	-2.548	&	-2.628	&	-13.18	&	-1.877	&	0.275	&	0.940	&	-1.537	&	-1.540	&	-1.602	&	-11.31	&	-0.531	&	0.206	&	0.767	&	-0.298	&	-0.299	&	-0.337	&	-5.47	\\\hline
			&	18	&	-4.132	&	0.255	&	0.792	&	-3.613	&	-3.645	&	-3.832	&	-8.97	&	-3.062	&	0.267	&	0.846	&	-2.507	&	-2.570	&	-2.746	&	-6.21	&	-1.543	&	0.279	&	0.858	&	-0.954	&	-1.027	&	-1.215	&	-3.77	\\
			&	45	&	-3.924	&	0.256	&	0.800	&	-3.412	&	-3.439	&	-3.624	&	-10.47	&	-2.842	&	0.270	&	0.860	&	-2.296	&	-2.356	&	-2.525	&	-8.23	&	-1.336	&	0.279	&	0.861	&	-0.768	&	-0.833	&	-1.011	&	-5.59	\\
		75	&	60	&	-3.684	&	0.257	&	0.811	&	-3.193	&	-3.211	&	-3.386	&	-13.51	&	-2.591	&	0.272	&	0.875	&	-2.071	&	-2.122	&	-2.277	&	-12.2	&	-1.106	&	0.275	&	0.861	&	-0.595	&	-0.646	&	-0.795	&	-8.82	\\
			&	72	&	-3.356	&	0.254	&	0.831	&	-2.933	&	-2.931	&	-3.075	&	-16.73	&	-2.261	&	0.271	&	0.895	&	-1.818	&	-1.849	&	-1.964	&	-16	&	-0.822	&	0.254	&	0.850	&	-0.436	&	-0.465	&	-0.559	&	-11.31	\\
			&	84	&	-2.762	&	0.222	&	0.850	&	-2.557	&	-2.508	&	-2.559	&	-10.44	&	-1.714	&	0.233	&	0.887	&	-1.500	&	-1.482	&	-1.502	&	-8.95	&	-0.430	&	0.183	&	0.740	&	-0.268	&	-0.259	&	-0.270	&	-4.25	\\\hline
			&	18	&	-3.838	&	0.243	&	0.773	&	-3.368	&	-3.386	&	-3.559	&	-8	&	-2.762	&	0.248	&	0.805	&	-2.266	&	-2.311	&	-2.474	&	-5.27	&	-1.324	&	0.270	&	0.853	&	-0.791	&	-0.847	&	-1.014	&	-2.61	\\
			&	45	&	-3.635	&	0.243	&	0.780	&	-3.185	&	-3.192	&	-3.358	&	-8.77	&	-2.541	&	0.248	&	0.820	&	-2.075	&	-2.108	&	-2.257	&	-6.89	&	-1.117	&	0.269	&	0.858	&	-0.625	&	-0.668	&	-0.814	&	-4.17	\\
		50	&	60	&	-3.404	&	0.242	&	0.788	&	-2.998	&	-2.989	&	-3.136	&	-10.7	&	-2.292	&	0.247	&	0.837	&	-1.883	&	-1.901	&	-2.020	&	-9.96	&	-0.892	&	0.261	&	0.859	&	-0.484	&	-0.510	&	-0.616	&	-6.87	\\
			&	72	&	-3.103	&	0.230	&	0.797	&	-2.799	&	-2.768	&	-2.870	&	-12.43	&	-1.980	&	0.235	&	0.855	&	-1.691	&	-1.687	&	-1.749	&	-12.6	&	-0.636	&	0.228	&	0.837	&	-0.362	&	-0.370	&	-0.420	&	-8.84	\\
			&	84	&	-2.630	&	0.172	&	0.743	&	-2.548	&	-2.476	&	-2.492	&	-7.01	&	-1.543	&	0.173	&	0.787	&	-1.481	&	-1.436	&	-1.412	&	-5.83	&	-0.318	&	0.148	&	0.684	&	-0.245	&	-0.222	&	-0.204	&	-2.53	\\\hline
			&	18	&	-3.329	&	0.228	&	0.743	&	-2.980	&	-2.954	&	-3.079	&	-5.76	&	-2.288	&	0.231	&	0.746	&	-1.910	&	-1.904	&	-2.030	&	-3.51	&	-0.937	&	0.247	&	0.806	&	-0.551	&	-0.561	&	-0.673	&	-0.61	\\
			&	45	&	-3.142	&	0.221	&	0.745	&	-2.867	&	-2.820	&	-2.915	&	-5.23	&	-2.088	&	0.222	&	0.753	&	-1.798	&	-1.772	&	-1.860	&	-3.84	&	-0.744	&	0.235	&	0.808	&	-0.452	&	-0.445	&	-0.516	&	-1.37	\\
		25	&	60	&	-2.945	&	0.204	&	0.741	&	-2.765	&	-2.700	&	-2.759	&	-5.41	&	-1.881	&	0.200	&	0.755	&	-1.698	&	-1.656	&	-1.700	&	-5.07	&	-0.560	&	0.206	&	0.795	&	-0.375	&	-0.355	&	-0.383	&	-2.98	\\
			&	72	&	-2.736	&	0.165	&	0.700	&	-2.669	&	-2.589	&	-2.606	&	-5.39	&	-1.665	&	0.157	&	0.722	&	-1.608	&	-1.552	&	-1.547	&	-5.93	&	-0.383	&	0.156	&	0.730	&	-0.311	&	-0.281	&	-0.268	&	-4.06	\\
			&	84	&	-2.476	&	0.096	&	0.541	&	-2.561	&	-2.460	&	-2.427	&	-2.52	&	-1.437	&	0.090	&	0.557	&	-1.519	&	-1.445	&	-1.395	&	-1.6	&	-0.210	&	0.087	&	0.525	&	-0.253	&	-0.209	&	-0.163	&	0.29	\\\hline
			&	18	&	-2.700	&	0.162	&	0.697	&	-2.734	&	-2.632	&	-2.612	&	-2.04	&	-1.733	&	0.165	&	0.699	&	-1.703	&	-1.626	&	-1.628	&	-0.61	&	-0.451	&	0.158	&	0.705	&	-0.397	&	-0.346	&	-0.349	&	1.61	\\
			&	45	&	-2.597	&	0.126	&	0.653	&	-2.698	&	-2.587	&	-2.547	&	-1.17	&	-1.634	&	0.130	&	0.646	&	-1.669	&	-1.585	&	-1.566	&	-0.09	&	-0.358	&	0.121	&	0.643	&	-0.369	&	-0.312	&	-0.293	&	1.78	\\
		10	&	60	&	-2.514	&	0.090	&	0.574	&	-2.668	&	-2.551	&	-2.496	&	-0.7	&	-1.553	&	0.093	&	0.561	&	-1.641	&	-1.552	&	-1.517	&	-0.06	&	-0.283	&	0.084	&	0.553	&	-0.346	&	-0.285	&	-0.250	&	1.45	\\
			&	72	&	-2.445	&	0.057	&	0.456	&	-2.639	&	-2.519	&	-2.452	&	-0.18	&	-1.486	&	0.059	&	0.440	&	-1.616	&	-1.525	&	-1.477	&	0.11	&	-0.225	&	0.051	&	0.429	&	-0.326	&	-0.263	&	-0.217	&	1.33	\\
			&	84	&	-2.367	&	0.020	&	0.296	&	-2.601	&	-2.477	&	-2.401	&	1.09	&	-1.417	&	0.025	&	0.287	&	-1.587	&	-1.493	&	-1.434	&	1.97	&	-0.178	&	0.023	&	0.275	&	-0.307	&	-0.242	&	-0.189	&	3.33	\\\hline
			&	18	&	-2.458	&	0.044	&	0.504	&	-2.679	&	-2.554	&	-2.482	&	0.03	&	-1.504	&	0.054	&	0.501	&	-1.651	&	-1.554	&	-1.505	&	1.2	&	-0.278	&	0.056	&	0.482	&	-0.367	&	-0.299	&	-0.266	&	3.05	\\
			&	45	&	-2.419	&	0.023	&	0.415	&	-2.665	&	-2.537	&	-2.459	&	0.75	&	-1.468	&	0.033	&	0.411	&	-1.639	&	-1.540	&	-1.484	&	1.77	&	-0.244	&	0.036	&	0.391	&	-0.356	&	-0.287	&	-0.247	&	3.42	\\
		5	&	60	&	-2.391	&	0.007	&	0.325	&	-2.653	&	-2.524	&	-2.441	&	1.23	&	-1.441	&	0.018	&	0.322	&	-1.629	&	-1.528	&	-1.468	&	2.06	&	-0.220	&	0.020	&	0.301	&	-0.348	&	-0.278	&	-0.233	&	3.49	\\
			&	72	&	-2.369	&	-0.004	&	0.244	&	-2.641	&	-2.511	&	-2.426	&	1.73	&	-1.422	&	0.008	&	0.242	&	-1.620	&	-1.518	&	-1.455	&	2.41	&	-0.202	&	0.010	&	0.222	&	-0.341	&	-0.270	&	-0.222	&	3.67	\\
			&	84	&	-2.342	&	-0.015	&	0.149	&	-2.623	&	-2.492	&	-2.405	&	2.5	&	-1.400	&	-0.003	&	0.155	&	-1.607	&	-1.505	&	-1.439	&	3.34	&	-0.187	&	0.001	&	0.146	&	-0.332	&	-0.261	&	-0.211	&	4.8	\\\hline
			&	18	&	-2.373	&	-0.017	&	0.208	&	-2.659	&	-2.527	&	-2.439	&	1.43	&	-1.423	&	-0.002	&	0.216	&	-1.633	&	-1.529	&	-1.464	&	2.43	&	-0.217	&	0.007	&	0.212	&	-0.355	&	-0.283	&	-0.238	&	4.11	\\
			&	45	&	-2.361	&	-0.023	&	0.153	&	-2.653	&	-2.520	&	-2.431	&	1.94	&	-1.411	&	-0.009	&	0.162	&	-1.628	&	-1.524	&	-1.457	&	2.9	&	-0.207	&	0.001	&	0.161	&	-0.352	&	-0.279	&	-0.232	&	4.48	\\
		2.5	&	60	&	-2.352	&	-0.027	&	0.114	&	-2.648	&	-2.515	&	-2.424	&	2.32	&	-1.403	&	-0.013	&	0.125	&	-1.624	&	-1.520	&	-1.452	&	3.22	&	-0.200	&	-0.003	&	0.125	&	-0.349	&	-0.276	&	-0.228	&	4.68	\\
			&	72	&	-2.344	&	-0.030	&	0.087	&	-2.643	&	-2.509	&	-2.418	&	2.7	&	-1.397	&	-0.016	&	0.099	&	-1.620	&	-1.516	&	-1.447	&	3.56	&	-0.195	&	-0.006	&	0.101	&	-0.346	&	-0.273	&	-0.224	&	4.93	\\
			&	84	&	-2.334	&	-0.032	&	0.054	&	-2.634	&	-2.500	&	-2.409	&	3.16	&	-1.389	&	-0.018	&	0.070	&	-1.614	&	-1.509	&	-1.440	&	4.13	&	-0.190	&	-0.007	&	0.076	&	-0.342	&	-0.269	&	-0.220	&	5.64	\\\hline
			&	18	&	-2.344	&	-0.038	&	0.019	&	-2.650	&	-2.515	&	-2.423	&	2.5	&	-1.395	&	-0.023	&	0.040	&	-1.624	&	-1.519	&	-1.449	&	3.4	&	-0.194	&	-0.011	&	0.051	&	-0.349	&	-0.276	&	-0.226	&	4.97	\\
			&	45	&	-2.341	&	-0.039	&	0.007	&	-2.648	&	-2.513	&	-2.420	&	2.76	&	-1.393	&	-0.024	&	0.029	&	-1.623	&	-1.518	&	-1.447	&	3.68	&	-0.192	&	-0.012	&	0.041	&	-0.348	&	-0.274	&	-0.225	&	5.23	\\
		1	&	60	&	-2.338	&	-0.039	&	0.000	&	-2.646	&	-2.511	&	-2.418	&	2.97	&	-1.391	&	-0.024	&	0.022	&	-1.622	&	-1.517	&	-1.446	&	3.9	&	-0.190	&	-0.012	&	0.036	&	-0.347	&	-0.274	&	-0.224	&	5.42	\\
			&	72	&	-2.336	&	-0.040	&	-0.004	&	-2.644	&	-2.509	&	-2.416	&	3.19	&	-1.389	&	-0.025	&	0.019	&	-1.620	&	-1.515	&	-1.444	&	4.15	&	-0.189	&	-0.013	&	0.032	&	-0.346	&	-0.273	&	-0.223	&	5.64	\\
			&	84	&	-2.332	&	-0.040	&	-0.008	&	-2.640	&	-2.505	&	-2.412	&	3.31	&	-1.386	&	-0.025	&	0.015	&	-1.618	&	-1.512	&	-1.442	&	4.45	&	-0.187	&	-0.013	&	0.030	&	-0.345	&	-0.271	&	-0.221	&	6.08	\\
\enddata 
\end{deluxetable}

\floattable
\begin{deluxetable}{ c  c | c c c c c c c| c c c c c c c | c c c c c c c} 
\tabletypesize{\tiny}
\rotate
\tablewidth{0pt}
\tablecolumns{23}
\tablecaption{Synthetic infrared magnitudes and colours computed for n=3.5\label{Tab_n35}}
\tablehead{
    \colhead{$\rho_{0}/10^{-12}$} & \colhead{i} & \colhead{W1} &  \colhead{W12} &  \colhead{W23} &  \colhead{J} &  \colhead{H} &  \colhead{K} &  \colhead{EW$_{H\alpha}$} & \colhead{W1} &  \colhead{W12} &  \colhead{W23} &  \colhead{J} &  \colhead{H} &  \colhead{K} &  \colhead{EW$_{H\alpha}$} & \colhead{W1} &  \colhead{W12} &  \colhead{W23} &  \colhead{J} &  \colhead{H} &  \colhead{K} &  \colhead{EW$_{H\alpha}$} }
\startdata
	(gcm$^{-3}$)&($^o$)&(mag)&(mag)&(mag)&(mag)&(mag)&(mag)&($\AA$)&(mag)&(mag)&(mag)&(mag)&(mag)&(mag)&($\AA$)&(mag)&(mag)&(mag)&(mag)&(mag)&(mag)&($\AA$)\\
		 	&	 	&	 &	&	&B1	&	 	&	 	&	 	&	 	&		&	 	&	B3		 	&	 	&	 	&	 	&		&	 	& &	B7	&	 	&	 	&		\\ \hline		
			&	18	&	-3.890	&	0.199	&	0.633	&	-3.511	&	-3.521	&	-3.659	&	-4.01	&	-2.825	&	0.203	&	0.663	&	-2.422	&	-2.463	&	-2.585	&	-2.04	&	-1.420	&	0.239	&	0.748	&	-0.926	&	-0.988	&	-1.139	&	-0.34	\\
			&	45	&	-3.664	&	0.200	&	0.640	&	-3.294	&	-3.299	&	-3.433	&	-4.62	&	-2.581	&	0.203	&	0.676	&	-2.194	&	-2.230	&	-2.344	&	-2.97	&	-1.197	&	0.238	&	0.752	&	-0.725	&	-0.779	&	-0.920	&	-1.35	\\
		100	&	60	&	-3.406	&	0.198	&	0.649	&	-3.062	&	-3.057	&	-3.181	&	-5.98	&	-2.307	&	0.202	&	0.691	&	-1.953	&	-1.979	&	-2.077	&	-4.89	&	-0.949	&	0.233	&	0.753	&	-0.540	&	-0.578	&	-0.689	&	-3.14	\\
			&	72	&	-3.069	&	0.191	&	0.660	&	-2.811	&	-2.783	&	-2.870	&	-7.17	&	-1.958	&	0.194	&	0.707	&	-1.708	&	-1.709	&	-1.759	&	-6.59	&	-0.658	&	0.205	&	0.737	&	-0.384	&	-0.400	&	-0.453	&	-4.43	\\
			&	84	&	-2.582	&	0.135	&	0.596	&	-2.547	&	-2.473	&	-2.475	&	-3.81	&	-1.508	&	0.134	&	0.627	&	-1.483	&	-1.438	&	-1.405	&	-1.91	&	-0.318	&	0.129	&	0.586	&	-0.252	&	-0.231	&	-0.213	&	0.33	\\\hline
			&	18	&	-3.726	&	0.192	&	0.622	&	-3.378	&	-3.378	&	-3.506	&	-3.64	&	-2.663	&	0.193	&	0.639	&	-2.293	&	-2.322	&	-2.439	&	-1.67	&	-1.286	&	0.230	&	0.731	&	-0.829	&	-0.880	&	-1.020	&	0.2	\\
			&	45	&	-3.503	&	0.192	&	0.629	&	-3.169	&	-3.161	&	-3.285	&	-4.03	&	-2.422	&	0.192	&	0.652	&	-2.074	&	-2.095	&	-2.202	&	-2.47	&	-1.063	&	0.228	&	0.734	&	-0.637	&	-0.677	&	-0.802	&	-0.67	\\
		75	&	60	&	-3.251	&	0.190	&	0.635	&	-2.955	&	-2.934	&	-3.042	&	-4.98	&	-2.152	&	0.189	&	0.667	&	-1.855	&	-1.862	&	-1.944	&	-4.06	&	-0.820	&	0.220	&	0.734	&	-0.474	&	-0.497	&	-0.584	&	-2.21	\\
			&	72	&	-2.935	&	0.175	&	0.638	&	-2.747	&	-2.702	&	-2.765	&	-5.59	&	-1.823	&	0.171	&	0.678	&	-1.654	&	-1.636	&	-1.662	&	-5.24	&	-0.551	&	0.183	&	0.706	&	-0.346	&	-0.349	&	-0.378	&	-3.22	\\
			&	84	&	-2.529	&	0.111	&	0.533	&	-2.550	&	-2.465	&	-2.452	&	-2.66	&	-1.453	&	0.106	&	0.562	&	-1.488	&	-1.431	&	-1.385	&	-1.03	&	-0.264	&	0.106	&	0.532	&	-0.246	&	-0.218	&	-0.186	&	1.05	\\\hline
			&	18	&	-3.494	&	0.183	&	0.608	&	-3.185	&	-3.174	&	-3.290	&	-3.16	&	-2.451	&	0.182	&	0.609	&	-2.115	&	-2.127	&	-2.242	&	-1.19	&	-1.101	&	0.215	&	0.694	&	-0.700	&	-0.733	&	-0.857	&	0.93	\\
			&	45	&	-3.275	&	0.182	&	0.613	&	-2.999	&	-2.972	&	-3.077	&	-3.16	&	-2.216	&	0.179	&	0.621	&	-1.922	&	-1.920	&	-2.015	&	-1.66	&	-0.879	&	0.211	&	0.697	&	-0.534	&	-0.550	&	-0.647	&	0.36	\\
		50	&	60	&	-3.035	&	0.176	&	0.615	&	-2.828	&	-2.781	&	-2.856	&	-3.55	&	-1.959	&	0.171	&	0.632	&	-1.751	&	-1.729	&	-1.785	&	-2.69	&	-0.648	&	0.194	&	0.692	&	-0.407	&	-0.407	&	-0.457	&	-0.78	\\
			&	72	&	-2.765	&	0.146	&	0.593	&	-2.682	&	-2.613	&	-2.640	&	-3.57	&	-1.680	&	0.137	&	0.619	&	-1.612	&	-1.569	&	-1.568	&	-3.28	&	-0.422	&	0.147	&	0.639	&	-0.315	&	-0.297	&	-0.297	&	-1.45	\\
			&	84	&	-2.468	&	0.080	&	0.445	&	-2.558	&	-2.460	&	-2.427	&	-1.12	&	-1.416	&	0.074	&	0.465	&	-1.509	&	-1.438	&	-1.384	&	0.08	&	-0.212	&	0.076	&	0.444	&	-0.250	&	-0.210	&	-0.167	&	2	\\\hline
			&	18	&	-3.087	&	0.173	&	0.585	&	-2.887	&	-2.836	&	-2.907	&	-2	&	-2.093	&	0.178	&	0.578	&	-1.845	&	-1.816	&	-1.901	&	-0.26	&	-0.777	&	0.191	&	0.627	&	-0.509	&	-0.500	&	-0.578	&	1.97	\\
			&	45	&	-2.888	&	0.162	&	0.584	&	-2.779	&	-2.705	&	-2.741	&	-1.47	&	-1.888	&	0.163	&	0.576	&	-1.739	&	-1.689	&	-1.734	&	-0.09	&	-0.578	&	0.172	&	0.622	&	-0.416	&	-0.389	&	-0.423	&	1.94	\\
		25	&	60	&	-2.701	&	0.133	&	0.563	&	-2.695	&	-2.604	&	-2.603	&	-1.24	&	-1.698	&	0.130	&	0.555	&	-1.658	&	-1.594	&	-1.597	&	-0.34	&	-0.407	&	0.132	&	0.586	&	-0.351	&	-0.313	&	-0.307	&	1.47	\\
			&	72	&	-2.542	&	0.087	&	0.472	&	-2.631	&	-2.529	&	-2.497	&	-0.82	&	-1.537	&	0.081	&	0.467	&	-1.599	&	-1.524	&	-1.492	&	-0.3	&	-0.272	&	0.082	&	0.478	&	-0.306	&	-0.260	&	-0.224	&	1.28	\\
			&	84	&	-2.394	&	0.035	&	0.305	&	-2.579	&	-2.464	&	-2.403	&	0.72	&	-1.407	&	0.034	&	0.308	&	-1.555	&	-1.469	&	-1.411	&	1.65	&	-0.174	&	0.037	&	0.299	&	-0.277	&	-0.222	&	-0.169	&	3.34	\\\hline
			&	18	&	-2.600	&	0.113	&	0.547	&	-2.710	&	-2.599	&	-2.557	&	-0.04	&	-1.645	&	0.119	&	0.551	&	-1.684	&	-1.599	&	-1.581	&	1.26	&	-0.390	&	0.116	&	0.551	&	-0.390	&	-0.332	&	-0.322	&	3.19	\\
			&	45	&	-2.501	&	0.074	&	0.485	&	-2.677	&	-2.558	&	-2.497	&	0.62	&	-1.549	&	0.080	&	0.482	&	-1.653	&	-1.561	&	-1.524	&	1.74	&	-0.301	&	0.077	&	0.472	&	-0.362	&	-0.299	&	-0.270	&	3.49	\\
		10	&	60	&	-2.434	&	0.040	&	0.383	&	-2.653	&	-2.529	&	-2.457	&	1.05	&	-1.485	&	0.048	&	0.377	&	-1.631	&	-1.535	&	-1.486	&	1.95	&	-0.241	&	0.044	&	0.364	&	-0.344	&	-0.278	&	-0.236	&	3.5	\\
			&	72	&	-2.389	&	0.016	&	0.273	&	-2.634	&	-2.508	&	-2.429	&	1.54	&	-1.442	&	0.024	&	0.267	&	-1.614	&	-1.517	&	-1.460	&	2.27	&	-0.203	&	0.021	&	0.254	&	-0.331	&	-0.264	&	-0.214	&	3.67	\\
			&	84	&	-2.346	&	-0.007	&	0.157	&	-2.615	&	-2.486	&	-2.402	&	2.31	&	-1.403	&	0.003	&	0.159	&	-1.599	&	-1.500	&	-1.437	&	3.15	&	-0.176	&	0.004	&	0.156	&	-0.321	&	-0.252	&	-0.199	&	4.69	\\\hline
			&	18	&	-2.423	&	0.020	&	0.376	&	-2.670	&	-2.542	&	-2.464	&	1.14	&	-1.473	&	0.032	&	0.378	&	-1.645	&	-1.545	&	-1.489	&	2.26	&	-0.257	&	0.037	&	0.368	&	-0.363	&	-0.294	&	-0.257	&	4.02	\\
			&	45	&	-2.388	&	-0.001	&	0.278	&	-2.658	&	-2.527	&	-2.443	&	1.69	&	-1.439	&	0.012	&	0.280	&	-1.633	&	-1.531	&	-1.470	&	2.72	&	-0.226	&	0.018	&	0.271	&	-0.354	&	-0.283	&	-0.240	&	4.35	\\
		5	&	60	&	-2.366	&	-0.013	&	0.196	&	-2.648	&	-2.517	&	-2.429	&	2.09	&	-1.418	&	0.000	&	0.199	&	-1.625	&	-1.522	&	-1.457	&	3.01	&	-0.207	&	0.007	&	0.191	&	-0.347	&	-0.276	&	-0.229	&	4.5	\\
			&	72	&	-2.352	&	-0.019	&	0.135	&	-2.640	&	-2.508	&	-2.419	&	2.51	&	-1.406	&	-0.006	&	0.143	&	-1.618	&	-1.515	&	-1.449	&	3.35	&	-0.196	&	0.000	&	0.137	&	-0.342	&	-0.270	&	-0.222	&	4.75	\\
			&	84	&	-2.337	&	-0.027	&	0.073	&	-2.631	&	-2.498	&	-2.408	&	2.96	&	-1.393	&	-0.013	&	0.086	&	-1.611	&	-1.508	&	-1.440	&	3.9	&	-0.187	&	-0.005	&	0.088	&	-0.338	&	-0.266	&	-0.216	&	5.41	\\\hline
			&	18	&	-2.363	&	-0.024	&	0.143	&	-2.656	&	-2.523	&	-2.433	&	1.99	&	-1.412	&	-0.010	&	0.155	&	-1.629	&	-1.526	&	-1.458	&	2.99	&	-0.210	&	0.001	&	0.158	&	-0.353	&	-0.281	&	-0.234	&	4.64	\\
			&	45	&	-2.351	&	-0.030	&	0.090	&	-2.650	&	-2.517	&	-2.426	&	2.4	&	-1.402	&	-0.016	&	0.104	&	-1.625	&	-1.521	&	-1.452	&	3.36	&	-0.201	&	-0.005	&	0.111	&	-0.350	&	-0.277	&	-0.229	&	4.94	\\
		2.5	&	60	&	-2.345	&	-0.033	&	0.060	&	-2.646	&	-2.512	&	-2.421	&	2.71	&	-1.397	&	-0.019	&	0.075	&	-1.622	&	-1.518	&	-1.448	&	3.63	&	-0.195	&	-0.008	&	0.083	&	-0.348	&	-0.274	&	-0.226	&	5.13	\\
			&	72	&	-2.339	&	-0.035	&	0.041	&	-2.642	&	-2.508	&	-2.416	&	3.01	&	-1.393	&	-0.020	&	0.058	&	-1.620	&	-1.515	&	-1.445	&	3.92	&	-0.192	&	-0.009	&	0.067	&	-0.346	&	-0.272	&	-0.223	&	5.37	\\
			&	84	&	-2.334	&	-0.036	&	0.020	&	-2.638	&	-2.504	&	-2.411	&	3.23	&	-1.388	&	-0.021	&	0.040	&	-1.616	&	-1.511	&	-1.441	&	4.27	&	-0.189	&	-0.010	&	0.051	&	-0.343	&	-0.270	&	-0.221	&	5.82	\\\hline
			&	18	&	-2.341	&	-0.039	&	0.005	&	-2.649	&	-2.514	&	-2.421	&	2.69	&	-1.393	&	-0.024	&	0.027	&	-1.624	&	-1.519	&	-1.448	&	3.61	&	-0.193	&	-0.012	&	0.040	&	-0.349	&	-0.275	&	-0.226	&	5.18	\\
			&	45	&	-2.338	&	-0.040	&	-0.006	&	-2.646	&	-2.512	&	-2.419	&	2.91	&	-1.391	&	-0.025	&	0.017	&	-1.622	&	-1.517	&	-1.446	&	3.84	&	-0.190	&	-0.013	&	0.032	&	-0.348	&	-0.274	&	-0.224	&	5.39	\\
		1	&	60	&	-2.337	&	-0.040	&	-0.011	&	-2.645	&	-2.511	&	-2.417	&	3.08	&	-1.390	&	-0.025	&	0.012	&	-1.621	&	-1.516	&	-1.445	&	4.03	&	-0.190	&	-0.013	&	0.027	&	-0.347	&	-0.273	&	-0.223	&	5.55	\\
			&	72	&	-2.335	&	-0.040	&	-0.014	&	-2.644	&	-2.509	&	-2.416	&	3.22	&	-1.388	&	-0.025	&	0.010	&	-1.620	&	-1.515	&	-1.444	&	4.22	&	-0.189	&	-0.013	&	0.025	&	-0.346	&	-0.273	&	-0.223	&	5.74	\\
			&	84	&	-2.333	&	-0.041	&	-0.016	&	-2.642	&	-2.507	&	-2.414	&	3.26	&	-1.387	&	-0.026	&	0.008	&	-1.619	&	-1.513	&	-1.442	&	4.34	&	-0.187	&	-0.014	&	0.023	&	-0.345	&	-0.272	&	-0.222	&	6.03 \\
\enddata
\end{deluxetable}

\floattable
\begin{deluxetable}{ c  c | c c c c c c c| c c c c c c c | c c c c c c c} 
\tabletypesize{\tiny}
\rotate
\tablewidth{0pt}
\tablecolumns{23}
\tablecaption{Synthetic infrared magnitudes and colours computed for n=4.\label{Tab_n40}}  	
\tablehead{
    \colhead{$\rho_{0}/10^{-12}$} & \colhead{i} & \colhead{W1} &  \colhead{W12} &  \colhead{W23} &  \colhead{J} &  \colhead{H} &  \colhead{K} &  \colhead{EW$_{H\alpha}$} & \colhead{W1} &  \colhead{W12} &  \colhead{W23} &  \colhead{J} &  \colhead{H} &  \colhead{K} &  \colhead{EW$_{H\alpha}$} & \colhead{W1} &  \colhead{W12} &  \colhead{W23} &  \colhead{J} &  \colhead{H} &  \colhead{K} &  \colhead{EW$_{H\alpha}$} }
\startdata   			
	(gcm$^{-3}$)&($^o$)&(mag)&(mag)&(mag)&(mag)&(mag)&(mag)&($\AA$)&(mag)&(mag)&(mag)&(mag)&(mag)&(mag)&($\AA$)&(mag)&(mag)&(mag)&(mag)&(mag)&(mag)&($\AA$)\\
		 	&	 	&	 &	&	&B1	&	 	&	 	&	 	&	 	&		&	 	&	B3		 	&	 	&	 	&	 	&		&	 	& &	B7	&	 	&	 	&		\\ \hline
			&	18	&	-3.601	&	0.157	&	0.520	&	-3.341	&	-3.327	&	-3.424	&	-1.85	&	-2.555	&	0.154	&	0.523	&	-2.274	&	-2.289	&	-2.376	&	-0.11	&	-1.220	&	0.196	&	0.621	&	-0.832	&	-0.875	&	-0.991	&	1.59	\\
			&	45	&	-3.365	&	0.156	&	0.525	&	-3.120	&	-3.100	&	-3.192	&	-2.03	&	-2.301	&	0.152	&	0.535	&	-2.042	&	-2.051	&	-2.127	&	-0.53	&	-0.984	&	0.193	&	0.621	&	-0.626	&	-0.659	&	-0.762	&	1.16	\\
		100	&	60	&	-3.103	&	0.153	&	0.530	&	-2.896	&	-2.862	&	-2.939	&	-2.52	&	-2.021	&	0.148	&	0.547	&	-1.813	&	-1.807	&	-1.860	&	-1.41	&	-0.728	&	0.183	&	0.617	&	-0.453	&	-0.468	&	-0.533	&	0.35	\\
			&	72	&	-2.790	&	0.132	&	0.520	&	-2.700	&	-2.640	&	-2.670	&	-2.61	&	-1.698	&	0.123	&	0.547	&	-1.623	&	-1.592	&	-1.590	&	-1.85	&	-0.462	&	0.140	&	0.574	&	-0.330	&	-0.324	&	-0.333	&	-0.04	\\
			&	84	&	-2.464	&	0.070	&	0.385	&	-2.555	&	-2.460	&	-2.427	&	-0.33	&	-1.406	&	0.064	&	0.407	&	-1.501	&	-1.433	&	-1.377	&	0.89	&	-0.221	&	0.071	&	0.392	&	-0.251	&	-0.215	&	-0.174	&	2.88	\\\hline
			&	18	&	-3.465	&	0.151	&	0.511	&	-3.232	&	-3.210	&	-3.298	&	-1.61	&	-2.433	&	0.149	&	0.507	&	-2.172	&	-2.177	&	-2.262	&	0.12	&	-1.106	&	0.186	&	0.596	&	-0.752	&	-0.785	&	-0.891	&	1.93	\\
			&	45	&	-3.233	&	0.150	&	0.516	&	-3.019	&	-2.988	&	-3.070	&	-1.67	&	-2.184	&	0.145	&	0.517	&	-1.952	&	-1.947	&	-2.021	&	-0.19	&	-0.871	&	0.182	&	0.596	&	-0.559	&	-0.578	&	-0.666	&	1.62	\\
		75	&	60	&	-2.977	&	0.145	&	0.518	&	-2.822	&	-2.772	&	-2.830	&	-1.94	&	-1.912	&	0.139	&	0.527	&	-1.752	&	-1.729	&	-1.770	&	-0.84	&	-0.625	&	0.167	&	0.589	&	-0.411	&	-0.411	&	-0.456	&	0.98	\\
			&	72	&	-2.698	&	0.114	&	0.489	&	-2.667	&	-2.594	&	-2.605	&	-1.79	&	-1.626	&	0.104	&	0.508	&	-1.604	&	-1.558	&	-1.544	&	-1.03	&	-0.389	&	0.118	&	0.527	&	-0.312	&	-0.294	&	-0.287	&	0.75	\\
			&	84	&	-2.435	&	0.054	&	0.339	&	-2.561	&	-2.459	&	-2.417	&	0.28	&	-1.394	&	0.049	&	0.357	&	-1.515	&	-1.440	&	-1.381	&	1.32	&	-0.196	&	0.055	&	0.343	&	-0.255	&	-0.213	&	-0.166	&	3.26	\\\hline
			&	18	&	-3.273	&	0.144	&	0.500	&	-3.074	&	-3.042	&	-3.120	&	-1.26	&	-2.268	&	0.144	&	0.489	&	-2.028	&	-2.020	&	-2.106	&	0.44	&	-0.950	&	0.172	&	0.560	&	-0.645	&	-0.661	&	-0.755	&	2.37	\\
			&	45	&	-3.046	&	0.142	&	0.502	&	-2.888	&	-2.838	&	-2.900	&	-1.08	&	-2.028	&	0.139	&	0.495	&	-1.839	&	-1.813	&	-1.877	&	0.36	&	-0.718	&	0.165	&	0.559	&	-0.482	&	-0.479	&	-0.542	&	2.26	\\
		50	&	60	&	-2.807	&	0.129	&	0.498	&	-2.740	&	-2.668	&	-2.694	&	-1.06	&	-1.778	&	0.123	&	0.496	&	-1.692	&	-1.646	&	-1.666	&	0.02	&	-0.494	&	0.139	&	0.542	&	-0.371	&	-0.351	&	-0.367	&	1.85	\\
			&	72	&	-2.588	&	0.087	&	0.433	&	-2.636	&	-2.547	&	-2.531	&	-0.69	&	-1.555	&	0.079	&	0.438	&	-1.594	&	-1.531	&	-1.504	&	0.08	&	-0.308	&	0.088	&	0.452	&	-0.302	&	-0.268	&	-0.243	&	1.77	\\
			&	84	&	-2.401	&	0.034	&	0.278	&	-2.571	&	-2.461	&	-2.406	&	0.98	&	-1.392	&	0.032	&	0.289	&	-1.538	&	-1.456	&	-1.396	&	1.9	&	-0.176	&	0.037	&	0.279	&	-0.266	&	-0.216	&	-0.165	&	3.73	\\\hline
			&	18	&	-2.937	&	0.137	&	0.480	&	-2.834	&	-2.767	&	-2.805	&	-0.44	&	-1.971	&	0.144	&	0.476	&	-1.807	&	-1.763	&	-1.821	&	1.06	&	-0.676	&	0.155	&	0.509	&	-0.484	&	-0.464	&	-0.519	&	3.07	\\
			&	45	&	-2.734	&	0.122	&	0.475	&	-2.732	&	-2.641	&	-2.641	&	0.07	&	-1.766	&	0.125	&	0.468	&	-1.707	&	-1.642	&	-1.659	&	1.36	&	-0.477	&	0.129	&	0.496	&	-0.398	&	-0.357	&	-0.369	&	3.24	\\
		25	&	60	&	-2.568	&	0.084	&	0.431	&	-2.664	&	-2.560	&	-2.526	&	0.4	&	-1.599	&	0.085	&	0.422	&	-1.642	&	-1.565	&	-1.546	&	1.41	&	-0.327	&	0.086	&	0.437	&	-0.343	&	-0.295	&	-0.272	&	3.14	\\
			&	72	&	-2.454	&	0.045	&	0.320	&	-2.621	&	-2.508	&	-2.452	&	0.88	&	-1.485	&	0.044	&	0.311	&	-1.601	&	-1.517	&	-1.474	&	1.68	&	-0.229	&	0.044	&	0.316	&	-0.311	&	-0.257	&	-0.213	&	3.25	\\
			&	84	&	-2.362	&	0.007	&	0.186	&	-2.592	&	-2.471	&	-2.398	&	1.88	&	-1.401	&	0.012	&	0.189	&	-1.575	&	-1.483	&	-1.423	&	2.81	&	-0.167	&	0.015	&	0.188	&	-0.294	&	-0.234	&	-0.179	&	4.45	\\\hline
			&	18	&	-2.543	&	0.082	&	0.447	&	-2.697	&	-2.580	&	-2.527	&	0.87	&	-1.590	&	0.090	&	0.454	&	-1.671	&	-1.582	&	-1.552	&	2.08	&	-0.354	&	0.092	&	0.454	&	-0.385	&	-0.323	&	-0.307	&	3.91	\\
			&	45	&	-2.449	&	0.041	&	0.371	&	-2.666	&	-2.542	&	-2.471	&	1.42	&	-1.501	&	0.051	&	0.372	&	-1.644	&	-1.547	&	-1.500	&	2.51	&	-0.270	&	0.052	&	0.363	&	-0.359	&	-0.292	&	-0.257	&	4.22	\\
		10	&	60	&	-2.396	&	0.013	&	0.263	&	-2.647	&	-2.520	&	-2.440	&	1.82	&	-1.451	&	0.023	&	0.263	&	-1.626	&	-1.527	&	-1.470	&	2.77	&	-0.223	&	0.024	&	0.253	&	-0.344	&	-0.276	&	-0.231	&	4.33	\\
			&	72	&	-2.366	&	-0.004	&	0.175	&	-2.634	&	-2.505	&	-2.421	&	2.26	&	-1.422	&	0.007	&	0.177	&	-1.615	&	-1.515	&	-1.453	&	3.1	&	-0.197	&	0.009	&	0.168	&	-0.335	&	-0.266	&	-0.216	&	4.56	\\
			&	84	&	-2.338	&	-0.020	&	0.091	&	-2.623	&	-2.492	&	-2.404	&	2.73	&	-1.396	&	-0.008	&	0.101	&	-1.605	&	-1.504	&	-1.438	&	3.68	&	-0.179	&	-0.003	&	0.102	&	-0.328	&	-0.259	&	-0.206	&	5.23	\\\hline
			&	18	&	-2.404	&	0.006	&	0.296	&	-2.666	&	-2.536	&	-2.454	&	1.67	&	-1.452	&	0.019	&	0.301	&	-1.639	&	-1.538	&	-1.478	&	2.75	&	-0.244	&	0.029	&	0.297	&	-0.361	&	-0.291	&	-0.252	&	4.47	\\
			&	45	&	-2.371	&	-0.013	&	0.195	&	-2.654	&	-2.522	&	-2.434	&	2.14	&	-1.422	&	0.000	&	0.202	&	-1.629	&	-1.526	&	-1.461	&	3.15	&	-0.216	&	0.009	&	0.201	&	-0.353	&	-0.281	&	-0.235	&	4.77	\\
		5	&	60	&	-2.355	&	-0.022	&	0.125	&	-2.646	&	-2.513	&	-2.424	&	2.49	&	-1.407	&	-0.009	&	0.134	&	-1.623	&	-1.519	&	-1.452	&	3.42	&	-0.202	&	0.000	&	0.135	&	-0.347	&	-0.275	&	-0.227	&	4.94	\\
			&	72	&	-2.345	&	-0.027	&	0.081	&	-2.640	&	-2.507	&	-2.417	&	2.84	&	-1.399	&	-0.013	&	0.093	&	-1.618	&	-1.514	&	-1.446	&	3.73	&	-0.194	&	-0.004	&	0.097	&	-0.344	&	-0.271	&	-0.222	&	5.19	\\
			&	84	&	-2.335	&	-0.032	&	0.038	&	-2.635	&	-2.501	&	-2.410	&	3.08	&	-1.391	&	-0.018	&	0.054	&	-1.614	&	-1.510	&	-1.441	&	4.08	&	-0.188	&	-0.008	&	0.062	&	-0.341	&	-0.268	&	-0.218	&	5.63	\\\hline
			&	18	&	-2.357	&	-0.028	&	0.106	&	-2.654	&	-2.521	&	-2.430	&	2.27	&	-1.408	&	-0.013	&	0.120	&	-1.629	&	-1.524	&	-1.456	&	3.26	&	-0.206	&	-0.003	&	0.127	&	-0.353	&	-0.280	&	-0.232	&	4.9	\\
			&	45	&	-2.347	&	-0.034	&	0.056	&	-2.649	&	-2.515	&	-2.423	&	2.63	&	-1.398	&	-0.019	&	0.073	&	-1.624	&	-1.519	&	-1.450	&	3.58	&	-0.197	&	-0.007	&	0.082	&	-0.350	&	-0.276	&	-0.228	&	5.16	\\
		2.5	&	60	&	-2.341	&	-0.036	&	0.032	&	-2.645	&	-2.511	&	-2.419	&	2.89	&	-1.394	&	-0.021	&	0.050	&	-1.622	&	-1.517	&	-1.447	&	3.83	&	-0.193	&	-0.010	&	0.062	&	-0.348	&	-0.274	&	-0.225	&	5.34	\\
			&	72	&	-2.337	&	-0.037	&	0.019	&	-2.643	&	-2.508	&	-2.416	&	3.13	&	-1.391	&	-0.022	&	0.039	&	-1.620	&	-1.515	&	-1.444	&	4.07	&	-0.191	&	-0.011	&	0.051	&	-0.346	&	-0.273	&	-0.223	&	5.56	\\
			&	84	&	-2.334	&	-0.038	&	0.004	&	-2.640	&	-2.505	&	-2.413	&	3.22	&	-1.388	&	-0.023	&	0.025	&	-1.618	&	-1.513	&	-1.442	&	4.24	&	-0.189	&	-0.011	&	0.039	&	-0.344	&	-0.271	&	-0.221	&	5.87	\\\hline
			&	18	&	-2.341	&	-0.040	&	-0.003	&	-2.649	&	-2.514	&	-2.421	&	2.79	&	-1.393	&	-0.025	&	0.020	&	-1.624	&	-1.519	&	-1.448	&	3.72	&	-0.192	&	-0.012	&	0.034	&	-0.349	&	-0.276	&	-0.226	&	5.29	\\
			&	45	&	-2.338	&	-0.040	&	-0.012	&	-2.646	&	-2.512	&	-2.418	&	2.99	&	-1.390	&	-0.025	&	0.011	&	-1.622	&	-1.517	&	-1.446	&	3.93	&	-0.190	&	-0.013	&	0.026	&	-0.348	&	-0.274	&	-0.224	&	5.47	\\
		1	&	60	&	-2.336	&	-0.041	&	-0.016	&	-2.645	&	-2.510	&	-2.417	&	3.13	&	-1.389	&	-0.026	&	0.007	&	-1.621	&	-1.516	&	-1.445	&	4.09	&	-0.189	&	-0.014	&	0.023	&	-0.347	&	-0.273	&	-0.223	&	5.61	\\
			&	72	&	-2.335	&	-0.041	&	-0.018	&	-2.644	&	-2.509	&	-2.416	&	3.21	&	-1.388	&	-0.026	&	0.006	&	-1.621	&	-1.515	&	-1.444	&	4.23	&	-0.189	&	-0.014	&	0.022	&	-0.347	&	-0.273	&	-0.223	&	5.77	\\
			&	84	&	-2.333	&	-0.041	&	-0.020	&	-2.642	&	-2.508	&	-2.414	&	3.23	&	-1.387	&	-0.026	&	0.004	&	-1.619	&	-1.514	&	-1.443	&	4.27	&	-0.188	&	-0.014	&	0.020	&	-0.346	&	-0.272	&	-0.222	&	5.94	\\
\enddata
\end{deluxetable}

\begin{table}
\caption{Disk Masses}\label{Tab_Masses}
\centering
\begin{tabular}{| c c | c c c|} \hline\hline 
n &$\rho_0/10^{-12}$&{\small B1 (13.2M$\odot$)} &  {\small B3 (7.6M$\odot$)} &  {\small B7 (4.2M$\odot$)} \\
  &g\,cm$^{-3}$& \multicolumn{3}{|c|}{M$_d$[M$_{*}$]}\\\hline
  2 &  100  &  2.74$\times$10$^{-8}$  &  1.95$\times$10$^{-8}$  &  9.97$\times$10$^{-9}$\\
2 &   75  &  2.05$\times$10$^{-8}$  &  1.46$\times$10$^{-8}$  &  7.47$\times$10$^{-9}$\\
2 &   50  &  1.37$\times$10$^{-8}$  &  9.74$\times$10$^{-9}$  &  4.98$\times$10$^{-9}$\\
2 &   25  &  6.84$\times$10$^{-9}$  &  4.87$\times$10$^{-9}$  &  2.49$\times$10$^{-9}$\\
2 &   10  &  2.74$\times$10$^{-9}$  &  1.95$\times$10$^{-9}$  &  9.97$\times$10$^{-10}$\\
2 &    5  &  1.37$\times$10$^{-9}$  &  9.74$\times$10$^{-10}$  &  4.98$\times$10$^{-10}$\\
2 &  2.5  &  6.84$\times$10$^{-10}$  &  4.87$\times$10$^{-10}$  &  2.49$\times$10$^{-10}$\\
2 &   1   &  2.74$\times$10$^{-10}$  &  1.95$\times$10$^{-10}$  &  9.97$\times$10$^{-11}$\\\hline
2.5 &  100  &  5.71$\times$10$^{-9}$  &  4.06$\times$10$^{-9}$  &  2.08$\times$10$^{-9}$\\
2.5 &   75  &  4.28$\times$10$^{-9}$  &  3.05$\times$10$^{-9}$  &  1.56$\times$10$^{-9}$\\
2.5 &   50  &  2.85$\times$10$^{-9}$  &  2.03$\times$10$^{-9}$  &  1.04$\times$10$^{-9}$\\
2.5 &   25  &  1.43$\times$10$^{-9}$  &  1.02$\times$10$^{-9}$  &  5.19$\times$10$^{-10}$\\
2.5 &   10  &  5.71$\times$10$^{-10}$  &  4.06$\times$10$^{-10}$  &  2.08$\times$10$^{-10}$\\
2.5 &   5   &  2.85$\times$10$^{-10}$  &  2.03$\times$10$^{-10}$  &  1.04$\times$10$^{-10}$\\
2.5 &  2.5  &  1.43$\times$10$^{-10}$  &  1.02$\times$10$^{-10}$  &  5.19$\times$10$^{-11}$\\
2.5 &    1  &  5.71$\times$10$^{-11}$  &  4.06$\times$10$^{-11}$  &  2.08$\times$10$^{-11}$\\\hline
3 &  100  &  1.41$\times$10$^{-9}$  &  1.01$\times$10$^{-9}$  &  5.15$\times$10$^{-10}$\\
3 &   75  &  1.06$\times$10$^{-9}$  &  7.55$\times$10$^{-10}$  &  3.86$\times$10$^{-10}$\\
3 &   50  &  7.07$\times$10$^{-10}$  &  5.03$\times$10$^{-10}$  &  2.57$\times$10$^{-10}$\\
3 &   25  &  3.54$\times$10$^{-10}$  &  2.52$\times$10$^{-10}$  &  1.29$\times$10$^{-10}$\\
3 &   10  &  1.41$\times$10$^{-10}$  &  1.01$\times$10$^{-10}$  &  5.15$\times$10$^{-11}$\\
3 &    5  &  7.07$\times$10$^{-11}$  &  5.03$\times$10$^{-11}$  &  2.57$\times$10$^{-11}$\\
3 &  2.5  &  3.54$\times$10$^{-11}$  &  2.52$\times$10$^{-11}$  &  1.29$\times$10$^{-11}$\\
3 &    1  &  1.41$\times$10$^{-11}$  &  1.01$\times$10$^{-11}$  &  5.15$\times$10$^{-12}$\\\hline
3.5 &  100  &  4.56$\times$10$^{-10}$  &  3.24$\times$10$^{-10}$  &  1.66$\times$10$^{-10}$\\
3.5 &   75  &  3.42$\times$10$^{-10}$  &  2.43$\times$10$^{-10}$  &  1.24$\times$10$^{-10}$\\
3.5 &   50  &  2.28$\times$10$^{-10}$  &  1.62$\times$10$^{-10}$  &  8.29$\times$10$^{-11}$\\
3.5 &   25  &  1.14$\times$10$^{-10}$  &  8.11$\times$10$^{-11}$  &  4.15$\times$10$^{-11}$\\
3.5 &   10  &  4.56$\times$10$^{-11}$  &  3.24$\times$10$^{-11}$  &  1.66$\times$10$^{-11}$\\
3.5 &    5  &  2.28$\times$10$^{-11}$  &  1.62$\times$10$^{-11}$  &  8.29$\times$10$^{-12}$\\
3.5 &  2.5  &  1.14$\times$10$^{-11}$  &  8.11$\times$10$^{-12}$  &  4.15$\times$10$^{-12}$\\
3.5 &    1  &  4.56$\times$10$^{-12}$  &  3.24$\times$10$^{-12}$  &  1.66$\times$10$^{-12}$\\\hline
4 &  100  &  2.00$\times$10$^{-10}$  &  1.42$\times$10$^{-10}$  &  7.28$\times$10$^{-11}$\\
4 &   75  &  1.50$\times$10$^{-10}$  &  1.07$\times$10$^{-10}$  &  5.46$\times$10$^{-11}$\\
4 &   50  &  1.00$\times$10$^{-10}$  &  7.12$\times$10$^{-11}$  &  3.64$\times$10$^{-11}$\\
4 &   25  &  5.00$\times$10$^{-11}$  &  3.56$\times$10$^{-11}$  &  1.82$\times$10$^{-11}$\\
4 &   10  &  2.00$\times$10$^{-11}$  &  1.42$\times$10$^{-11}$  &  7.28$\times$10$^{-12}$\\
4 &    5  &  1.00$\times$10$^{-11}$  &  7.12$\times$10$^{-12}$  &  3.64$\times$10$^{-12}$\\
4 &  2.5  &  5.00$\times$10$^{-12}$  &  3.56$\times$10$^{-12}$  &  1.82$\times$10$^{-12}$\\
4 &    1  &  2.00$\times$10$^{-12}$  &  1.42$\times$10$^{-12}$  &  7.28$\times$10$^{-13}$\\  \hline\hline
\end{tabular}
\end{table}

\subsection{Allowed $n$ and $\rho_0$ for Be stars}

From Figure~\ref{5Clusters_a}, we see that the late-type Be stars and candidates have W1-W2$<$0.25, whereas the  mid- and early B-type objects have W1-W2$<$0.35. We now explore our models to see which combinations of $\rho_0$ and $n$ give W1-W2 compatible with these limits.

\begin{figure}
\figurenum{11}
\gridline{\fig{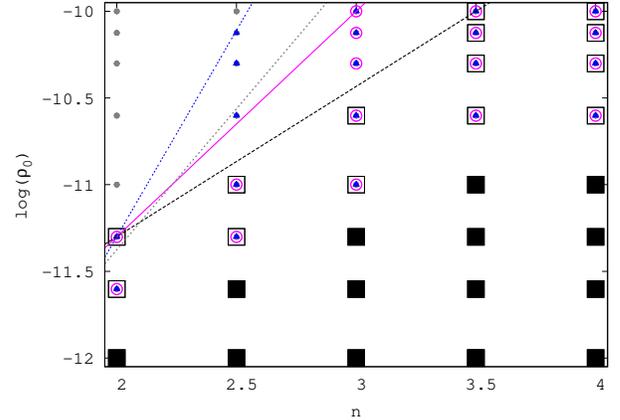}{0.465\textwidth}{}
          }
\caption{$\rho_0$ versus $n$ plot for our disk models for an intermediate inclination angle of 60 degrees. Small gray full circles correspond to models that do not describe the observed W1-W2 colours for any spectral type, blue triangles show the 
combinations of parameters that describe early-type Be stars, magenta open circles correspond to intermediate-type stars and black open squares correspond to late-type Be stars. The black full squares correspond to models with W1-W2$<$0.05. The blue, magenta and black lines describe the limits for each spectral type. The thin gray line indicates the limit of the {\it forbidden zone} as described by \citet{Vieira2017}. \label{rho_n}}
\end{figure}

Figure~\ref{rho_n} shows, for an inclination of 60$^o$, which combinations of the disk density parameters $\rho_0$ and $n$ are consistent with the above limits. The blue, magenta and black lines describe the limits for early, mid, and late spectral types, respectively. There is a clear trend with spectral type: larger $\rho_0$ base densities are allowed for smaller $n$ for earlier spectral types, indicating that earlier spectra types can have more massive disks. Changing the inclination angle changes these limits somewhat, particularly larger inclination angles where the disk is projected against the stellar surface, but the overall trend with spectral type is not modified. The limits here agree with disk density {\it forbidden zone} as described by \citet{Vieira2017} (shown in the figure as a gray line), obtained for a large sample of Be stars with different inclination angles. Finally, the open squares indicate the combinations of parameters that lead to W1-W2$<$0.05, thus appearing as normal B stars.
\begin{figure}
\figurenum{12}
\gridline{\fig{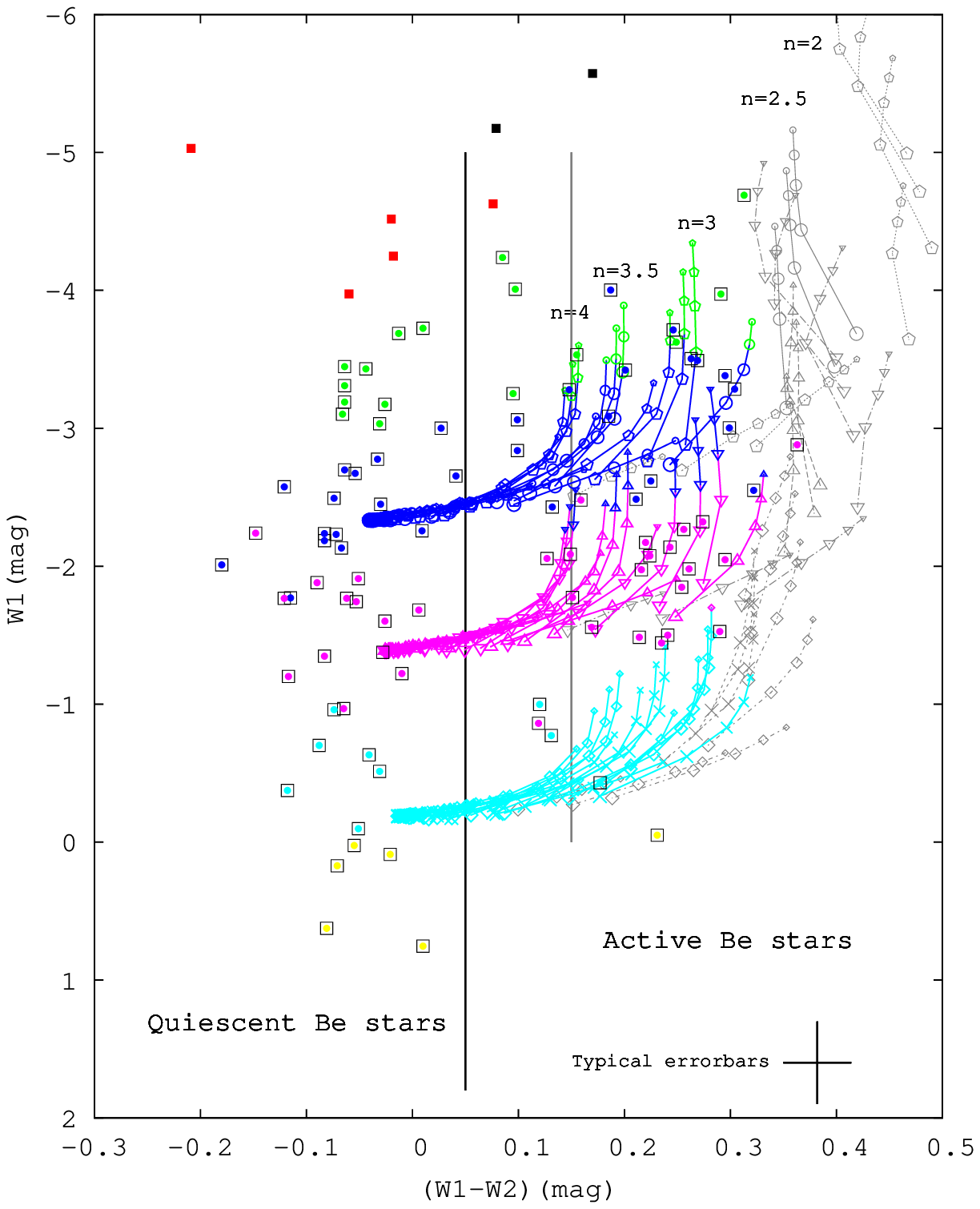}{0.465\textwidth}{}}
\caption{Emission line stars along with synthetic points in the WISE CMD obtained with {\sc bedisk}/{\sc beray}. Full coloured symbols indicate emission-line stars, those surrounded by a black square indicate classical Be stars. The black line separates active Be stars (with W1-W2$\geq$0.05) from quiescent Be stars. Circles, hexagons, triangles, crosses and diamonds indicate models  with different combinations of $n$ and $\rho$ computed with {\sc bedisk}/{\sc beray}. The gray symbols correspond to  combinations of $n$ and $\rho$ that are unlikely to be found in Be stars, following \citet{Vieira2017}. Families of models with different $n$ for early B stars are indicated in the plot. \label{CMD_Models}}
\end{figure}

In Figure~\ref{CMD_Models}, we plot the known Be stars (squares with a black central point) for the five open clusters, together with the synthetic colours and magnitudes in the W1 versus W1-W2 colour magnitude diagram (gray symbols and coloured circles). Once more, the different colours indicate different J magnitude ranges. The gray symbols correspond to models with combinations of $n$ and $\rho_{0}$ that are outside the usual range of parameters found when fitting Be star disk models, called the {\it forbidden region} by \citet{Vieira2017}. Circles and hexagons correspond to models with an early-type central stars, triangles to mid B central stars and crosses and diamonds to late B type stars. The different shapes help to distinguish models with different $n$, and different symbol sizes correspond to different viewing angles. 
We see that there are no Be stars in the region of the CMD occupied by the forbidden models, indicating that 
indeed these kinds of more massive disks do not correspond to MS Be stars. The computed models are seen to nicely cover the region occupied by Be stars and to correctly describe their corresponding $J$ magnitudes. The models with the smallest disks (lowest $\rho_{0}$) are located in the region of {\it normal} B stars or very close to it. With increasing $\rho_{0}$, the effect of inclination becomes more prominent, as discussed in the next section.

\subsection{Effects of the inclination angle}

In Figure~\ref{CMD_Models}, identical disk models but with different viewing inclination angles of 18$^o$ (smallest symbols), 45$^o$, 60$^o$, 72$^o$ and 84$^o$ (largest symbols) are connected with a continuous line.  For a fixed stellar mass, the models with the smallest $n$ have more massive disks for the same $\rho_0$ and have redder colours. 
Interestingly, the analysis of near-IR photometry of Be stars has the potential of being useful to study whether there is a preferential viewing angle in stellar clusters, as has already been found in certain clusters \citep{Corsaro2017}. According to our results, we expect that clusters with preferred pole-on inclinations and Be stars with developed disks, with W1-W2$\geq$0.15, have W1 magnitudes around one magnitude brighter than those clusters with preferential equator-on inclinations. 

\subsection{Mass of the disk and its relation to the H$\alpha$ equivalent width}

\begin{figure}
\figurenum{13}
\gridline{\fig{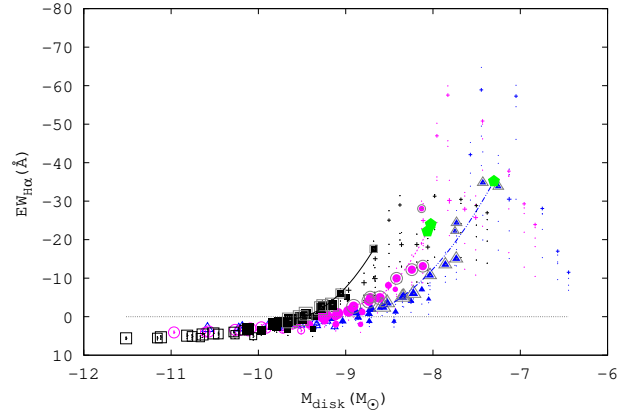}{0.465\textwidth}{}      }
\caption{H$\alpha$ equivalent width versus disk mass. {\bf Small} dots indicate models for all five inclinations given in this article. All bigger symbols correspond to an inclination angle of 60$^o$. Blue triangles show the values for early B type models, magenta circles for intermediate mass and black squares for late B-type stars.  Open symbols indicate forbidden models with W1-W2$<$0.05 and small crosses indicate  models that produce large colours not observed for cluster Be stars. Gray symbols indicate those models with W1-W2$\geq$0.15, and the coloured curves are second order polynomial fits to these points. The gray horizontal line separates emission (negative values) and absorption (positive values) H$\alpha$ lines. The green symbols identify the three stars with available H$\alpha$ spectroscopy.\label{EW_M}}
\end{figure}

By integrating our model disk densities over the volume of the disk to radial distance of 50 stellar radii, we obtain an estimate of the mass of the disk. Table~\ref{Tab_Masses} provides the disk mass for each model in units of the stellar mass (M$_{*}$). The assumed stellar mass for each spectral type is indicated on top of each column, so conversion to solar masses is straightforward. Figure~\ref{EW_M} shows the H$\alpha$ equivalent width (EW) versus the corresponding model disk mass (in M$_{\odot}$) for different spectral types and disk density parameters, all for an inclination angle of 60$^o$. Models with other inclination angles are indicated with dots, for ease of comparison. Small crosses correspond to combinations of parameters that belong to the disk density forbidden region. The colour coding in Figure~\ref{EW_M} is the same as in Figure~\ref{rho_n}: black squares correspond to late B stars, magenta circles to intermediate B stars, and blue triangles to early B stars. The point size is proportional to the value of $n$, the smallest correspond to $n$=2 and the largest, to $n$=4.

We find that the early-type models can have more massive disks and produce larger H$\alpha$ equivalent widths than the later B-type models , in agreement with \citet{Arcos2017}. 
Open symbols indicate objects with W1-W2$<$0.05, the region typically occupied by normal B stars. Interestingly, all these models have positive EW, indicating that these objects do not have a significant disk emission contribution to the photospheric H$\alpha$ line. Full symbols correspond to models with W1-W2 below the forbidden region limit and W1-W2$\geq$0.05. It is remarkable that for a fixed inclination angle and spectral type of the central star, there is a one to one relation between EW and the mass of the disk, particularly for those models with large colour excess. We discuss this later in the context of stars with stable, developed disks. 
The one-to-one relation for the three spectral types and an inclination angle of 60$^o$ is represented by coloured lines (second-order polynomial fits) in Figure~\ref{EW_M}.

We see that some of the models with large $n$ (the largest symbols) and small disk masses are very close to the EW=0 line. Moreover, for late-type Be stars, many of the models with W1-W2$\geq$0.05 have a positive ${H\alpha}$ EW. Interestingly, for late-type B stars, many objects with a small disk mass (smaller than 10$^{-9.5}$M$_{\sun}$ for the inclination of 60$^o$), there is a clear IR excess, indicating the presence of a circumstellar disk, but no significant H$\alpha$: such objects could easily elude the Be star classification.

\begin{figure*}
\figurenum{14}
\gridline{\fig{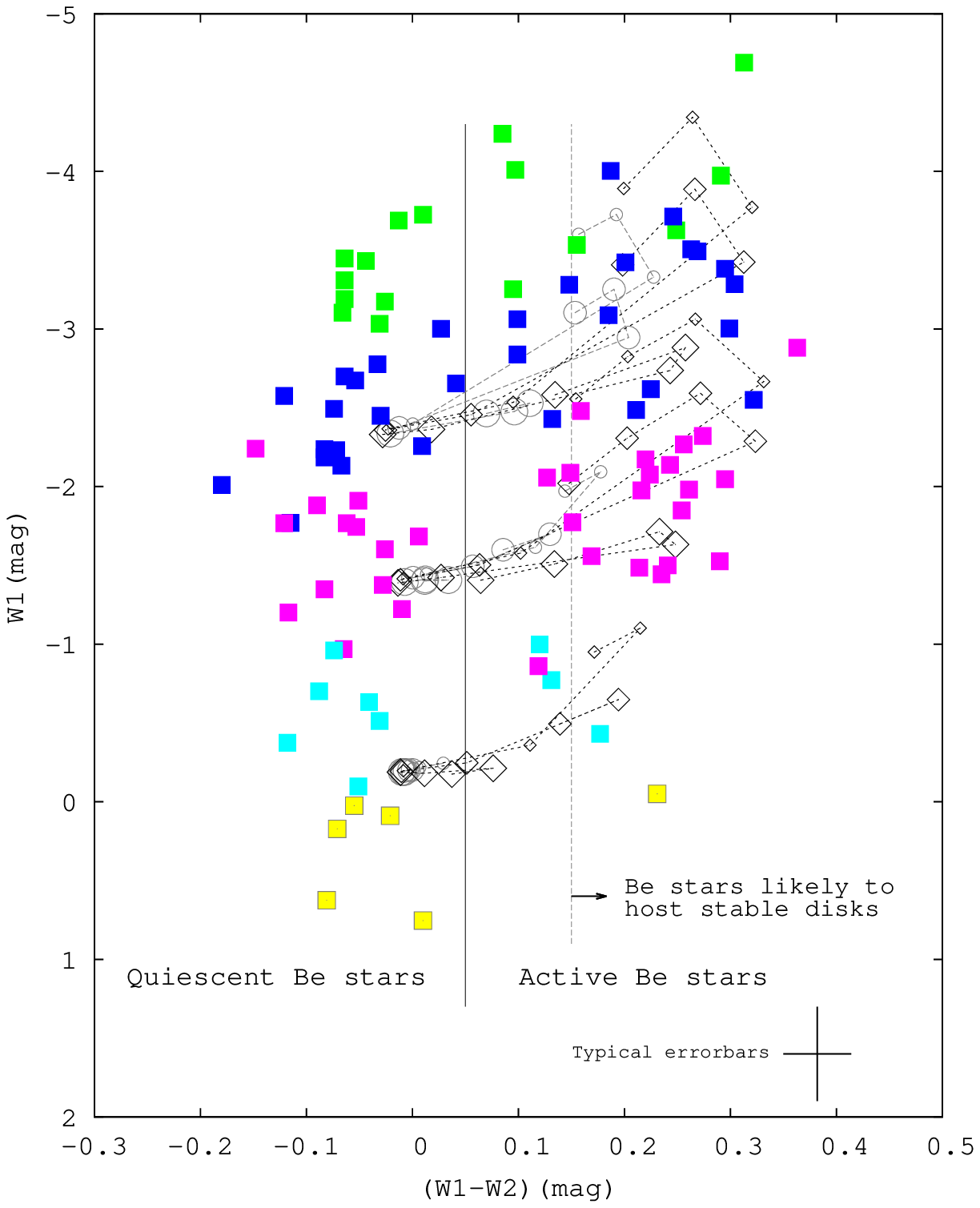}{0.465\textwidth}{(a)}
          \fig{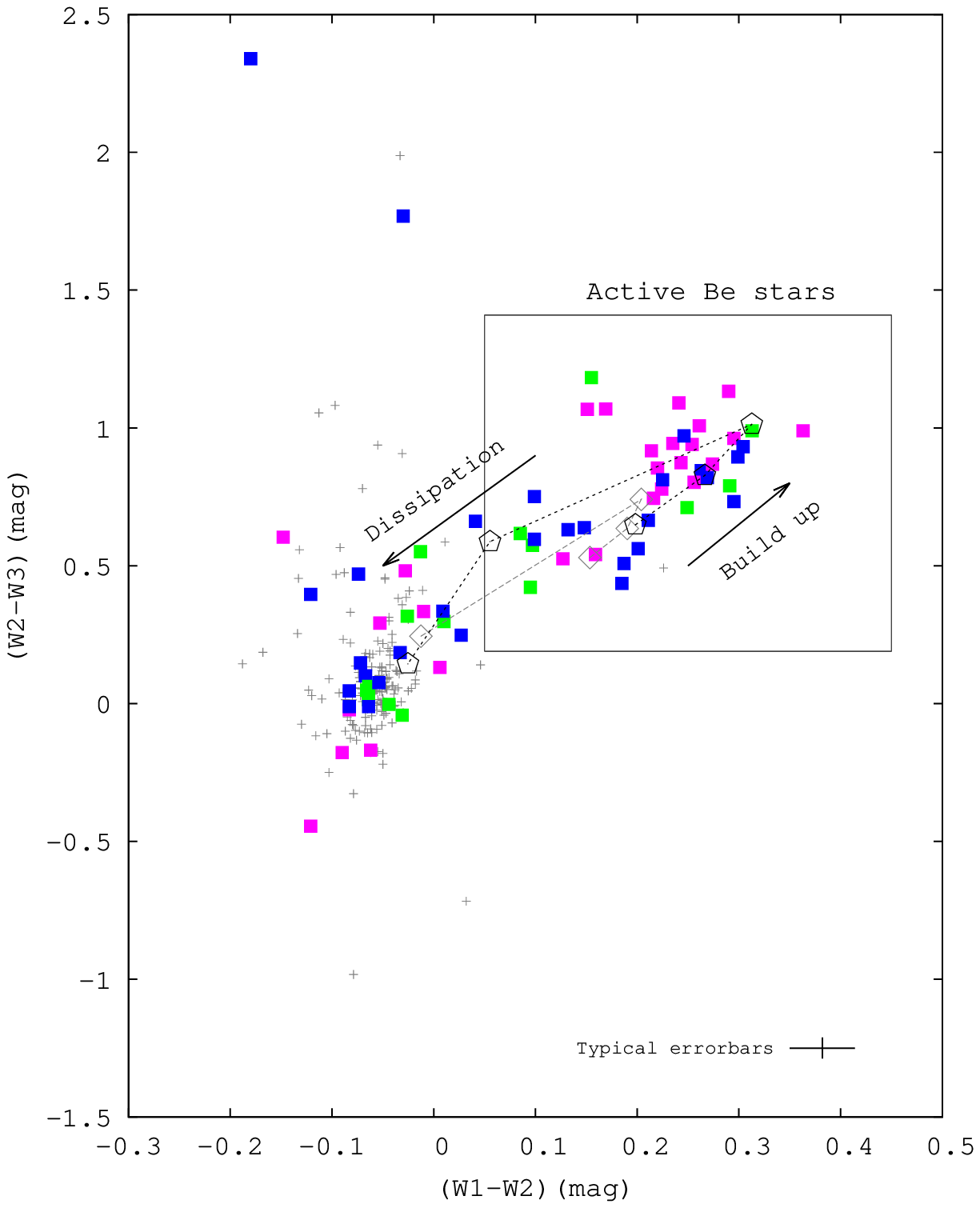}{0.465\textwidth}{(b)}
          }
\caption{Be stars shown together with the modeled loops predicted during disk formation and dissipation phases in the colour-magnitude (a) and colour-colour (b) diagrams. The colour coding is the same as in previous plots. We see in (a) that except for systems with a large inclination angle, the direction of the loop is clockwise. Our modelling shows that active Be stars with W1-W2$\geq$0.15 are likely to have stable developed disks. The loops shown correspond to the three different spectral types and three inclination angles. Panel (b) shows only early- and mid-type B stars in the colour-colour diagram,  and the predicted loop for mid B-type stars with an inclination angle of 60$^o$. In this case, the direction of the loop is counterclockwise. The black crosses indicate all non-Be objects. \label{CMD_loop}}
\end{figure*}

\subsection{Disk growth and dissipation phases}

As suggested by \citet{Vieira2017}, disk growth and dissipation phases can be represented with different combinations of the parameters $n$ and $\rho_0$.  During these formation/dissipation phases, the star is expected to describe a loop in the CMD \citep{Dougherty1994,deWit2006,Sigut2013}.

Figure~\ref{CMD_loop} shows our confirmed Be stars together with models of disk growth and dissipation phases, similar to those presented by \citet{Vieira2017}: first the disk forms and grows in size, represented by a decrease in $n$ from 4 to 3.5 for a constant $\rho_0$, and then subsequently dissipates, represented by a decrease in $\rho_0$ and decreasing $n$. 
For each stellar model, two different values of $\rho_0$ at the beginning of the disk growth were considered in an attempt to mimmick two different disk mass loss rates. 

Models represented by gray circles correspond to a logarithm of the mass loss rates (in M$_{\sun}$yr$^{-1}$) of -9.1 for the B1 star, -10.1 for the B3 star, and -11.1 for the B7 star, as in \citet{Vieira2017}. Models indicated with black diamonds
represent mass loss rates of an order of magnitude larger and are consistent with the mass loss rates predicted by \citet{Granada2013a} from stellar angular momentum loss rates for critically rotating B-type stellar models. The models joined with dashed and dotted lines correspond to disk evolutionary tracks through a formation/dissipation phase for three different inclination angles, 18$^{\circ}$, 60$^{\circ}$ and 84$^{\circ}$. The size of the points increases from pole to equatorial viewing angle.  For most of the models, once the disk appears, the star very quickly reaches a large colour excess and becomes brighter in W1 band.

We see that for the confirmed early and mid Be stars, both sets of disk evolutionary tracks qualitatively describe the location of the stars having an excess in WISE colours. The observed late Be stars are better described by disks with the larger density presented here.  The large number of Be candidates in this spectral type range that have not been confirmed yet (see Figure~\ref{5Clusters_a} (b)) shows that smaller disks may be frequent but hard to detect, but this conclusion requires further investigation.

We can see in Figure~\ref{CMD_loop} that W1-W2$>$0.15 is predicted for models with combinations of $\rho_0$ and $n$ that describe disks being continuously fed by the central star. Therefore, we propose that most Be stars with W1-W2$>$0.15 (observed in particular among mid and early Be stars) host developed, stable disks. Most Be stars with W1-W2$\leq$0.05 are likely in a diskless phase, and objects with W1-W2 between 0.05 and 0.15 are either stars with small mass loss rates or objects with dissipating disks.

Following this scheme, and as mentioned before, 53.1$\%$ of the early Be stars in our sample and 39.4$\%$ of the mid Be stars are in a quiescent or diskless phase. Among active stars 14.3$\%$ of all early Be stars and 9.1$\%$ of all mid Be stars have a dissipating or small disk, and 32.7$\%$ of early Be stars and 51.5$\%$ of mid Be stars have a developed disk. For the overall sample of 80 early and mid Be stars, 46.3$\%$ are in a diskless phase, 12.5$\%$ have a small or dissipating disk, and 41.3$\%$ have a developed stable disk. This last value is consistent with the fraction (37$\%$) of Be stars with long term photometric variability \citep{Labadie2017}.

For objects with W1-W2$\geq$0.15, our modelling shows that the allowed combinations of $\rho$ and $n$ that correspond to stable, developed disks, lead to a tight relation between the corresponding H$\alpha$ equivalent width and the mass of the disk when the inclination angle and the spectral type of the star are fixed. Therefore, in the frame of this simple disk modelling, we suggest that for very stable disks, we could spectroscopically estimate the mass of the disk using the spectral type of the star, an estimate of the viewing inclination (via the morphology of the H$\alpha$ emission line), and the EW$_{\rm H \alpha}$.

 For three Be stars in our sample with W1-W2$>$0.15 and flagged as stable that are likely to host stable developed disks, there is available data of H$\alpha$ spectroscopy from the BeSS catalogue: there is one early Be star BD+60 341 (EW$_{\rm H\alpha}$=-35.2$\AA$), and two mid Be stars: V* V984 Cas (EW$_{\rm H\alpha}$=-24.0), and EM* GGA 93 (EW$_{\rm H\alpha}$=-22.2). The Bess catalogue gathers the information of classical Be stars and Herbig Ae/Be stars, and assembles spectra obtained by professional and amateur astronomers of these stars. For the three objects with available spectra, using the relations shown in Figure \ref{EW_M}, we obtained disk masses (log(M$_{disk}$)) of -7.302, -8.025, -8.064 respectively. The green symbols in Figure \ref{EW_M}, indicate the values derived from observations, and in all three cases correspond to the upper limit of disk mass. 

\section{Conclusions}

Open clusters provide a unique laboratory to study stellar populations and, in particular, the Be stars. The five young open clusters studied in this work, NGC 663, NGC 869, NGC 884, NGC 3766 and NGC 4755, have long been known to host numerous Be  stars that have been broadly studied in the literature. WISE near-IR photometry allows identification of Be stars and the detection of new Be star candidates which could be confirmed spectroscopically. In these clusters, virtually all mid- and early-type Be stars with W1-W2$>$0.05 are known Be stars, which leads us to conclude that in this spectral range, almost all of Be stars in these clusters have been identified. Conversely, many late-type Be stars may not yet have been identified as such. 

{\sc bedisk/beray} models with typical disk density structures derived for Be stars correctly describe the global near-IR photometric characteristics of Be stars in our sample. 

For small and intermediate inclination angles, we obtain that stars with W1-W2$\geq$0.15 have values of $n\leq$3.5 and intermediate values of $\rho_0$. Models with very large disk density do not lead to colours observed in the Be stars of our sample, and the limits we derive are coincident with the disk density ``forbidden region" defined by \citet{Vieira2017}. For models with large inclination angles (nearly equator-on), the near-IR excesses are rather small, particularly when the central star is of a late spectral type.

The location of the mid- and early-type Be stars with fully-developed disks in the CMD, W1-W2$\geq$0.15, requires mass loss rates in agreement with those of \citet{Vieira2017}, obtained for a large sample of observed Be stars, and those of \citet{Granada2013a} predicted from the stellar angular momentum loss rates obtained for critically rotating models. We find for these stars, if the spectral type and inclination of the central star is known, the disk mass can be estimated from the tight relation between ${\rm H \alpha}$ EW and the mass of the disk.

The location of Be stars in WISE CMD and CC diagrams provides a convenient method to separate them into active (Be stars hosting a developed circumstellar disk with W1-W2$\geq$0.05) and quiescent stages (Be stars in a diskless phase with W1-W2$<$0.05). This can be used as a tool to explore the frequency of these different activity states.

 From the analysis of Be stars in five open clusters between 15 and 30 Myrs, if we understand our observed sample as a “snapshot” of the behaviour of these objects, we deduce that half of the time, early Be stars are in an active phase, while mid  Be stars are in an active phase 60\% of the time.  In particular, 34\% of the time, early Be stars seem to host a large, developed disk with W1-W2$>$0.15. For mid Be stars the fraction grows to 51.5\%. 

Among early and mid Be stars, 15\% and 9\% of the objects respectively, have W1-W2 between 0.05 and 0.15, which points towards the finding that the formation of small short lived disks (perhaps from outbursts) is more common among early type stars, in agreement with other authors  \citep[e.g.][]{Hubert1998,Barnsley2013,Labadie2017}.

In the late B-type group there is a considerable number of unconfirmed Be candidates. This is why we consider that the sample of late Be stars is not complete enough to make an analysis as the one performed for early and mid Be stars: a large fraction of objects with small infrared excesses might have disks hard to detect from H$\alpha$ spectroscopy. Near IR observations in the H band such as those presented by \citet{Chojnowski2015, Chojnowski2017}, or the K and L bands (\citep{Lenorzer2002,Mennickent2009,Granada2010, Sabogal2017}), trace the innermost regions of the circumstellar disk, and are more appropriate to study the disks around these late Be stars.

The WISE colours of the sample of active Be stars show that early-type objects have more massive disks than late-type Be stars. Some of the models of late-type Be stars with low mass disks in an active phase (W1-W2$>$0.05) have positive equivalent widths, indicating that many late-type stars hosting a disk might be more difficult to detect spectroscopically. This could explain the large number of late Be candidates in our sample.

Finally, the analysis of near-IR photometry of Be stars in open clusters could be useful to explore whether there is a preferential viewing angle in open stellar clusters (i.e., preferential alignment of the stellar rotation axes), as has already  been  found  in  certain  clusters \citep{Corsaro2017}. Further analysis of WISE photometric observations may prove to be a valuable diagnostics for setting constraints on rotating stellar models. Following this work, we plan to analyze Be stars in open clusters of different ages and metallicities. 

\vspace{5mm}

\acknowledgments
AG acknowledges the Swiss National Science Foundation, Advanced Postdoc Mobility Grant number P300P2$\_$158443. CEJ and TAAS wish to acknowledge support though the Natural Sciences and Engineering Research Council of Canada. This work is sponsored the Swiss National Science Foundation (project number 200020-172505). This publication makes use of data products from the Wide-field Infrared Survey Explorer, which is a joint project of the University of California, Los Angeles, and the Jet Propulsion Laboratory/California Institute of Technology, funded by the National Aeronautics and Space Administration. This work has made use of the BeSS database, operated at LESIA, Observatoire de  Meudon, France: {\it http://basebe.obspm.fr}. We thank Christian Buil, the amateur astronomer who obtained the three H$\alpha$ spectra used in this article and made them available through the BeSS database. We also thank the referee of this paper for his/her careful reading of the manuscript and a very constructive report.

\facility{WISE}
\software{SYCLIST \citep{Georgy2014}, {\sc Bedisk} \citep{Sigut2007}, {\sc Beray} \citep{Sigut2011a}}

\bibliography{./citas.bib}

\end{document}